\documentclass[12pt]{article}
\pdfoutput=1

\usepackage[DIV13]{typearea}
\usepackage{indentfirst}
\usepackage{amsmath}
\usepackage{amsfonts}
\usepackage{mathrsfs}
\usepackage{amssymb}
\usepackage{mathtools}
\usepackage{bbm}
\usepackage{epsfig}
\usepackage{graphicx}
\usepackage{subfigure}
\usepackage{slashed}
\usepackage{multicol}
\usepackage[usenames,dvipsnames]{color}
\usepackage{cite}
\RequirePackage[colorlinks=true,urlcolor=blue,anchorcolor=blue,citecolor=blue,filecolor=blue,
               linkcolor=blue,menucolor=blue,linktocpage=true,pdfproducer=medialab]{hyperref}
\usepackage{cancel}

\usepackage{feynmp}
\makeatletter
\def\endfmffile{%
  \fmfcmd{\p@rcent\space the end.^^J%
          end.^^J%
          endinput;}%
  \if@fmfio
    \immediate\closeout\@outfmf
  \fi
  \ifnum\pdfshellescape=\@ne
    \immediate\write18{mpost \thefmffile}%
  \fi}
\makeatother

\usepackage{lscape}
\usepackage{multirow}
\usepackage{array}
\usepackage{pdfpages}
\usepackage{fancyhdr}
\usepackage{pifont}

\usepackage[left=.9in, right=.9in]{geometry}

\newcommand{\email}[1]{\href{mailto:#1}{\tt #1}}

\numberwithin{equation}{section}
\newcommand{\LL}{\mathscr{L}}
\newcommand{\cG}{\mathscr{G}}
\def\cB{{\cal B}}

\def\cG{{\cal G}}

\def\cJ{{\cal J}}
\def\cM{{\cal M}}
\def\cN{{\cal N}}
\def\cO{{\cal O}}

\def\cT{{\cal T}}

\def\cW{{\cal W}}

\def\cw{c_{\textrm w}}
\def\sw{s_{\textrm w}}

\def\be{\begin{equation}}
\def\ee{\end{equation}}
\def\beq{\begin{equation}}
\def\eeq{\end{equation}}
\def\bc{\begin{center}}
\def\ec{\end{center}}
\def\bea{\begin{eqnarray}}
\def\eea{\end{eqnarray}}

\def\bry{\begin{array}}
\def\ery{\end{array}}

\def\nt{\noindent}

\newcommand{\hc}{\text{h.c.}}
\def\fA{\mbox{{\bf M$_{4+5}$}}}
\def\fB{\mbox{{\bf M$_{4+14}$}}}
\def\oA{\mbox{{\bf M$_{1+5}$}}}
\def\oB{\mbox{{\bf M$_{1+14}$}}}

\def\lra#1{\overset{\text{\scriptsize$\leftrightarrow$}}{#1}}

\def\fourpletL{\Psi_{\textbf{4}L}}
\def\fourpletR{\Psi_{\textbf{4}R}}
\def\fourplet{\Psi_{\bf{4}}}

\def\Barfourplet{\overline{\Psi}_{\bf{4}}}

\def\singletL{\Psi_{\textbf{1}L}}
\def\singletR{\Psi_{\textbf{1}R}}
\def\singlet{\Psi_{\bf{1}}}

\def\5qplet{q_L^{\mathbf{5}}}
\def\qBar5plet{\overline{q}^{\bf{5}}_L}
\def\14qplet{q_L^{\bf{14}}}
\def\q14Barplet{\overline{q}^{\bf{14}}_L}

\def\u5plet{u_R^{\bf 5}}
\def\uBar5plet{\overline{u}^{\bf{5}}_R}
\def\usinglet{u_R^{\bf 1}}

\def\d5plet{d_R^{\bf 5}}
\def\dBar5plet{\overline{d}^{\bf{5}}_R}

\def\Xtt{{X_{\hspace{-0.09em}\mbox{\scriptsize2}\hspace{-0.06em}{\raisebox{0.1em}{\tiny\slash}}\hspace{-0.06em}\mbox{\scriptsize3}}}}
\def\Xft{{X_{\hspace{-0.09em}\mbox{\scriptsize5}\hspace{-0.06em}{\raisebox{0.1em}{\tiny\slash}}\hspace{-0.06em}\mbox{\scriptsize3}}}}
\def\Tt{\widetilde{T}}

\newcommand{\cJmuuichi}{\cJ^{\mu}_{i\chi}}

\newcommand{\cJmuuq}{\cJ^\mu_q}
\newcommand{\cJmuuauchi}{\cJ^{\mu a}_{u\chi}}

\newcommand{\cJmuupsi}{\cJ^\mu_\psi}
\newcommand{\cJmuuqpsi}{\cJ^\mu_{q\psi}}
\newcommand{\cJmuuupsi}{\cJ^\mu_{u\psi}}

\newcommand{\rhochi}{\rho_\chi}

\newcommand{\rhochimuu}{\rho^\mu_\chi}
\newcommand{\rhochimud}{\rho_{\mu\chi}}

\newcommand{\rhoLmu}{\rho^\mu_L}
\newcommand{\rhoRmu}{\rho^\mu_R}

\newcommand{\echimuu}{e^\mu_\chi}
\newcommand{\echimud}{e_{\mu\chi}}

\newcommand{\rhomunuchiu}{\rho^{\mu\nu}_{\chi}}
\newcommand{\rhomunuchid}{\rho_{\mu\nu\chi}}

\newcommand{\BarT}{\overline{T}}

\newcommand{\gL}{g_L}
\newcommand{\gR}{g_R}

\newcommand{\grhochi}{g_{\rho_\chi}}

\newcommand{\grhoL}{\mathit{g}_{\rho_L}}
\newcommand{\grhoR}{\mathit{g}_{\rho_R}}

\newcommand{\mrhochi}{m_{\rho_\chi}}
\newcommand{\mrhoL}{\mathit{m}_{\rho_L}}
\newcommand{\mrhoR}{\mathit{m}_{\rho_R}}

\newcommand{\aichi}{\alpha^\chi_i}

\newcommand{\eg}{e.g.\,\,}
\newcommand{\ie}{i.e.\,\,}

\newcommand{\hhref}[1]{\href{http://arxiv.org/abs/#1}{arXiv:#1}}
\newcommand{\hhep}[1]{\href{http://xxx.lanl.gov/abs/hep-ph/#1}{hep-ph/#1}}

\begin{document}

\begin{titlepage}
\vspace*{-1cm}
\phantom{hep-ph/***} 
\vskip 1cm
\begin{center}
\mathversion{bold}
{\LARGE\bf Modelling top partner-vector resonance phenomenology}\\
\mathversion{normal}
\vskip .3cm
\end{center}
\vskip 0.5  cm
\begin{center}
{\large Juan Yepes}~$^{a)}$ and
{\large Alfonso Zerwekh}~$^{a)}$
\\
\vskip .7cm
{\footnotesize
$^{a)}$~
\emph{Department of Physics and Centro Cient\'{i}fico-Tecnol\'{o}gico de Valpara\'{i}so\\
Universidad T\'{e}cnica Federico Santa Mar\'{i}a, Valpara\'{i}so, Chile}\\
\vskip .1cm
\vskip .3cm
\begin{minipage}[l]{.9\textwidth}
\begin{center} 
\textit{E-mail:} 
\email{juan.yepes@usm.cl},
\email{alfonso.zerwekh@usm.cl}
\end{center}
\end{minipage}
}
\end{center}
\vskip 0.5cm
\begin{abstract}
\nt We have analysed the observable consequences of the interactions of spin-1 resonances coupled to the invariant fermionic currents that arise in an $SO(5)$ Composite Higgs set-up. The phenomenology entailed by such interactions is thoroughly analysed by studying heavy vector resonances production and decay modes in the viable resonance mass range. Additionally, the production of double and single-composite fermion final states has been scanned along the fermion mass scale. Such production is mediated by the SM gauge and Higgs interactions, and also by charged and neutral vector resonances. The coupling between the new fermions and vector resonances induces a sizeable effect in the production rates. We use the recent 13 TeV LHC searches for vector-like quarks to constrain our parameter space. Specifically, we explore the allowed regions by analysing the decays of a heavy vector-like quark in the $Wb$-channel. We conclude that generically the impact of the couplings between the spin-1 and spin-1/2 resonances will substantially reduce the permitted regions, leading us to test the sensitivity of the parametric dependence in the light of exotic matter interactions.
\end{abstract}
\end{titlepage}
\setcounter{footnote}{0}

\tableofcontents

%
%

\newpage

\section{Introduction}

\nt Despite the Higgs discovery at the LHC~\cite{Aad:2012tfa,Chatrchyan:2012xdj}, the long-standing Hierarchy Problem is still pending to be solved. Healing such UV sensitivity of the Higgs mass demands new dynamics beyond the Standard Model (BSM), characterized by an energy close to the electroweak (EW) scale. The stabilization of the EW scale may be achieved by postulating the existence of new particles, that may involve other quantum numbers different to the SM top quark. Exact cancellations among the virtual contributions of the new particles and those from the top quarks will restore the UV insensitivity of the Higgs mass. New physics (NP) states exhibiting this property are generically named as \emph{top partners}. In some BSM frameworks such partners might be scalar quarks, as in the well known supersymmetry, or vector-like fermions~\cite{Aguilar-Saavedra:2013qpa,Panizzi:2014dwa} as in composite Higgs~\cite{Kaplan:1983fs,Kaplan:1983sm,Georgi:1984ef,Banks:1984gj,Georgi:1984af,Dugan:1984hq,Contino:2003ve,Agashe:2004rs,Contino:2010rs,Bellazzini:2014yua} models\footnote{Top partners introduced via larger theory group representations, \eg in a $SO(5)$-multiplet, would imply top partners with different quantum numbers with respect to the SM top quark, that must propagate in the loop for the  EW stabilization~\cite{Contino:2006qr,Matsedonskyi:2012ym,Marzocca:2012zn,Pomarol:2012qf,Redi:2012ha,Panico:2012uw}.} (CHMs). Vector-like quarks are hypothetical spin-1/2 particles whose left- and right-handed components transform in the same way under the SM symmetries. They are the simplest example of coloured fermions still allowed by experimental data, extensively analysed in the literature~\cite{Contino:2006qr,Matsedonskyi:2012ym,Marzocca:2012zn,Pomarol:2012qf,Redi:2012ha,Panico:2012uw}. Complementarily, these models often contain exotic spin-0 and spin-1 resonances at the TeV scale, whose impact on the pseudo Nambu-Goldstone bosons (PNGBs) scattering, and then on the high-energy vector boson scattering, have been thoroughly studied~\cite{Contino:2011np}.

The aim of this work is to explore the low energy implications from the interplay among three matter sectors: elementary states, composite fermionic partners and spin-1 resonances in a $SO(5)$ CHM, within the two-site model approximation~\cite{Contino:2006nn,Panico:2015jxa,Panico:2011pw}. We will consider the interactions of the vector resonances $\rho$, here assumed to be triplets of $SU(2)_L\times SU(2)_R$, with a set of $SO(5)$-invariant fermionic currents. Such invariants cover all the structures built using the elementary sector fields together with the top partners $\Psi$ transforming in the unbroken $SO(4)$. Concretely, we consider an $SO(4)$ fourplet $\fourplet$ and a singlet $\singlet$, encoded through 
\be
\fourplet={1\over \sqrt{2}}\left(\begin{matrix}
i\cB-i\Xft\\
\cB+\Xft\\
i\cT+i\Xtt\\
-\cT+\Xtt
\end{matrix}\right),\qquad\qquad \singlet=\widetilde{\cT}\,.
\label{fourplet-singlet}
\ee

\nt The fourplet $\fourplet$ is decomposable into two doublets $(\cT,\cB)$ and $(\Xft,\Xtt)$ of hypercharge $1/6$ and $7/6$ 
respectively. The former has the same quantum numbers as the SM quark doublet, whilst the latter contains a state of exotic charge $5/3$ plus another top-like quark $\Xtt$. The singlet representation $\singlet$ contains only one exotic top-like state, denoted in here as $\widetilde{\cT}$. On the other hand, the elementary sector will be shaped according to the partial compositeness mechanism instead~\cite{KerenZur:2012fr,Gherghetta:2000qt,Grossman:1999ra,ArkaniHamed:1999dc,Agashe:2004cp}, via the Goldstone symmetry breaking Lagrangian
\be
{\cal L}_{\text{mix}}= \sum_q y\,\bar q\,\cO_q .
\label{UV-mix}
\ee

\nt The  strong sector operator $\cO_q$ transforms in one of the $SO(5)$-representations, determining thus two choices for the elementary sector embeddings: either as a fundamental ${\bf 5}$ or ${\bf 14}$ representation. In the former scenario, both fermion chiralities have elementary representatives coupled to the strong sector through ${\bf 5}$-plets
\be
\hspace*{0.5cm}
\5qplet={1\over \sqrt{2}}\left(
i d_L,\,\,
 d_L,\,\,
i u_L,\,\,
- u_L,\,\,
0
\right)^T,
\qquad\quad
\u5plet = \left(
0,\,\,
0,\,\,
0,\,\,
0,\,\,
u_R
\right)^T,
\label{emb}
\ee

\nt whereas in the latter the right-handed $q$ quark enters as a totally composite state arising itself from the operator ${\cal Q}_q$ at low energies, so that\footnote{In both cases the representations $q_L$ and $u_R$ have the same $X$-charge $2/3$, allowing to reproduce the correct electric charge of the top. The doublet $q^T_L=(u_L,d_L)$ has an isospin $T_R^3=-1/2$, providing thus a protection from large deformations of the $b_L$-couplings~\cite{Agashe:2006at,Mrazek:2011iu}.}
\be
\hspace*{0.5cm}
\14qplet={1\over \sqrt{2}}\left(\begin{matrix}
0 & 0 & 0 & 0 & i d_L\\
0 & 0 & 0 & 0 & d_L\\
0 & 0 & 0 & 0 & i u_L\\
0 & 0 & 0 & 0 & -u_L\\
i d_L & d_L & i u_L & -u_L &0\\
\end{matrix}\right),
\quad
\usinglet\,.
\label{emb}
\ee

\nt  All in all, the previous matter content will frame four models each of them  generically described at the Lagrangian level through
\be
\LL = \LL_{\text{elem}}\,\,+\,\,\,\LL_{\text{comp}}\,\,+\,\,\,\LL_{\text{mix}}.
\label{Lagrangian}
\ee

\nt This picture will be coupled later on to the vector resonances $\rho$, in the triplet representations $\rhoLmu=(\mathbf{3},\mathbf{1})$ and $\rhoRmu =(\mathbf{1},\mathbf{3})$ of $SU(2)_L\times SU(2)_R$, whose description will follow the well known vector formalism~\cite{Ecker:1989yg}. All these Lagrangians will be thoroughly analysed along the text. The coupling to a hypothetical scalar field is postponed for a future analysis~\cite{Norero:2018dfg}. 

Top quark physics in CHMs has been extensively studied~\cite{Marzocca:2012zn,DeSimone:2012fs,ewpt}, with general flavour physics analyses~\cite{Barbieri:2012tu,Redi:2012uj,KerenZur:2012fr} considered in the context of top partner sectors~\cite{Matsedonskyi:2014iha}, whilst spin-0 and spin-1 resonances have been considered in CHMs~\cite{Contino:2011np} with updated analysis~\cite{CarcamoHernandez:2017pei,Hernandez:2015xka}. Our discussion will be based on the previous studies~\cite{Matsedonskyi:2014iha,DeSimone:2012fs}, extended up to a simple approach for effective top partners-vector resonances interplay proposed in~\cite{Yepes:2017pjr}. The phenomenology of this model is thoroughly analysed in here, where the heavy spin-1 resonances production and their decays modes are explored along a viable range for the resonance mass $M_\rho$. Likewise, the production of  double and single-partner final states has been scanned along the partner mass scale $M_\Psi$ in this work, and we provide its dependence on our model parameters. QCD drives the double production, as well as SM gauge, Higgs, and $\rho^0$-mediated processes. The $\rho^\pm$-mediated processes also appear for the single production in the case of charged final states. QCD pair production is completely model-independent, and non-zero model-dependent modifications are induced as soon as extra fermion-vector resonance effects are accounted for. Non-zero contributions arise in all the scenarios, but they are especially large in the fourplet models. Such corrections arise due to the mass mixings and the presence of additional couplings between the composite vectors and fermionic currents. 

Finally, the recent LHC searches for vector-like quarks production in $pp$-collisions at 13 TeV~\cite{Sirunyan:2017pks} have been imposed to exclude regions of the parameter spaces underlying our framework. Specifically, we consider the decay channels $\cT\to Wb$ and $\widetilde{\cT}\to Wb$ to put bounds on our parameter space according with the latest experimental limits. Generically, the impact of the extra fermion-vector resonance couplings treated here will substantially reduce the permitted regions, leading us to roughly estimate the sensitivity of the parametric dependence in the shed light of new exotic matter interactions.

This manuscript is divided in: introduction of the PNGBs for the assumed CHM, vector resonance sector and its generic interplay with the elementary-composite sector in Section~\ref{Interplay-model}. Heavy resonances production and their decays are analysed in Section~\ref{Resonance-production-decay}. Top partners production mechanism are introduced in Section~\ref{Top-partner-production} and discussed in detail in~\ref{Double-Partner-production}-\ref{Single-Partner-production}. The latest LHC searches for vector-like quark production are translated into parameter spaces associated to our models in Section~\ref{Some-parameter-spaces}. The impact of the additional fermion-vector resonance interactions is thoroughly studied along the text. The concluding summary is presented in Section~\ref{Summary}.

\section{Assumptions and set-up}
\label{Interplay-model}

\nt One matter sector of our framework is a \emph{composite sector}, entailing a composite Higgs boson and other composite resonances. The CCWZ formalism~\cite{ccwz} postulates the Higgs as a PNGB of the minimal global symmetry $\cG=SO(5)$~\cite{Agashe:2004rs} which is spontaneously broken to $SO(4)$ by the strong sector at the scale $f$. Four massless PNGBs are generated, yielding thus an $SU(2)_L$ Higgs doublet\footnote{Hence the Higgs is exactly massless unless the strong sector is coupled to some source of an explicit $\cal G$-breaking.}. An additional $U(1)_X$ factor is introduced in order to reproduce the proper SM hypercharge $Y=T_R^3+X$, then $\cG=SO(5)\times U(1)_X$. The PNGBs enter through the  $5\times 5$ Goldstone matrix
\be
U=\exp\left[i \frac{\sqrt{2}}{f}\,\Pi^i\,T^i\right]=\,
\left(\begin{array}{ccccc}
& & \vspace{-3mm}& & \\
 & \mathbb{I}_{3} & & &  \\
  & & \vspace{-3mm}& &  \\ 
   & & & \cos \frac{h + \langle h \rangle}{f} & \sin \frac{h + \langle h \rangle}{f} \\
    & & & -\sin \frac{h + \langle h \rangle}{f} & \cos \frac{h + \langle h \rangle}{f} 
\end{array}\right)\,,
\label{GB-matrix}
\ee

\nt where $T^i$ are the coset $SO(5)/SO(4)$-generators, whilst $\Pi^i$ and $f$ are the PNGB fields and the decay constant respectively, all them defined in~Appendix~\ref{CCWZ}.

Additionally, the \emph{elementary sector}, containing copies of all the SM fields except for the Higgs transforming under the SM gauge symmetry group ${\cal G}_{\text{SM}} \subset {\cal G}$. This sector is not $\cal G$ invariant, therefore the one-loop effective potential triggered by the elementary-composite interactions allows the Higgs to pick a mass, fixing thus its vacuum expectation value (VEV) in a ${\cal G}_{\text{SM}}$-breaking direction. The unbroken $SO(4)\times U(1)_X$ contains the SM  symmetry ${\cal G}_{\text{SM}}=SU(2)_L \times U(1)_Y$ whose breaking will be triggered via a non-zero Higgs VEV $\langle h \rangle \simeq v=246$~GeV, measuring together with the $SO(5)$ breaking scale $f$ the degree of tuning of the scalar potential through the ratio~\cite{Agashe:2004rs}
\be
\xi=\frac{v^2}{f^2}.
\label{Xi}
\ee

\nt Generically, the value of $f$ must be large to suppress NP effects, but not too far from $v$ to maintain a tolerable tuning. Since $\xi$ controls low energies SM departures, then it cannot be too large. Electroweak precision tests suggest $\xi \simeq 0.2$ or $\xi \simeq 0.1$ which corresponds to $f\simeq 550$ GeV and $f\simeq 800$ GeV, and also $\xi \simeq 0.25$ ($f\simeq 490$ GeV) as it is near the upper bound $\xi \simeq 0.30$ ($f\simeq 470$ GeV) allowed by the EWPT parameters~\cite{Agashe:2005dk}. More stringent constraints on $\xi$ have been reported previously, following the current 95\% combined limit from direct production of either the charged $\rho^\pm$ or the neutral $\rho^0$ at the LHC~\cite{Contino:2013gna}. Those limits\footnote{Direct vector resonance searches could merely imply a higher $g_\rho$, as the vector resonances contributions would enter through powers of $g_\rho$. In that sense, direct vector resonance should not set bounds on $\xi$. Nonetheless, in~\cite{Contino:2013gna} it has been fixed $m_\rho/(g_\rho f) =1$ in their analysis, and therefore via the relation $\xi=v^2/f^2$, it is possible to write $g_\rho$ as $g_\rho=m_\rho\sqrt{\xi}/v $.} allow $\xi\sim 0.02$, or even smaller, for a vector resonance mass $M_\rho\sim 2$ TeV. In fact, tree-level contributions to $\Delta \hat S= M_W^2/M_\rho^2$ from the $\rho$ exchange~\cite{Contino:2011np} and the 1-loop IR effect from the modified Higgs  couplings, it is possible to exclude at $95\%$ the region $\xi\gtrsim 0.03$, with $\xi$ tending to $\sim 0.02$ in the infinite $\rho$-mass case. Nonetheless, slight modifications to the EW parameter $\hat T$  shift and relax the 95\% exclusion boundary in such a manner that values as $\xi \simeq 0.30$ are achieved~\cite{Contino:2013gna}. For the present work we will test $\xi=\{0.1,\,0.2\}$, as they are compatible with the latter EWPT bounds, and with the vector resonance direct production bounds at LHC. In addition, those values are inside the domain of validity of the scenario, $g_\rho<4\pi$ and they will be assumed henceforth. 

In this work we  will cover, in a two-site model approximation~\cite{Contino:2006nn,Panico:2015jxa,Panico:2011pw},  all the possible couplings emerging from the interplay among the top partners sector and the composite operators sourced by the strong regime.  The $SO(5)$ invariance will fix the interactions of the following Lagrangian
\be
\LL_{\text{int}}=\LL_{\bf M}\,\,+\,\,\sum_{\chi=L,R}\left(\LL_{\rho_\chi}\,\,+\,\,\LL_{\bf M\,+\,\rho_\chi}\,\,+\,\,\dots\right)
\label{Interplay} 
\ee

\nt with $\bf M$ labelling each one of the models arising from the assumed fermionic matter content
\be
\bf M=\bf M_{\Psi +q}=\{\fA,\,\fB,\,\oA,\,\oB\}.
\label{Models}
\ee

\nt $\LL_{\bf M}$ is generically encoded by~\eqref{Lagrangian}, whilst $\LL_{\rhochi}$ is 
\be
\LL_{\rho_\chi}=-\frac{1}{4\,\grhochi^2}\,\rhomunuchiu\rhomunuchid\,\,\,\,+\,\,\,\,\frac{\mrhochi^2}{2\,\grhochi^2}\left(\rhochimuu-\echimuu\right)^2
\label{rho-Lagrangian}
\ee

\nt with the notation $\chi=L,\,R$ and the internal sum over the $SO(4)$ unbroken generators indices $T^a_\chi$ (defined in~\ref{CCWZ}) is assumed. The third Lagrangian in~\eqref{Interplay} encodes fermion currents coupled to the spin-1 resonances completely\footnote{Fermions-vector couplings have been thoroughly explored in the context of CHM and its holographic realizations, as well as in RS-GIM mechanism. Such couplings are the ground for the flavor bounds on CHM through the partial compositeness~\cite{KerenZur:2012fr}, the which allows to write $\Delta F = 1$ operators, some of them giving rise to fermions-vector couplings properly weighted by quantities measuring the compositeness of the elementary fields. In addition, in RS-GIM mechanism~\cite{Csaki:2008zd,Agashe:2004cp} those couplings emerge from the kinetic Lagrangian due to the mixing between the zero and Kaluza-Klein states of the $Z$ gauge field after EWSB. Finally, all the holographic realizations of the CHM naturally account for fermion-vector couplings (see~\cite{Agashe:2004rs,Medina:2007hz} and references therein). } provided in~\cite{Yepes:2017pjr}, and generically defined as
\be
\LL_{\bf M\,+\,\rho_\chi}=\frac{1}{\sqrt{2}}\,\aichi\,\cJmuuichi\left(\rhochimud-\echimud\right)\,\,+\,\,\hc
\,,
\label{Currents-L-R}
\ee

\nt with an implicit summation over the index $i$ spanning over all the possible currents and tensors that can be built upon the elementary $q$, top partner $\psi$ and elementary-composite pair $q\psi$ and $u\psi$, \eg $i$ can denote the set $i=\{q,\,\psi,\,q\psi,\,u\psi\}$. Generic coefficients $\aichi$ have been introduced and are correspondingly weighting each one of the fermion currents defined later on. The dots in~\eqref{Interplay} might account for higher dimensional operators (GB-scale suppressed), e.g.  2nd rank tensors made out of fermions and coupled to the resonance strength field, yielding contributions for the electric dipole moments at low energies (see~\cite{Yepes:2017pjr} for more details). Such operators have been  disregarded in here. In the next sections all the Lagrangians are explicitly provided.

\subsection{$\fA$ and $\oA$ coupled to $\rho$}

\nt The leading order Lagrangian corresponding to $\bf{5}$-elementary fermions is given by the kinetic terms
\be
\LL_{\text{elem}}= i\,\overline{q}_L \slashed{D}\,q_L\,\,+\,\,i\,\overline{u}_R\slashed{D}\,u_R,
\label{fA-oA-elem}
\ee

\nt whereas both of the top partners $\fourplet$ and $\singlet$ are introduced in $\LL_{\text{comp}}$~\eqref{Lagrangian} through the parametrization~\cite{DeSimone:2012fs} as
\be
\begin{aligned}
\LL_{\text{comp}}= i\,\Barfourplet\slashed{\nabla}\fourplet - M_{\bf{4}}\,\Barfourplet \fourplet\,+\,\left(\fourplet \leftrightarrow\singlet\right)\,+\,\frac{f^2}{4}d^2\,+\,\left(i\,c_{41}\, (\Barfourplet)^i \gamma^\mu d_\mu^i \singlet + {\rm h.c.}\right)
\label{fA-oA-comp}
\end{aligned}
\ee

\nt with $\nabla$ standing for $\nabla=\slashed{D}+i\slashed{e}$. Goldstone bosons kinetic terms are contained at the $d^2$-term, while the coefficient $c_{41}$ controls the strength of the interplaying fourplet-singlet partner term, and it is is expected to be order one by power counting~\cite{Giudice:2007fh}. The covariant derivatives through~\eqref{fA-oA-elem}-\eqref{fA-oA-comp}, together with the $d$ and $e$-symbols are defined in~\ref{CCWZ}. Finally, the mass terms mixing the elementary and top partners are described via 
\be
\begin{aligned}
\LL_{\text{mix}} =& y_L f \left(\qBar5plet\,U\right)_i \,\left(\fourpletR\right)^{i}\,\,+\,\,y_R f \left(\uBar5plet\,U\right)_i \,\left(\fourpletL\right)^{i}\,\,+\,\,\hc\,\,+\,\,,\\[5mm]
& + \tilde{y}_L f \left(\qBar5plet\,U\right)_5\singletR\,\,+\,\,\tilde{y}_R f \left(\uBar5plet\,U\right)_5\singletL\,\,+\,\,\hc
\label{fA-oA-mix}
\end{aligned}
\ee

\nt Suitable $U$ insertions have been done in order to guarantee the non-linear $SO(5)$ invariance. The small mixings $y_\chi$ and $\tilde{y}_\chi$ trigger the Goldstone symmetry breaking,  providing thus a proper low Higgs mass. The latter Lagrangian entails partially composite $u^5_R$ and it gives rise to quark mass terms as well as trilinear couplings contributing to the single production of top partners. Their mass spectrum, couplings, implied phenomenology, production mechanisms and relevant decay channels at LHC searches, are thoroughly analysed in~\cite{DeSimone:2012fs} for the case of a totally composite top quark $t_R$.

Altogether, the leading order composite and mixing Lagrangians contain seven parameters $\{M_{\bf{4}},\,M_{\bf{1}},\,c_{41},\, y_L,\, y_R,\, \tilde{y}_L,\, \tilde{y}_R\}$, aside from the Goldstone decay constant $f$. Six of them are arranged to reproduce the correct top mass plus the extra partner masses $\{m_{\Xft},\,m_{\Xtt},\,m_{\cT},\,m_{\cB},\,m_{\widetilde{\cT}}\}$. Their expressions are reported in Appendix~\ref{Physical-fermion-masses}. 

\nt The set of fermion currents  constructable for both of these models, firstly provided in~\cite{Yepes:2017pjr}, are listed in Table~\ref{Fermion-currents-set} (left column). It is worth to comment that no currents built upon elementary right handed quarks are allowed for these models as the current $\cJmuuauchi =\,\uBar5plet\,\,\gamma^\mu\,\BarT^a_\chi\,\u5plet$ turns out to be vanishing, with the definition $\BarT^a_\chi\,\equiv\,U\,T^a_\chi\,U^\dagger$. Check~\cite{Yepes:2017pjr} for more details on this and related issues concerning heavy vector resonances, their equation of motions, as well as analogous stuff for the top partner fields.
\begin{table}
\centering
\small{
\hspace*{-3mm}
\renewcommand{\arraystretch}{1.0}
\begin{tabular}{c||c}
\hline\hline
\\[-3mm]
$\fA$ & $\fB$
\\[0.5mm]
\hline\hline
\\[-2mm]
$\begin{array}{l}  
\cJmuuq=\,\qBar5plet\,\,\gamma^\mu\,\BarT\,\,\5qplet\\
\\[-1mm]
\cJmuupsi =\,\Barfourplet\,\gamma^\mu\,\tau\,\,\fourplet
\\[3mm]
\cJmuuupsi =\,\left(\uBar5plet\,\BarT\,U\right)_j\gamma^\mu\left(\fourpletR\right)^j
\\[3mm]
\cJmuuqpsi =\,\left(\qBar5plet\,\BarT\,U\right)_j\gamma^\mu\left(\fourpletL\right)^j\\[3mm]
\end{array}$  &  
$\begin{array}{l}  
\cJmuuq =\,\left(U^T\,\q14Barplet\,U\,T\right)_{5\,j}\,\gamma^\mu\,\left(U^T\,\14qplet\,U\right)_{j\,5}\\
\\[-1mm]
\cJmuupsi =\,\Barfourplet\,\gamma^\mu\,\tau\,\,\fourplet\\[4mm]
\cJmuuqpsi =\,\left(U^T\,\q14Barplet\,U\,T\,\right)_{5\,j}\gamma^\mu\,\left(\fourpletL\right)^j
\end{array}$\\[2mm]  
\hline\hline
\\[-3mm]
$\oA$ & $\oB$
\\[0.5mm]
\hline\hline
\\   
$\begin{array}{l}  
\cJmuuq =\,\qBar5plet\,\,\gamma^\mu\,\BarT\,\,\5qplet
\\[5mm]
\cJmuuqpsi =\,\left(\qBar5plet\,U\right)\,\gamma^\mu\,\singletL\\[4mm]
\end{array}$  &  
$\begin{array}{l}  
\\[-6mm]
\cJmuuq =\,\left(U^T\,\q14Barplet\,U\,T\right)_{5\,j}\,\gamma^\mu\,\left(U^T\,\14qplet\,U\right)_{j\,5}
\end{array}$
\\[5mm]  
\hline \hline
\end{tabular}
\caption{\sf Currents for all the models in a two-site model approximation. Definition $\BarT\,\equiv\,U\,T\,U^\dagger$ is involved through the currents, with $T$ standing for the $SO(4) \simeq SU(2)_L \times SU(2)_R$ unbroken generators $T^a_\chi$ ($\chi= L,\,R$ and $a = 1,2,3$), defined in Appendix~\ref{CCWZ} together with the matrices $\tau^a$ in~\eqref{taus}, introduced in order to keep the invariance under global $SO(5)$ transformations. }  
\label{Fermion-currents-set}
}
\end{table}

\subsection{$\fB$ and $\oB$ coupled to $\rho$}

\nt The elementary kinetic Lagrangian corresponding to this model is straightforwardly written
\be
\LL_{\text{elem}}= i\,\overline{q}_L \slashed{D}\,q_L\,,
\label{fB-oB-mass}
\ee

\nt  whereas the composite counterpart is reshuffled as
\be
\begin{aligned}
\LL_{\text{comp}}\quad\rightarrow\quad\LL_{\text{comp}}+i\,\overline{u}_R \slashed{D}\,u_R\,+\,\left(i\,c_{41}\, (\Barfourplet)^i \gamma^\mu d_\mu^i \singlet\,+\,i\,c_{4u}\, (\Barfourplet)^i \gamma^\mu d_\mu^i u_R + \hc\right),
\label{fB-oB-comp}
\end{aligned}
\ee

\nt where $\LL_{\text{comp}}$ corresponds to the strong sector Lagrangian of~\eqref{fA-oA-comp} augmented by the those terms mixing the fourplet $\fourplet$ with the singlet $\singlet$ and the totally composite $u_R$ through the coefficients $c_{41}$ and $c_{4u}$ respectively. The elementary and top partners sector are mixed via
\be
\begin{aligned}
\LL_{\text{mix}}&= y_L\,f\left(U^t\,\q14Barplet\,U\right)_{i\,5} \,\left(\fourpletR\right)^{i}\,\,+\,\,\tilde{y}_L\,f\left(U^t\,\q14Barplet\,U\right)_{5\,5} \,\singletR\,\,+\,\,y_R\,f\left(U^t\,\q14Barplet\,U\right)_{5\,5} \,\usinglet\,\,+\,\,\hc
\label{fB-oB-mix}
\end{aligned}
\ee

\nt This case also involves seven parameters $\{M_{\bf{4}},\,M_{\bf{1}},\,c_{41},\, c_{4u},\,y_L,\, y_R,\,\tilde{y}_L\}$, five of them are arranged  to reproduce the correct top mass, plus extra four partner masses as the degeneracy $m_{\Xft}=m_{\Xtt}$ is implied and also manifested at the previous two models. Notice that a direct mixing coupling $u_R$ and $\singlet$ has been removed by a field redefinition. Table~\ref{Fermion-currents-set} lists the associated fermion currents (right column). 

As we mentioned before, the assumption of spin-1 resonances brings us a mass scale $m_\rho$ below the cut-off of the theory at $\Lambda=4\pi f$, entailing thus the coupling $1< g_\rho < 4\pi$. Likewise, the top partner mass scales $M_{\bf{4(1)}}$, assumed here such that $M_{\bf{4(1)}} < m_{\rho}$, also brings the couplings $g_{\bf{4(1)}}$. Hereinafter the linking relations
\be
\grhochi\equiv \frac{\mrhochi}{f},\qquad\qquad g_{\bf{4(1)}}\equiv\frac{M_{\bf{4(1)}}}{f}
\label{grho-gpartner}
\ee

\nt will be used throughout. As it is commonly argued in the literature, the ranges 500~GeV$\lesssim M_{\bf{4(1)}} \lesssim 1.5$~TeV and $1 \lesssim g_{\bf{4(1)}}\lesssim 3$ are the most favoured by concrete models (see~\cite{DeSimone:2012fs} and references therein).

\section{Heavy spin-1 production and decays}
\label{Resonance-production-decay}

\nt Concerning the vector resonance production, the role of spin-0 and spin-1 resonances on the PNGBs scattering were studied in~\cite{Contino:2011np}. Their experimental searches~\cite{ATLAS:2013jma} were explored for $\xi=0.1$ in~\cite{Contino:2013gna,Pappadopulo:2014qza,Greco:2014aza}. Recently, the impact of heavy triplet resonances at the LHC in the final states $l^+l^-$ and $l\nu_l$ ($l=e,\mu$), $\tau^+\tau^-$, $jj$, $t\bar{t}$ as well as on the gauge and gauge-Higgs channels $WZ$, $WW$, $WH$ and $ZH$, has been analysed (see~\cite{CarcamoHernandez:2017pei,Shu:2016exh,Shu:2015cxm} and references therein), constraining the vector resonance mass in the range 2.1-3 TeV. The latest searches~\cite{Khachatryan:2016cfx} for heavy resonances decaying into a vector boson and a Higgs boson in final states with charged leptons, neutrinos and $b$ quarks have excluded resonance masses less than 2 TeV at 95\% confidence level. In order to explore the feasibility and potentiality of our scenarios, a broader mass range will be explored in here\footnote{Previous analysis of the relevant spin-1 decay channels in a CHM, based on the $SO(5)/SO(4)$ pattern with top partners in the fundamental of $SO(5)$, have been done in~\cite{Barducci:2012kk}. We will face similar treatment here, but analysing deeper the departures induced by our extra fermion-vector resonance couplings along several spin-1 decay channels.}. At the Lagrangian level, the vector resonance production is induced by the effective charged-neutral interactions
\be
\LL_{ud \rho^\pm_i}= -\frac{1}{\sqrt{2}}\,{\bar u}\,\slashed{\rho}^+_i\left(g_{u_L d_L \rho^+_i}\,P_L+ g_{u_R d_R \rho^+_i}\,P_R\right)d\,\,+\,\,\hc,
\label{rho-charged-currents}
\ee
\be
\hspace*{-1mm}
\LL_{ff\rho^0_i}= \sum_{f=u,d}{\bar f}\,\slashed{\rho}^0_i\left(g_{f_L f_L \rho^0_i}\,P_L+ g_{f_R f_R \rho^0_i}\,P_R\right)f.
\label{rho-neutral-currents}
\ee

\begin{figure}
\begin{center}
\hspace*{-0.8cm}
\vspace*{0.5cm}
\includegraphics[scale=0.46]{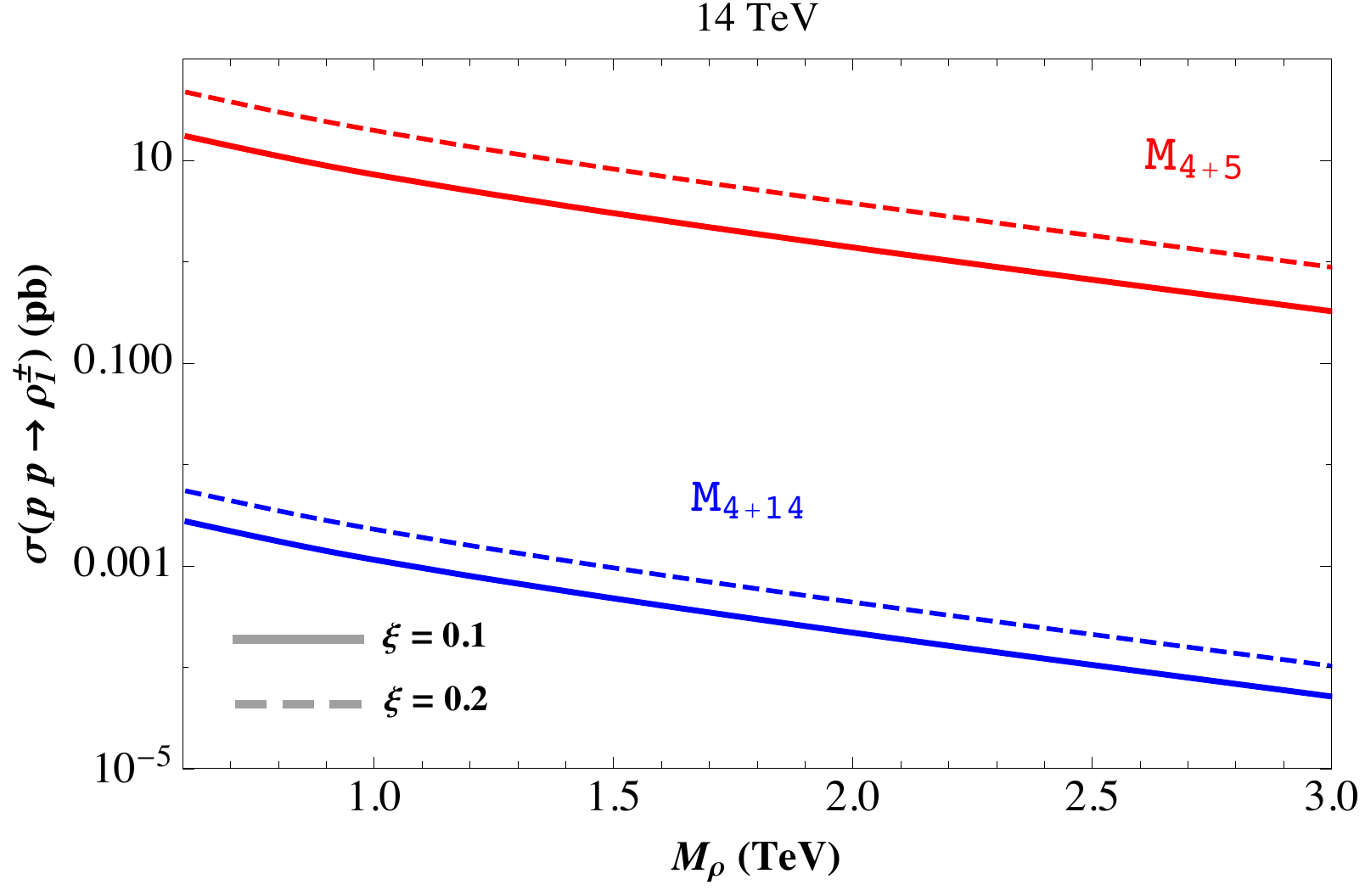}
\hspace*{0.4cm}
\includegraphics[scale=0.46]{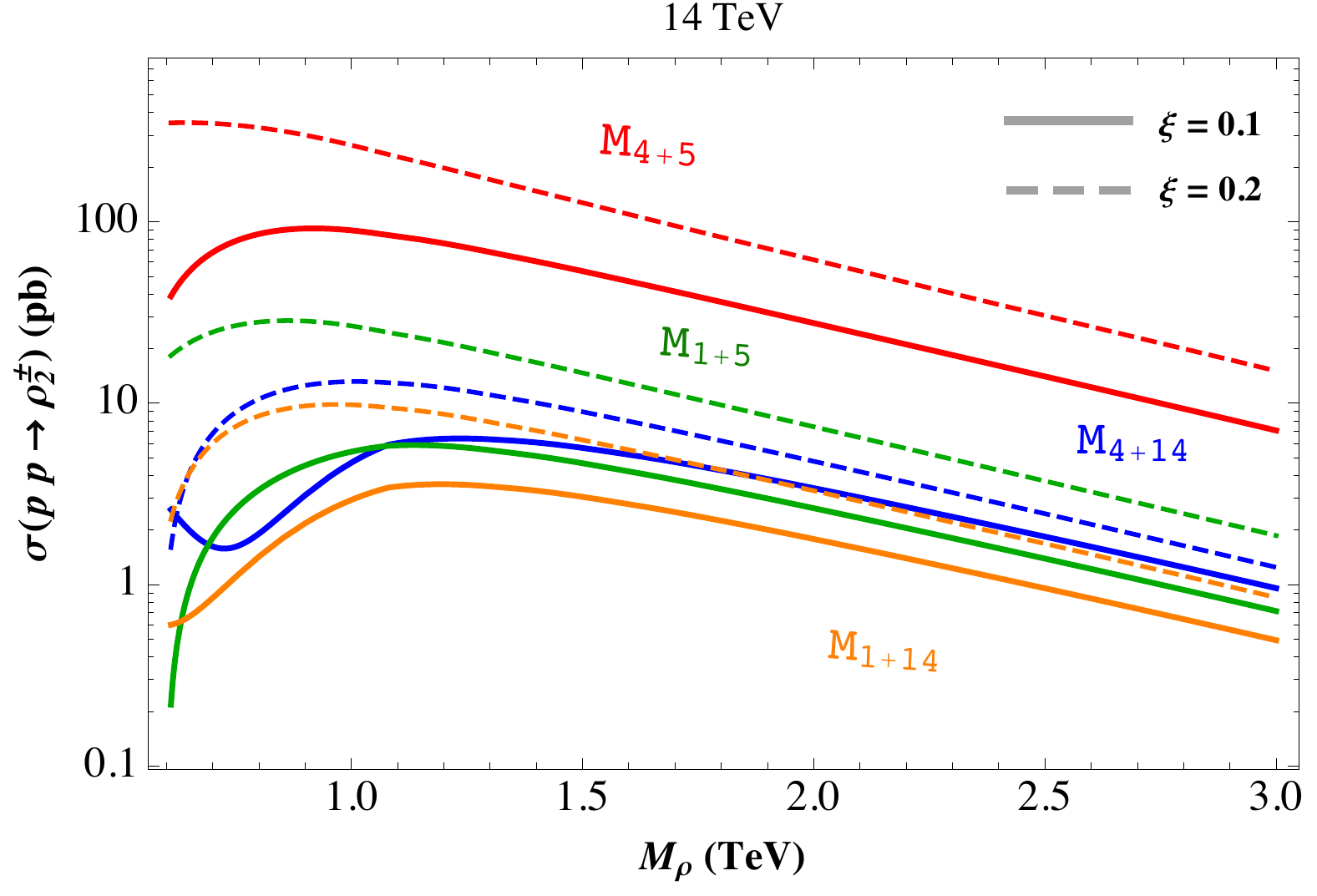}
\vspace*{0.5cm}
\hspace*{-0.5cm}
\includegraphics[scale=0.44]{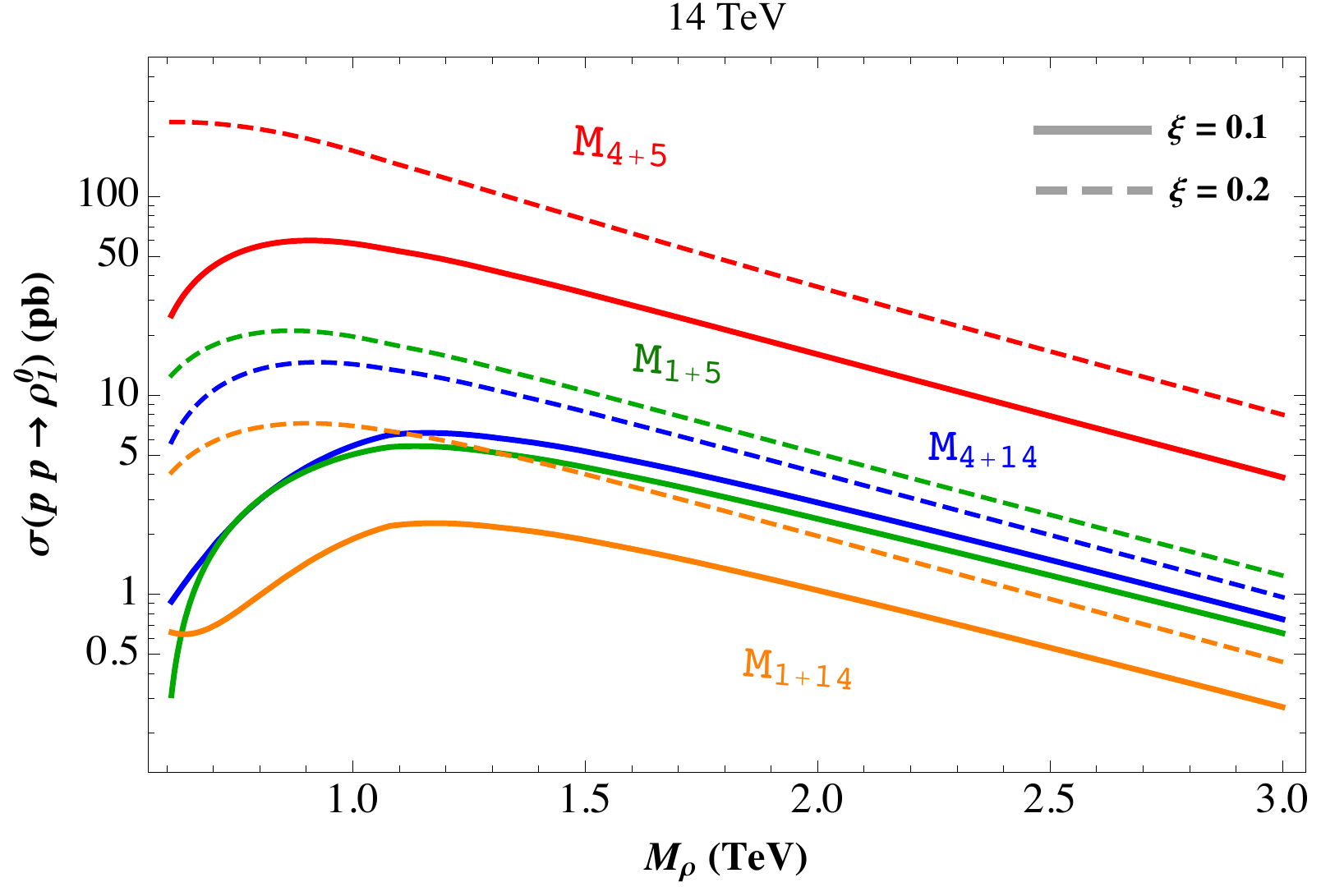}
\hspace*{0.7cm}
\includegraphics[scale=0.44]{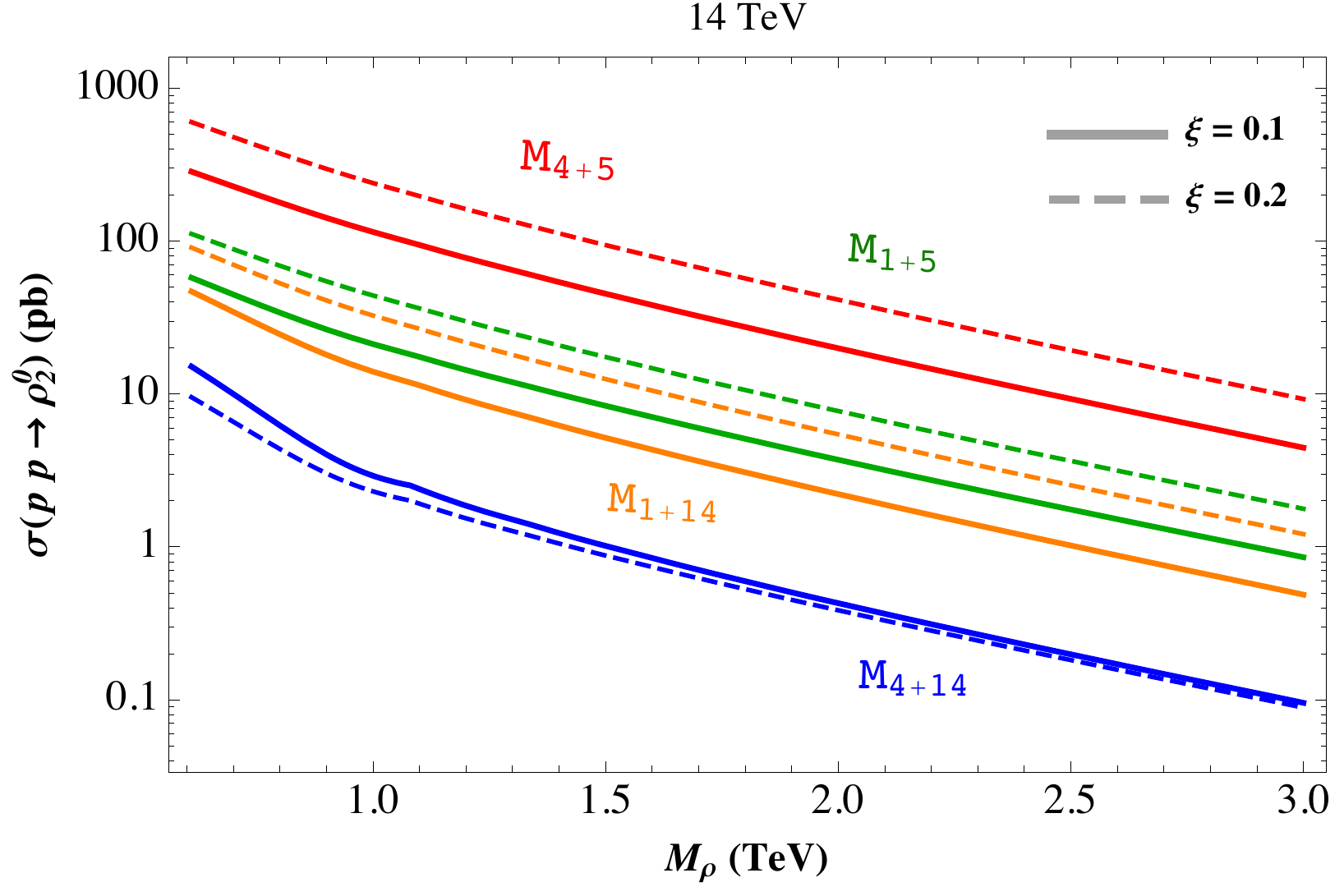}
\hspace*{-0.5cm}
\vspace*{0.5cm}
\includegraphics[scale=0.45]{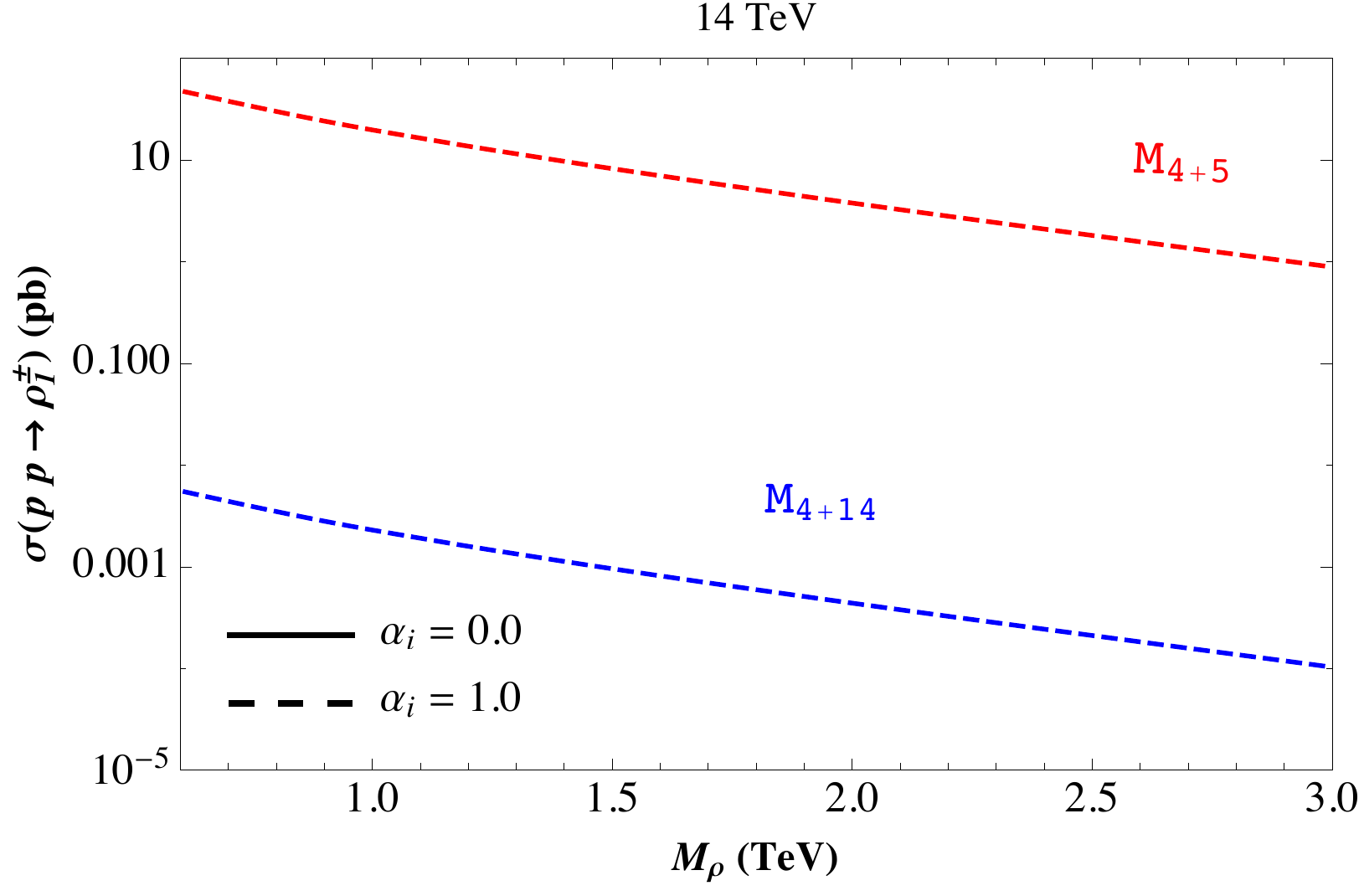}
\hspace*{0.5cm}
\includegraphics[scale=0.45]{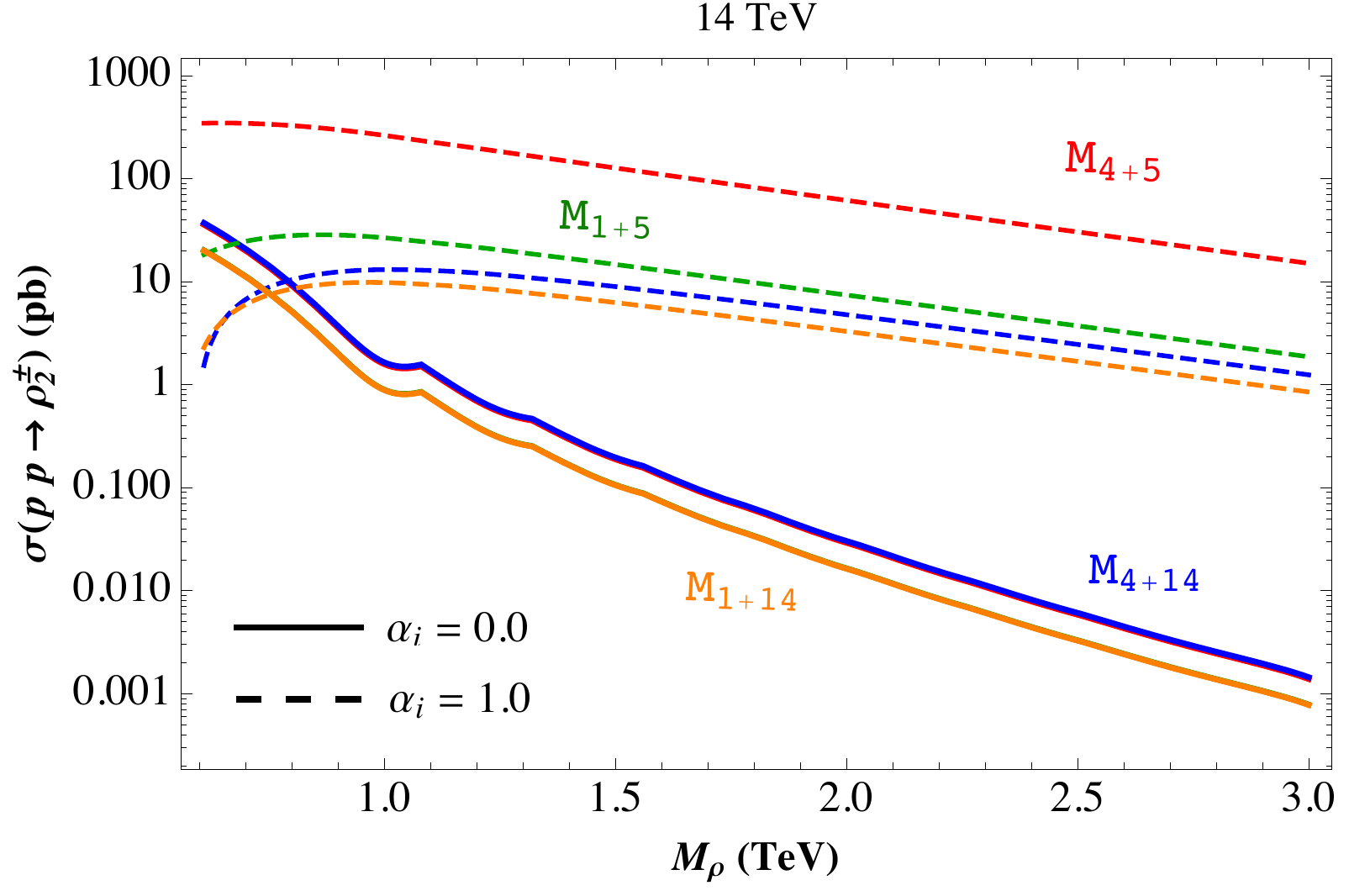}
\hspace*{-0.5cm}
\includegraphics[scale=0.45]{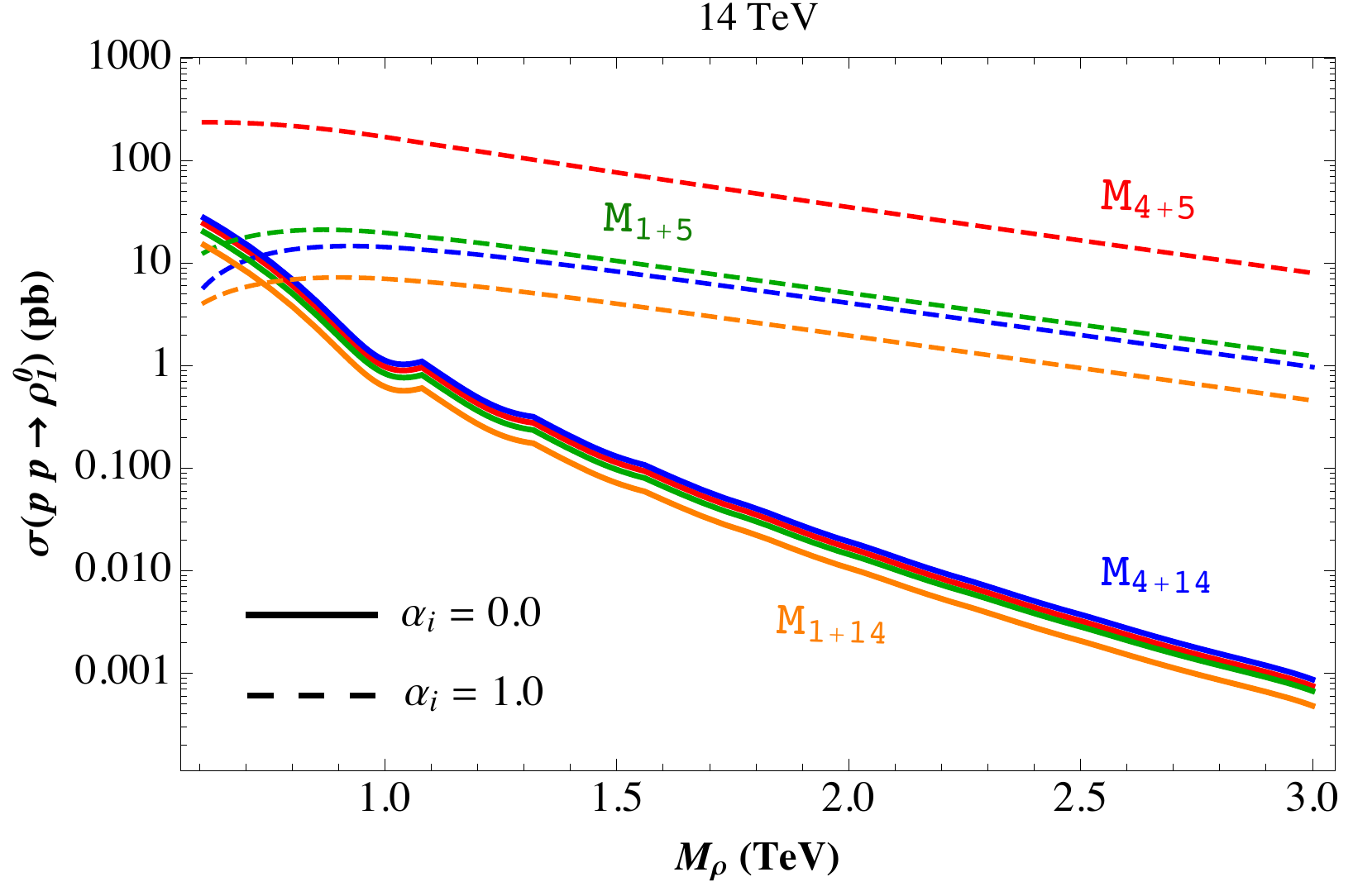}
\hspace*{0.7cm}
\includegraphics[scale=0.435]{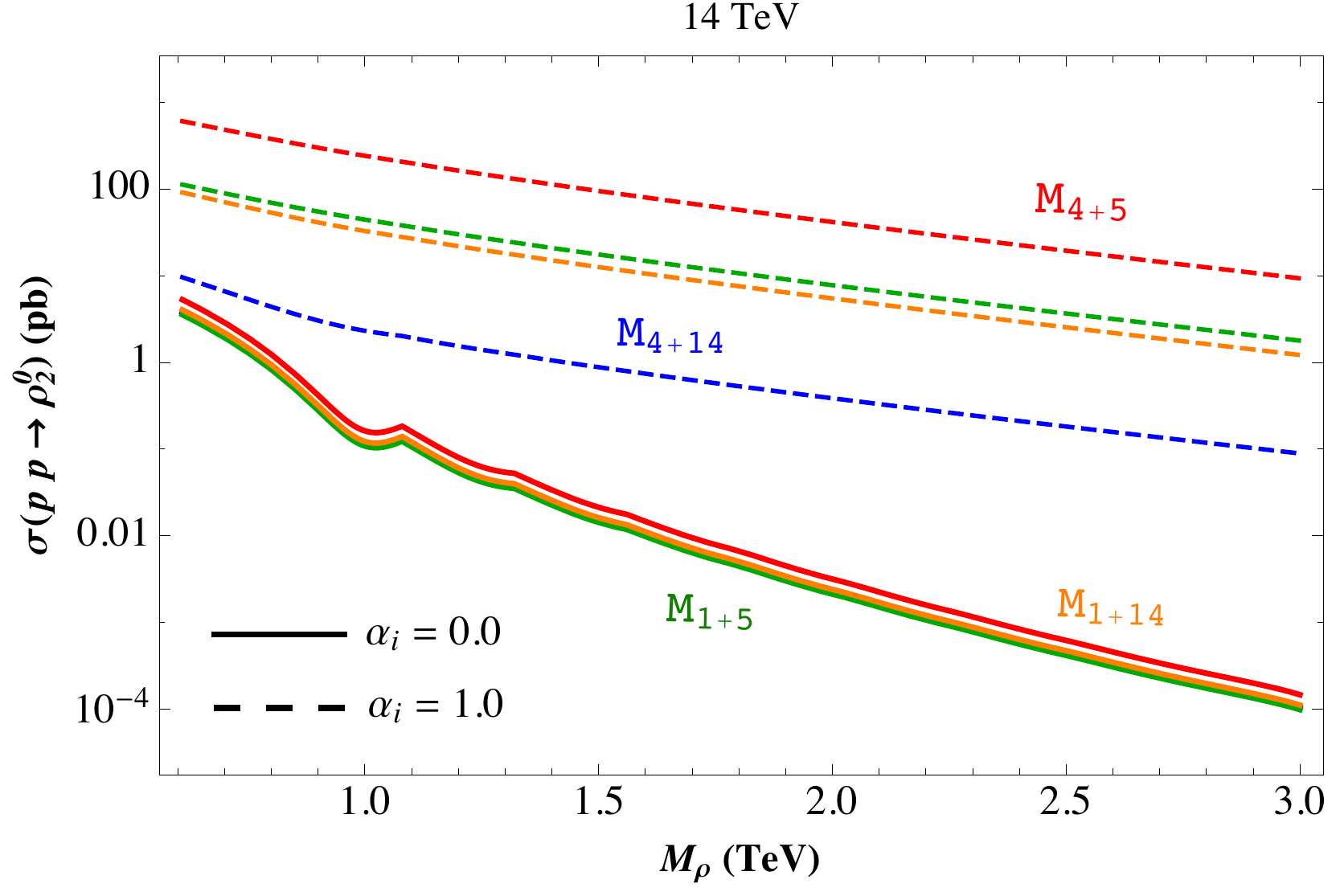}
\caption{\sf Production cross sections for the charged resonances $\rho^\pm_{1,2}$ (left-right 1st row) and for the neutral ones $\rho^0_{1,2}$ (2nd row) in all models at 14 TeV for the LHC, with $\xi=0.1,\,0.2$ (thick-dashed curves) and by setting $\alpha=1$. The impact from the fermion-vector resonance Lagrangian $\LL_{\bf M\,+\,\rho_\chi}$ in~\eqref{Currents-L-R} is displayed by comparing two different situations $\alpha=0,1$ (thick-dashed curves) for $\xi=0.2$ at the 3rd and 4th rows.}
\label{rho-production-cross-sections}
\end{center}
\end{figure}

\nt for $i=1,2$. The different involved couplings  directly depend on the weighting coefficients $\alpha$ of~\eqref{Currents-L-R} as well as on fermion-vector diagonalization effects, and are quite long to be reported here. Associated production cross sections through the processes $p\,p \to \rho^\pm_{1,2}$ and $p\,p \to \rho^0_{1,2}$ are computed from the latter Lagrangians by using MadGraph 5. Fig.~\ref{rho-production-cross-sections} displays all the spin-1 production cross sections as a functions of the parameter $\mrhoL=\mrhoR=M_\rho$ in the mass range $M_\rho \in [0.6,\,3]$\,TeV, for all the aforementioned top-partner models at $\sqrt{s}=14$ TeV, and by setting $\alpha=1$ for both $\xi=0.1,\,0.2$.  The   couplings $\gL$ and $\gR$ are fixed following the prescription in~\eqref{grho-gpartner}, whereas the corresponding Yukawa couplings $y_L,\, y_R,\, \tilde{y}_L,\, \tilde{y}_R$ in~\eqref{fA-oA-mix} and \eqref{fB-oB-mix} are suitably fixed to maintain the SM top quark mass at its experimental observed value, either through its predicted value in~\eqref{Masses-expanded-5} or~\eqref{Masses-expanded-14} and by implementing relations in~\eqref{eta-parameters}.  As it can be seen from Fig.~\ref{rho-production-cross-sections} (1st-2nd rows), the model $\fA$ is the most predominant one in yielding either charged or neutral heavy resonances. In addition, a higher $\xi$-value enhances all the productions, but the one for the $\rho^0_2$ at $\fB$ where its production is diminished. Notice that whether the top partner is a fourplet or singlet, the $\bf{5}$-elementary fermions scenario favours higher production values rather than the $\bf{14}$-ones. Among the charged and neutral resonances, $\rho^+_2$ and $\rho^0_2$ are predominantly yielded at $\fA$ and  reaching rough cross section values of $\sim 400$ pb (20 pb) and $\sim 600$ pb (10 pb)  at $M_\rho \sim 0.6$ TeV (3 TeV) for $\xi=0.2$. 

The resonance production is compared with respect the production with no fermion-vector resonance current interactions of~\eqref{Currents-L-R} in Fig.~\ref{rho-production-cross-sections} (3rd-4th rows). Notice how remarkably the cross section values are enhanced by the presence of the fermion-vector resonance current interactions of~\eqref{Currents-L-R}  by fixing $\alpha=1$ (dashed curves) with respect the situation  $\alpha=0$ (thick ones). In some cases such enhancement occurs by some orders of magnitude. The interactions of the heavy resonances to the SM fermions follow partly from the universal composite-elementary mixing, \ie from the elementary component of the heavy spin-1 mass eigenstate. They exhibit a strength of order $\sim g^2/g_\rho$, being thus extremely suppressed in the limit $g_\rho \gg g$ as it can be seen for larger $M_\rho$ in Fig.~\ref{rho-production-cross-sections}. Such scenario changes as soon as the $\cJ\cdot\rho$-coupings of~\eqref{Currents-L-R} are accounted for, as well as the  fermion-resonace diagonalization effects are considered in. All this clearly signals a feasible scenario for explaining future observations of heavy resonance production at higher energies, where the interactions encoded by~\eqref{Currents-L-R} and Table~\ref{Fermion-currents-set}, might help in determining the model and the strength for the involved effective terms.

Subsequent decays of the heavy resonance  may occur into final states containing single and double top partners\footnote{In~\cite{Barducci:2015vyf} was shown how the existing LHC searches can constrain decays of spin-1 resonances into a top partner pair, which generally make standard spin-1 searches, as dilepton resonant searches, ineffective. We will examine here how such top partner pair channels are altered-enhanced, once our additional fermion-vector resonance effects are switched on.}, as well as into gauge pair and gauge-Higgs final states. The fermionic decay channels will be triggered by the effective terms
\be
\begin{aligned}
&\LL_{Xf \rho^\pm}=\\ 
&\hspace*{-0.2cm}-\frac{1}{\sqrt{2}}\left[\sum_{f=u,d}{\bar X}\slashed{\rho}^+\left(g_{X_L f_L \rho^+}\,P_L+ g_{X_R f_R \rho^+}\,P_R\right)f\,+\,{\bar X}\slashed{\rho}^+\left(g_{X_L X'_L \rho^+}\,P_L+ g_{X_R X'_R \rho^+}\,P_R\right)X'\right]+\hc,
\label{top-parterners-rho-charged-currents}
\end{aligned}
\ee

\be
\hspace*{-1mm}
\LL_{Xf\rho^0}= \sum_{f=u,d}{\bar X}\,\slashed{\rho}^0\left(g_{X_L f_L \rho^0}\,P_L+ g_{X_R f_R \rho^0}\,P_R\right)f \,\,+\,\,\hc\,\,+\,\,{\bar X}\,\slashed{\rho}^0\left(g_{X_L X_L \rho^0}\,P_L+ g_{X_R X_R \rho^0}\,P_R\right)X,
\label{top-parterners-rho-neutral-currents}
\ee

\nt whilst the cubic interactions involving one heavy resonances are encoded by 
\be
\LL_{\rho^\pm W Z} = \\ i\left(\mathit{g}_{\text{\textit{$\rho^+ W Z$}}}^{(1)}
\,\rho^+_{\mu\nu}\,W^{-\mu}\,Z^\nu \,\, - \,\, \mathit{g}_{\text{\textit{$\rho^+ W Z$}}}^{(2)}
\,W^-_{\mu\nu}\,\rho^{+\mu}\,Z^\nu  \,\, + \,\, \mathit{g}_{\text{\textit{$\rho^+ W Z$}}}^{(3)}\,Z^{\mu\nu}\,\rho^+_\mu\,W^-_\nu \,\, + \,\,\hc\right),
\label{rho-W-Z}
\ee

\be
\LL_{\rho^0 W W} =i\left(\mathit{g}_{\text{\textit{$\rho^0 W W$}}}^{(1)}
\,W^+_{\mu\nu}\,W^{-\mu}\,\rho^{0\nu}\,\, + \,\,\hc\right)
\,\, + \,\,\frac{i}{2} \mathit{g}_{\text{\textit{$\rho^0 W W$}}}^{(2)}\,\rho^0_{\mu\nu}\,W^{+\mu}\,W^{-\nu},
\label{rho-WW}
\ee

\be
\LL_{\rho V h} =g_{\rho^+ W h}\left(\rho^+_{\mu}\,W^{-\mu}\,h
\,\,+\,\, \hc\right)\,\,+\,\, 
g_{\rho^0 Z h}\,\rho^0_\mu\,Z^\mu\,h,
\label{rho-W-Z-h}
\ee

\nt where the second term in~\eqref{top-parterners-rho-charged-currents} suitably couples two different top partners $X=X'=\{T,B,\Xtt,\Xft,\Tt\}$ to the spin-1 resonance. It is implied that the Lagrangians along~\eqref{top-parterners-rho-charged-currents}-\eqref{rho-W-Z-h} apply for both $\rho_1$ and $\rho_2$. The couplings involved in the fermionic decay channels entail both diagonalization effects from the gauge-resonance and elementary-composite sectors, being thus too long to be reported here. Nonetheless, the ones contributing to the resonance-gauge and  resonance-gauge-Higgs interactions only depend on a single diagonalization.  In fact, these couplings can be extracted by using the Equivalence Theorem for a heavy resonance field. In this limit the leading contribution to the interaction comes from the longitudinal polarizations of the SM vector fields, and the overall strength equals that of the coupling of one $\rho$ to two NG bosons up to small corrections $\cO(m^2_V/m^2_\rho)$. From~\eqref{rho-Lagrangian} the strength of the $\rho\pi\pi$-interaction is proportional to $g_\rho a^2_\rho$, with  $a_\rho=m_\rho/(g_\rho f)$, a quantity expected to be of order 1 according to naive dimensional analysis (NDA).

\begin{figure}
\begin{center}
\hspace*{-0.55cm}
\vspace*{1cm}
\includegraphics[scale=0.34]{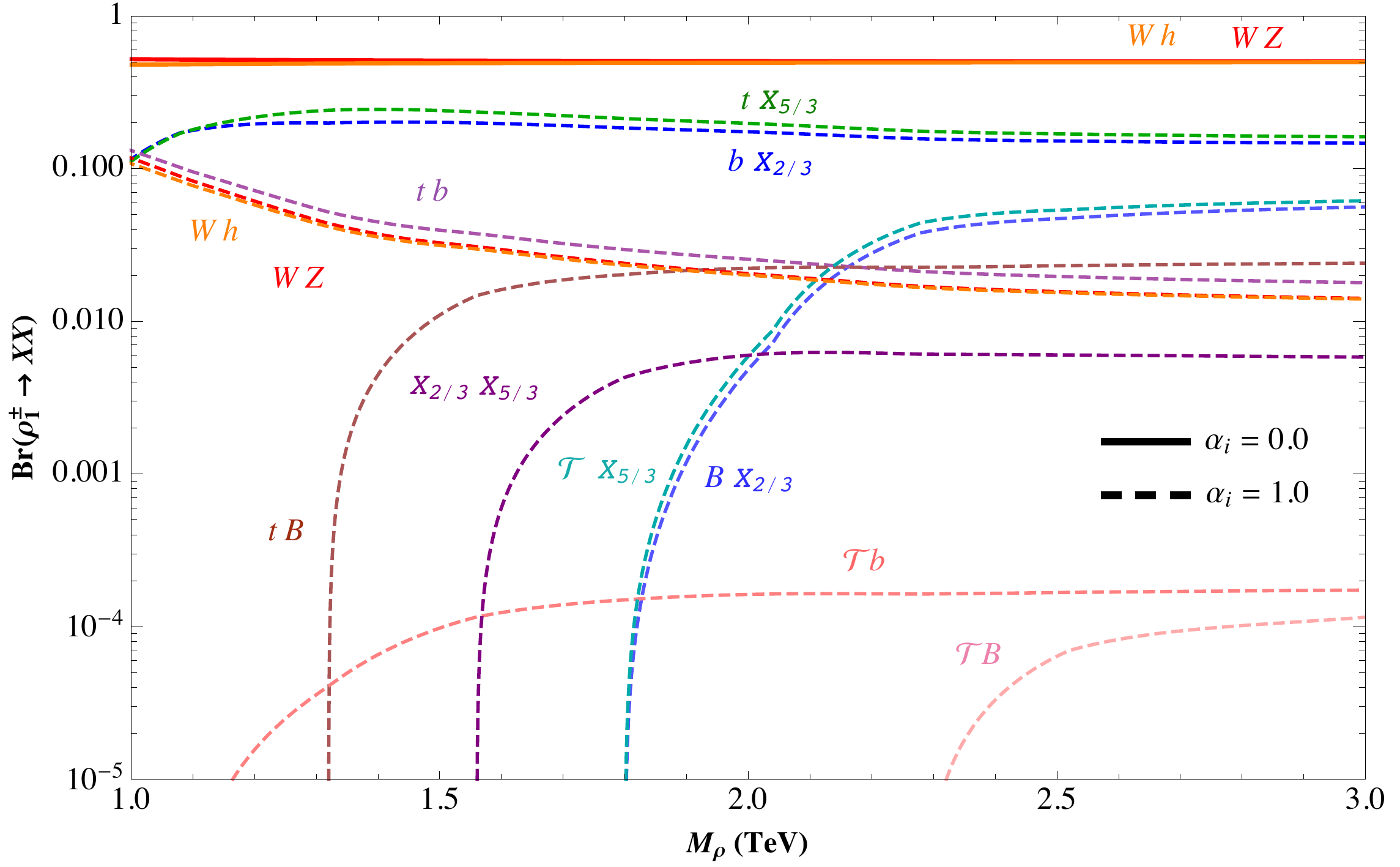}
\hspace*{0.1cm}
\includegraphics[scale=0.38]{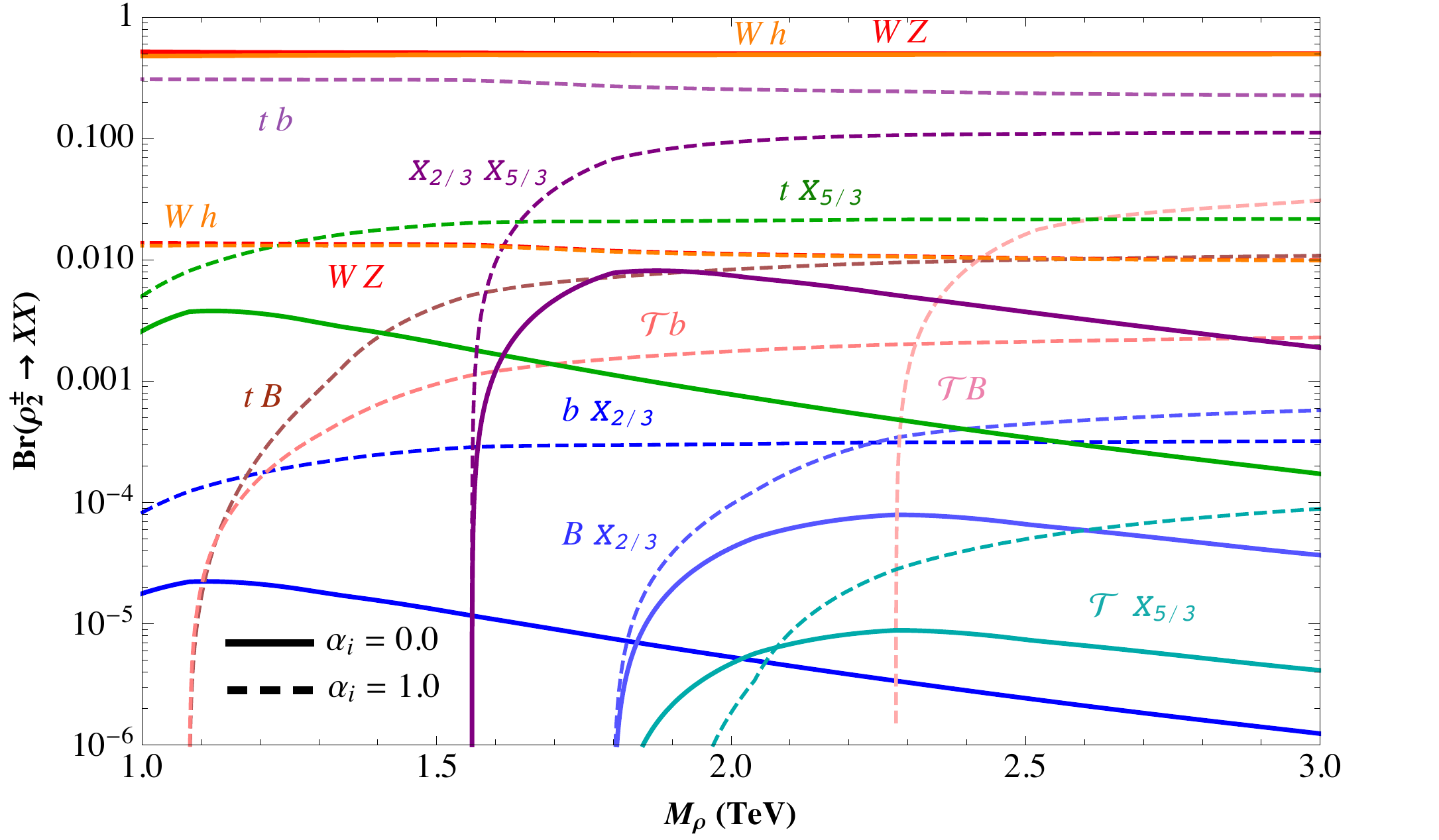}
\vspace*{0.5cm}
\hspace*{-0.7cm}
\includegraphics[scale=0.34]{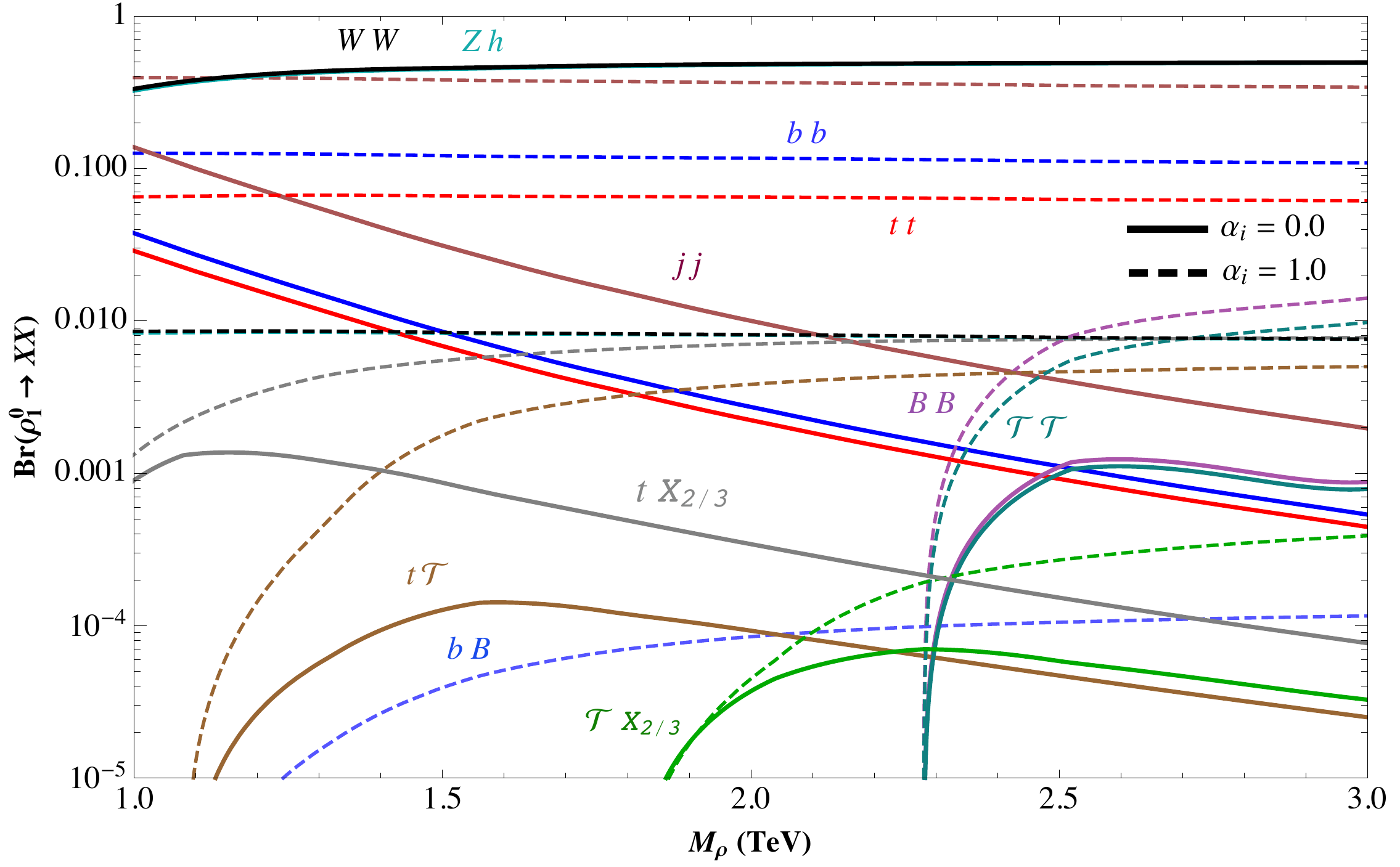}
\hspace*{0.1cm}
\includegraphics[scale=0.34]{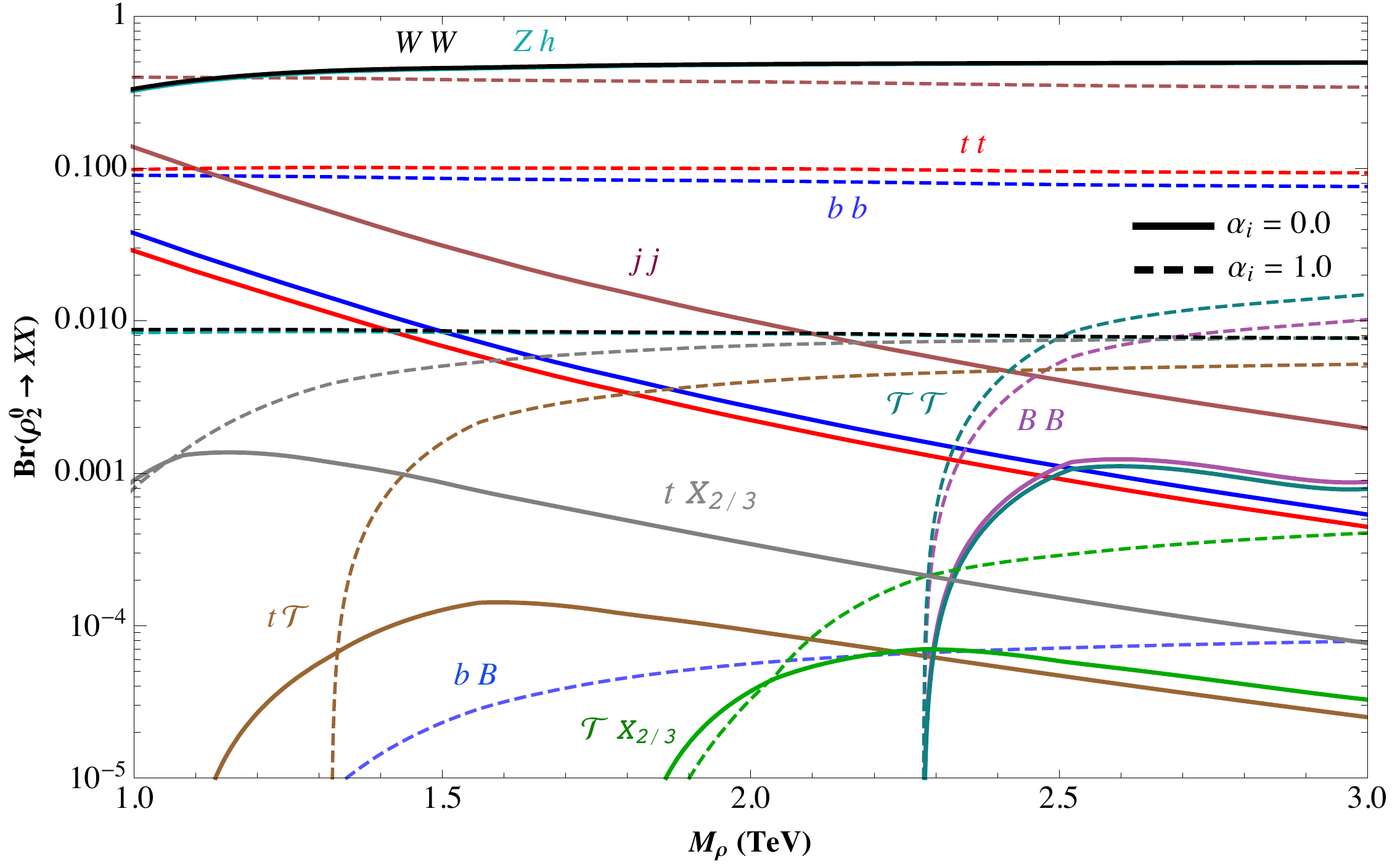}
\caption{\sf All branching ratios for the charged $\rho^\pm_{1,2}$-decay modes (upper panels left-right), and for the neutral ones $\rho^0_{1,2}$ (lower) at $\fA$ by setting $\xi=0.2$. The impact from the fermion-vector resonance Lagrangian $\LL_{\bf M\,+\,\rho_\chi}$ in~\eqref{Currents-L-R} is displayed by comparing two different situations $\alpha=0,1$ (thick-dashed curves). See text for details.}
\label{Branching-ratios}
\end{center}
\end{figure}

These features leads the heavy resonances to be strongly coupled to the composite states, i.e. the longitudinal polarizations of $W$, $Z$ and the Higgs boson, while their coupling strength to the SM fermions to be  extremely suppressed in the limit $g_\rho \gg g$. Nonetheless, such scenario changes as soon as the interactions of $\LL_{\bf M\,+\,\rho_\chi}$ in~\eqref{Currents-L-R} are considered. Indeed, the fermion-vector resonance couplings are augmented by $\LL_{\bf M\,+\,\rho_\chi}$ and are hence directly depending on the strength of the coefficients $\alpha$, as well as on the fermion and gauge-resonance diagonalization effects. Fig.~\ref{Branching-ratios} resumes all the previous remarks, where
the branching fractions are compared for two different cases $\alpha=0,1$ (thick-dashed curves) at $\fA$ by setting $\xi=0.2$. Notice that the branching fractions to $W Z$ and $W h$, as well as those to $WW$ and $Zh$, are equal to very good approximation. This is implied by the Equivalence
Theorem, which works well since $M_\rho \gg  M_{W,Z}$ for the chosen values of parameters. As expected, the branching ratios of the resonance to fermions are much smaller as a consequence of the suppressed couplings. Some remarks are in order:
\begin{itemize}

\item No fermion-vector resonance currents ($\alpha=0$) entails dominant pair gauge and gauge-Higgs decay channels, while extremely suppressed-subdominant fermionic channels for the charged-neutral resonances (upper-lower pannels). 

\item The scenario changes when fermion-vector resonance currents are switched on ($\alpha=1$). Indeed, the pair gauge $W W$, $W Z$ and the gauge-Higgs $Wh$, $Zh$ final states are still the relevant ones for $M_\rho\lesssim 1$ TeV, becoming subdominant with respect to $\rho^\pm_1\to\{\,t\Xft,\,b\Xtt\}$ and $\rho^\pm_2\to \{t\Xft,\,\Xft\Xtt\}$ at $M_\rho\gtrsim 1$ TeV. The channel $tb$ turns out to be dominant along the explored mass range for the charged resonance decays, as well as the modes $\rho^0_{1,2}\to\{bb,\,tt,\,jj\}$.

\item The higher regime $g_\rho \gg g$ ($M_\rho\gtrsim 2$ TeV) triggers other exotic channels dominant compared with the pair gauge and gauge-Higgs channels, \eg $\rho^\pm_1\to\{\cT\Xft,\,\cB\Xtt,\,\Xft\Xtt\}$ and $\rho^\pm_2\to \{tB,\,\cT\cB\}$, as well as $\rho^0_{1,2}\to\{\cB\cB,\,\cT\cT\}$. 

\item Even for no fermion-vector resonance currents ($\alpha=0$, thick curves) there will be exotic fermionic modes still active, although less relevant as the gauge and gauge-Higgs channels. Such fermionic exotic modes receive important contributions when the couplings $\cJ \cdot \rho$ in~\eqref{Currents-L-R} are included in, some of them being enhanced by one-two orders of magnitude, or even three orders as in the case for $\rho^\pm_2\to b\Xtt$ and  $\rho^0_{1,2}\to\{b\cB,\,t\cT\}$ for a higher regime mass $M_\rho$.

\end{itemize}

\nt Similar comments apply for the product of the resonance production cross section times the corresponding branching ratio, not displayed here for briefness purposes. Once the heavy resonance are produced, their decays can lead to the generation of either a single or double quark partner in the final states. A fuller top partner production mechanism is triggered by bringing  QCD, EW and Higgs-mediated interactions onto the stage. 

\section{Top partners production and decays}
\label{Top-partner-production}
\begin{figure}
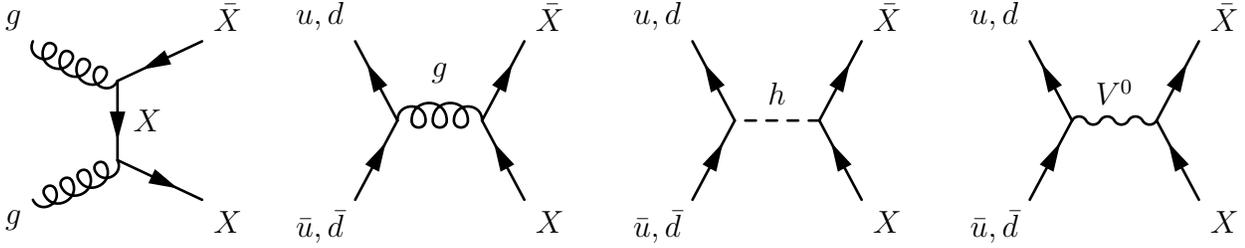

\hspace*{-0.7cm}
\begin{tabular}{cccc}
\input{DoubleProduction-1}&
\hspace*{0.5cm}
\input{DoubleProduction-3}&
\hspace*{0.7cm}
\input{DoubleProduction-2}&
\hspace*{0.7cm}
\input{DoubleProduction-4}
\vspace*{1cm}
\end{tabular}
\caption{\sf Diagrams contributing to the double partner production, where $V^0=Z,\gamma,\rho^0_{1,2}$ and with $X$ denoting any $X=\{T,B,\Xtt,\Xft,\Tt\}$. }
\label{Double-Production-diagrams}
\end{figure}

\nt Being colored all the quark partners, their production in pairs at hadron colliders is QCD-driven as it is shown in Fig.~\ref{Double-Production-diagrams}, furthermore, completely model-independent and insensitive on the degrees of compositeness of the associated SM quarks. Qualitatively, the top partner production is independent on whether both or only one multiplet is present in the effective theory.

\subsection{Double Partner production}
\label{Double-Partner-production}

\begin{figure}
\begin{center}
\hspace*{-0.6cm}
\vspace*{1cm}
\includegraphics[scale=0.42]{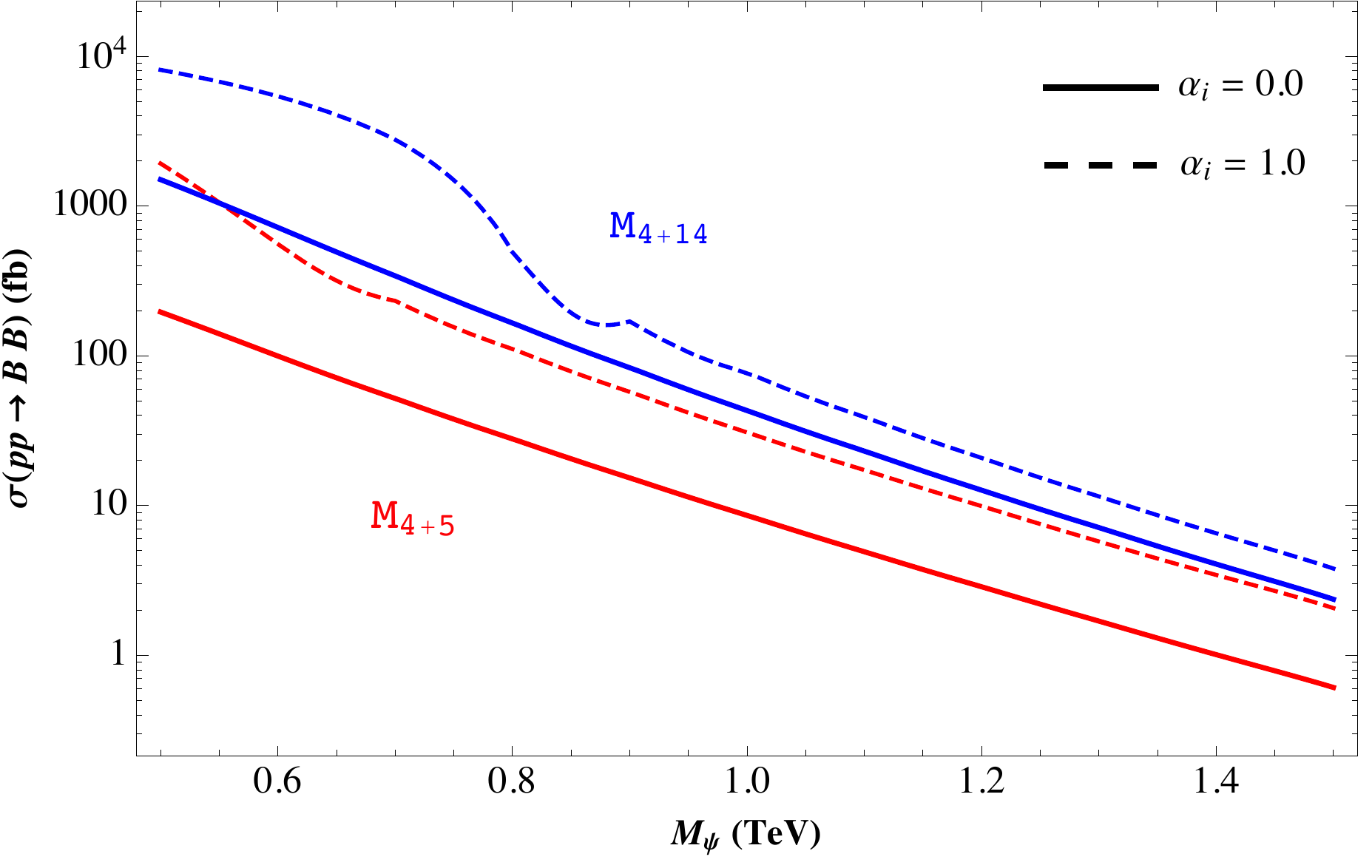}
\hspace*{0.6cm}
\includegraphics[scale=0.41]{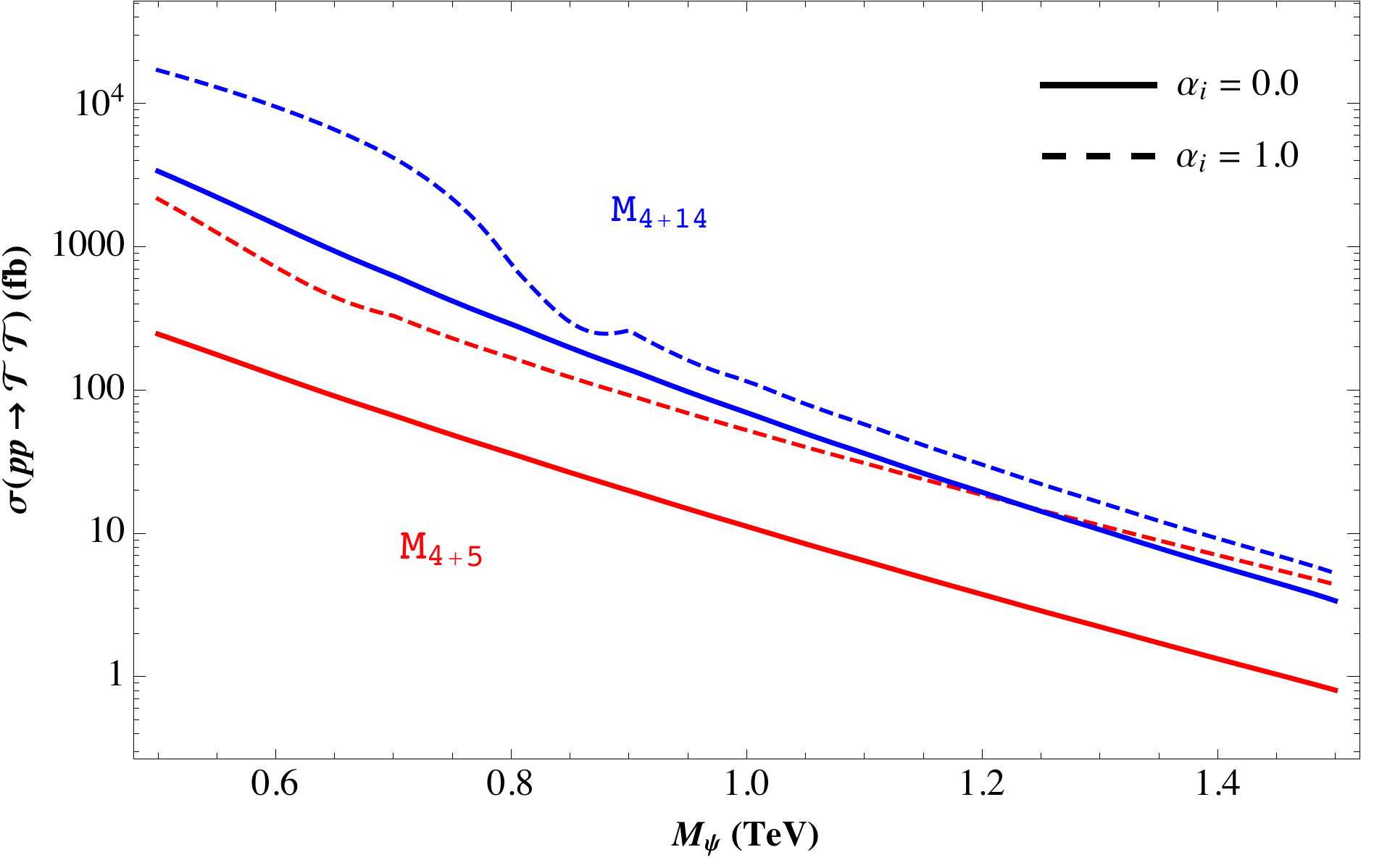}
\hspace*{-0.6cm}
\includegraphics[scale=0.42]{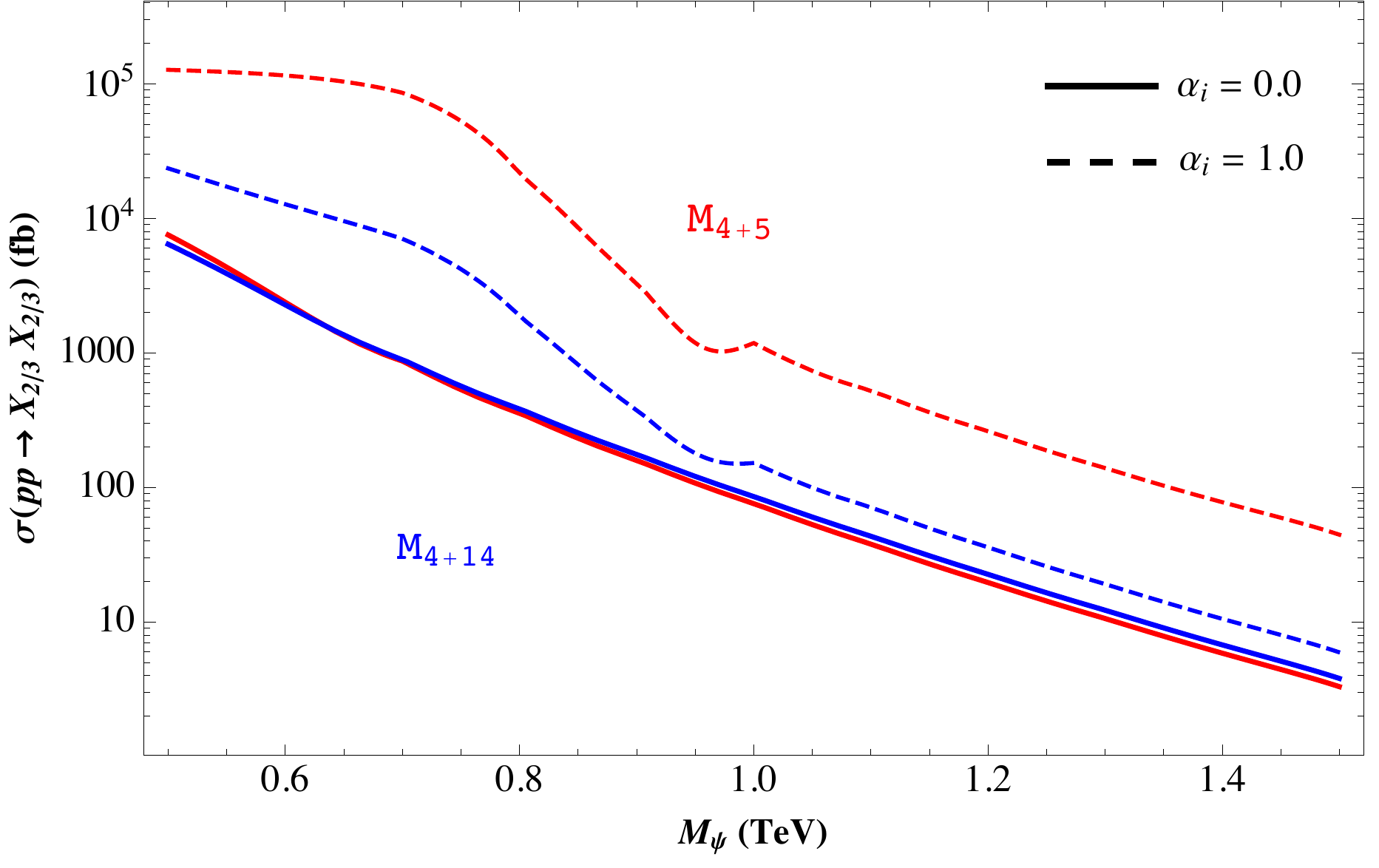}
\hspace*{0.7cm}
\includegraphics[scale=0.42]{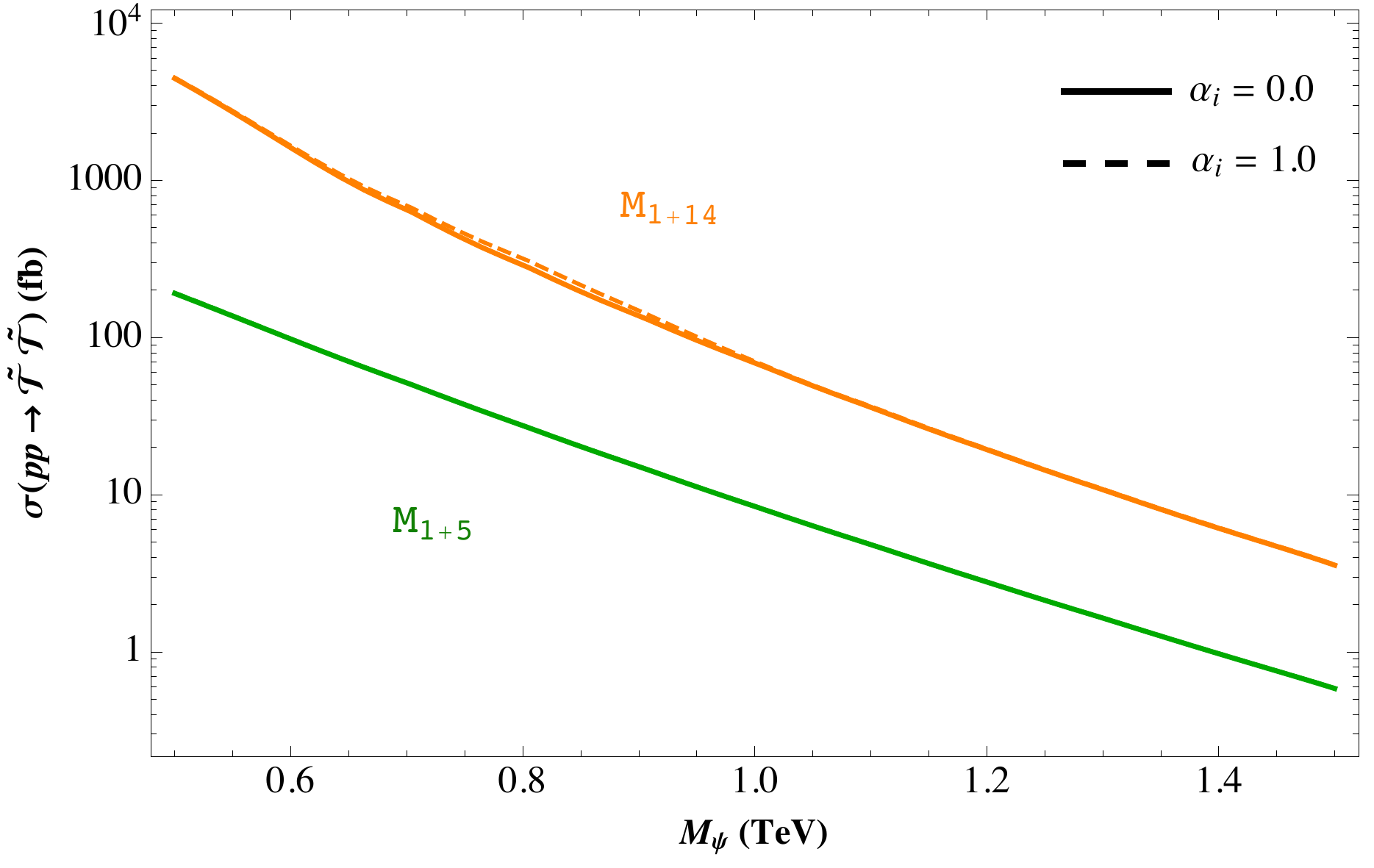}
\caption{\sf Double-partner production cross sections at 14 TeV  for $\xi=0.2$, only for neutral final states. Two different situations $\alpha=0,1$ (thick-dashed curves) are plotted to compare the impact on the production from the fermion-vector resonance Lagrangian $\LL_{\bf M\,+\,\rho_\chi}$ of~\eqref{Currents-L-R}.}
\label{Double-partner-production}
\end{center}
\end{figure}

\nt The production of double-partner final states is driven by QCD as well as SM gauge, Higgs, and $\rho^0$-mediated processes for the case of neutral final states as it is depicted in Fig.~\ref{Double-Production-diagrams}.  The double production mechanisms is controlled by the model-dependent couplings $g_{u d \rho^\pm},\,g_{f f \rho^0},\,g_{X f \rho^+},\,g_{X X' \rho^+},\,g_{X f \rho^0},\,g_{X X \rho^0}$ through~\eqref{rho-charged-currents}-\eqref{top-parterners-rho-neutral-currents} and by the analogous ones involving SM charged and neutral gauge fields correspondingly. QCD pair production is completely model-independent, although non-zero model-dependent modifications are induced as soon as extra fermion-vector resonance effects are accounted for. Fig.~\ref{Double-partner-production} gathers the double-partner production cross sections only for neutral final states, where we have constructed the pair cross sections for each value of the mass parameter $M_{\bf{4}}=M_{\bf{1}}=M_\Psi$ by interpolation using MadGraph 5 simulations, at 14 TeV LHC in all the models for $\xi=0.2$, and for a fixed resonance mass $\mrhoL=\mrhoR=M_\rho\sim 1.5$ TeV. The prescription in~\eqref{grho-gpartner} is assumed again for the couplings $\gL$ and $\gR$. Comparison of two different situations $\alpha=0,1$ (thick-dashed curves) reflects the impact on the production from the additional fermion-vector resonance effects regarded here. The latter effects may enhance double-partner production by one order of magnitude at the fourplet models, whereas vanishing contributions and tiny ones are obtained at the singlet scenarios. The combined effect of fermion-vector resonance rotation as well as the less number of additional fermionic currents determine such behaviour for the latter models. 

Notice how the final states $\cT\cT$ and $\cB\cB$ are mainly produced via proton-proton collision in $\fB$ as the involved quark partner masses are smaller than the corresponding ones at $\fA$ (see \eqref{Masses-expanded-5}-\eqref{Masses-expanded-14}  and Fig.~\ref{4plet-singlet-masses}). The $\Xtt\Xtt$-modes does not distinguish the elementary embeddings representation as the involved partner masses are equal at both models. Nonetheless, as soon as extra fermion-vector resonance effects of~\eqref{Currents-L-R} are regarded, the model $\fB$ gets disfavoured in turn compared with $\fA$, due to the implied fermion-vector resonance diagonalization effects and the different number of fermionic currents in each model as well. The same comments apply qualitatively and quantitatively for the channel $\Xft\Xft$ as the involved partner masses are degenerate with the corresponding one for $\Xtt$ (see Appendix~\ref{Physical-fermion-masses}). Generically, producing pairs either of $\Xtt$ or $\Xft$ will be kinematically favoured with respect double production of both $\cT$ and $\cB$, because their relatively higher masses. Similar arguments alike to the fourplet case work for the pair production of the singlet $\widetilde{\cT}$ (Fig.~\ref{Double-partner-production}), where the involved masses result smaller at $\bf{14}$-elementary embeddings compared with the one at $\bf{5}$-scenario (see Fig.~\ref{4plet-singlet-masses}, right plot), favouring the former scenario for its production in pairs.

\subsection{Single Partner production}
\label{Single-Partner-production}

\begin{figure}
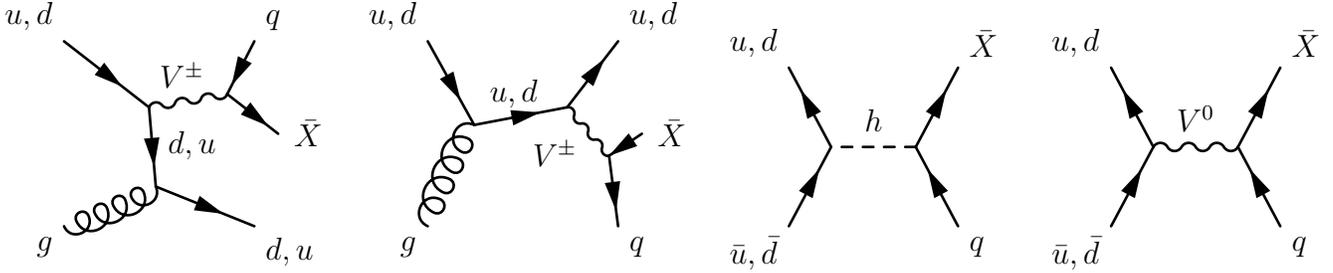

\hspace*{-0.8cm}
\begin{tabular}{cccc}
\input{SingleProduction-1}&
\hspace*{0.7cm}
\input{SingleProduction-3}&
\hspace*{0.7cm}
\input{SingleProduction-2}&
\hspace*{0.5cm}
\input{SingleProduction-4}
\vspace*{1cm}
\end{tabular}
\caption{\sf Diagrams contributing to the single partner production, where $V^\pm=W^\pm,\rho^\pm_{1,2}$ and $V^0=Z,\gamma,\rho^0_{1,2}$ and with $q$ standing for any up/down-like quark conveniently couple to $X=\{T,B,\Xtt,\Xft,\Tt\}$.}
\label{Single-Production-diagrams}
\end{figure}

\begin{figure}
\begin{center}
\includegraphics[scale=0.4]{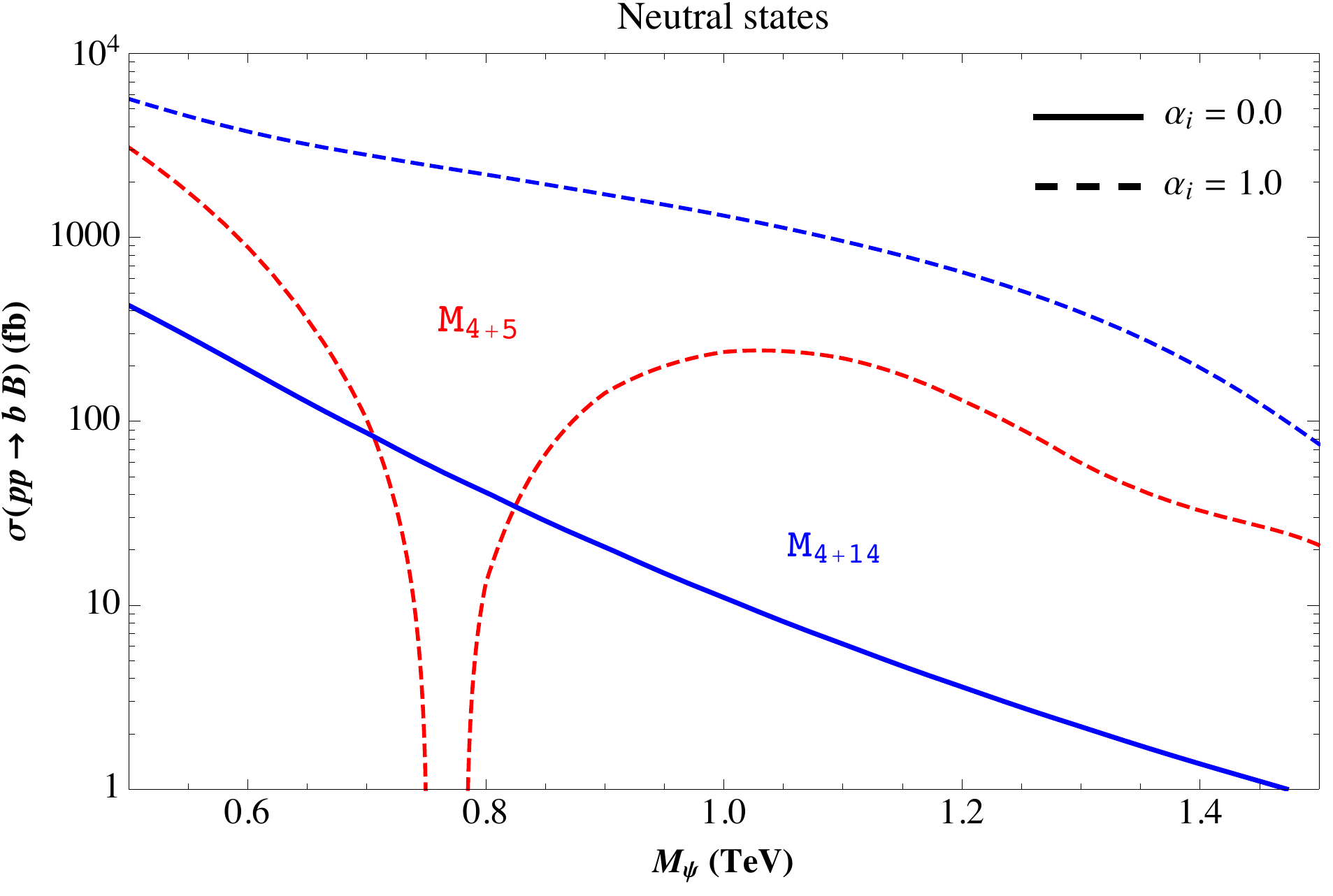}
\vspace*{0.5cm}
\hspace*{0.24cm}
\includegraphics[scale=0.385]{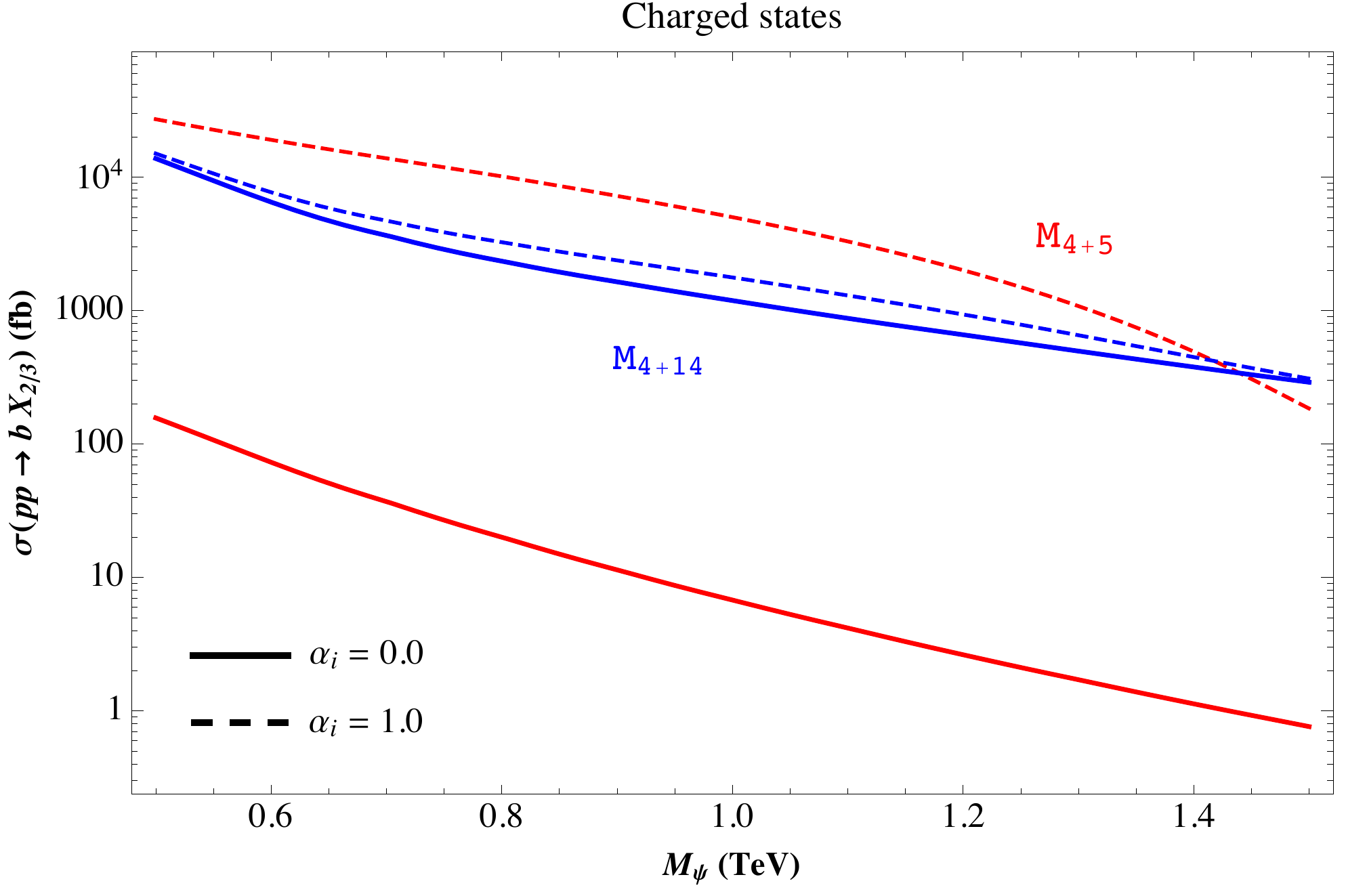}
\includegraphics[scale=0.41]{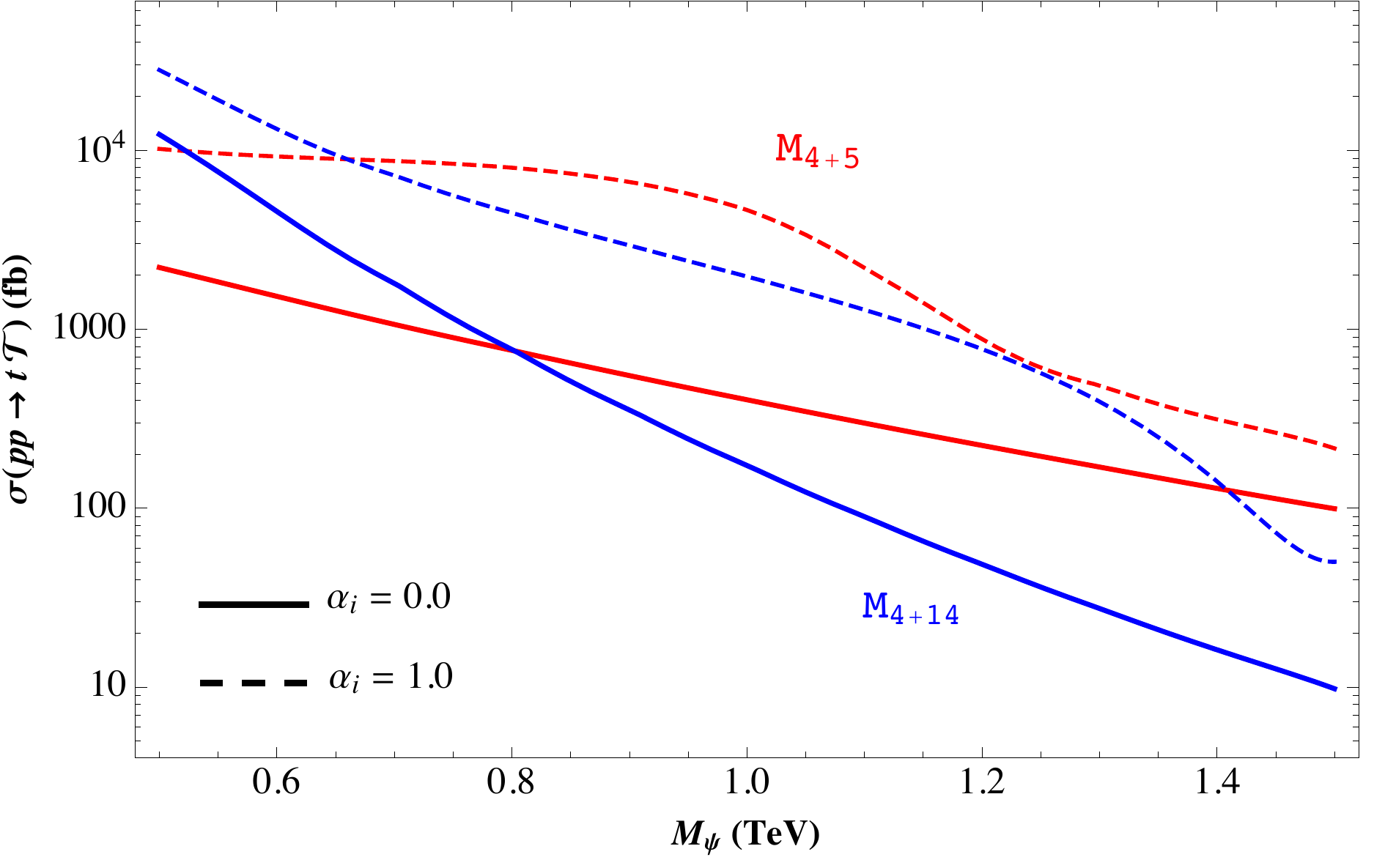}
\vspace*{0.5cm}
\hspace*{0.24cm}
\includegraphics[scale=0.385]{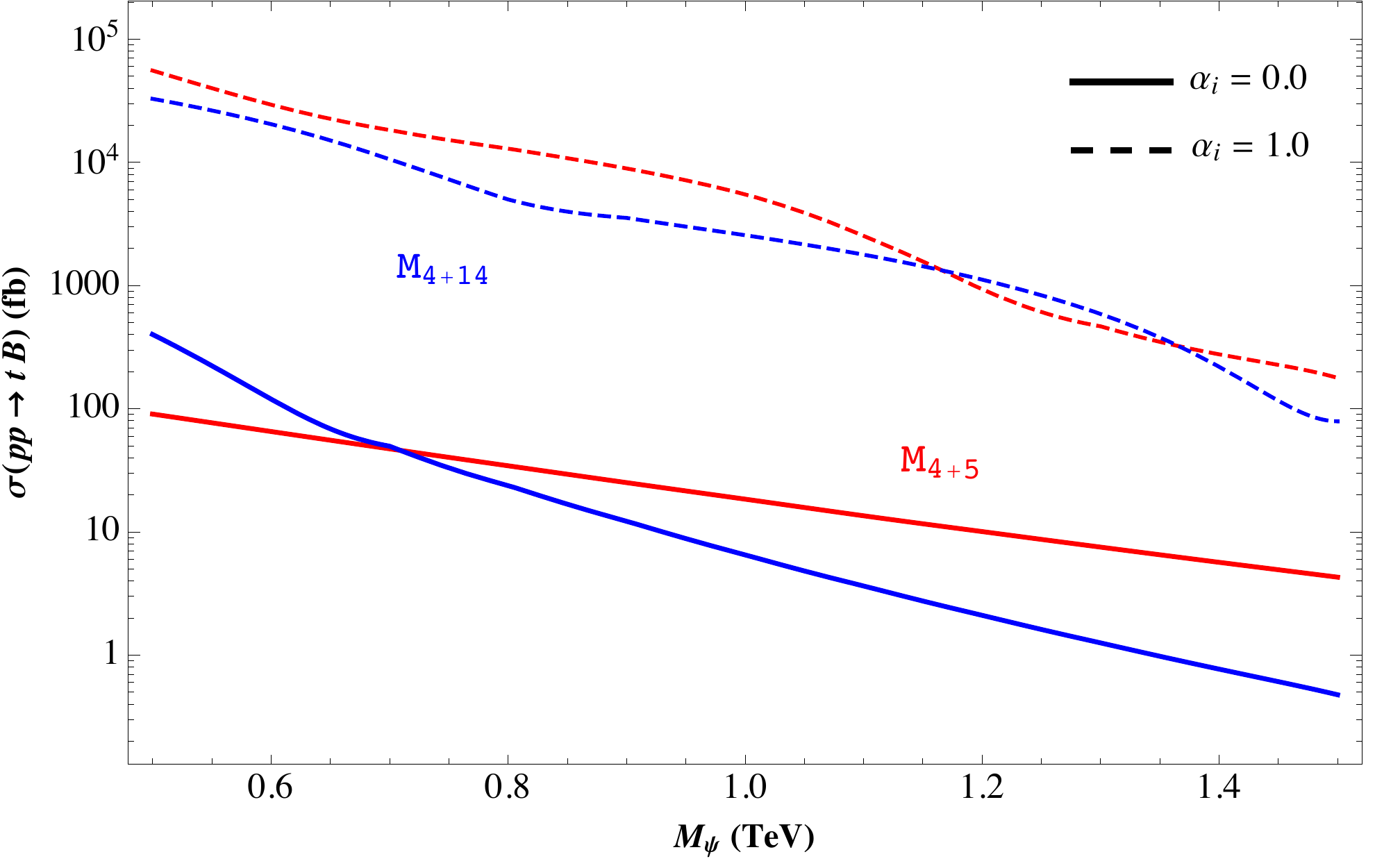}
\includegraphics[scale=0.4]{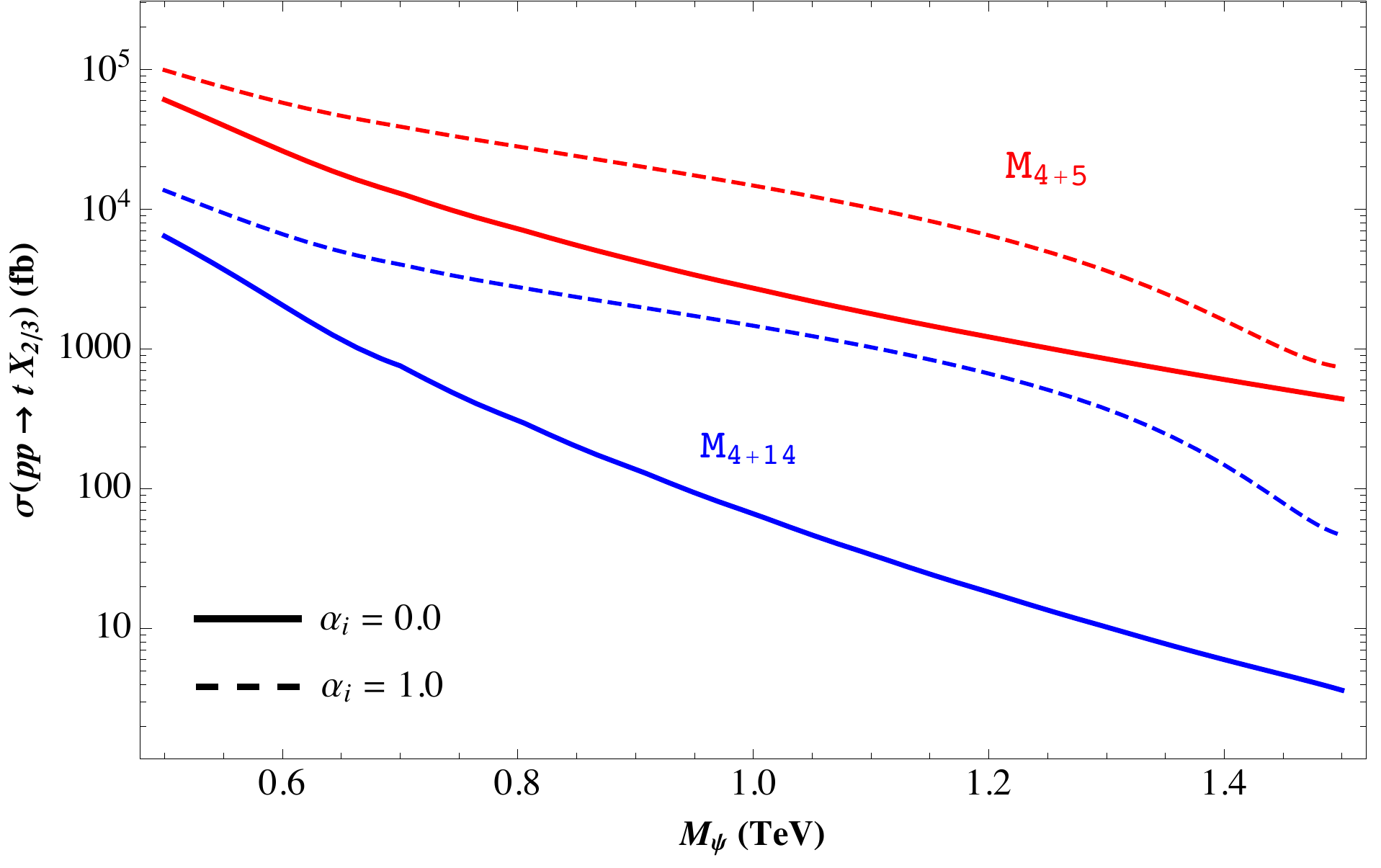}
\vspace*{0.5cm}
\hspace*{0.24cm}
\includegraphics[scale=0.385]{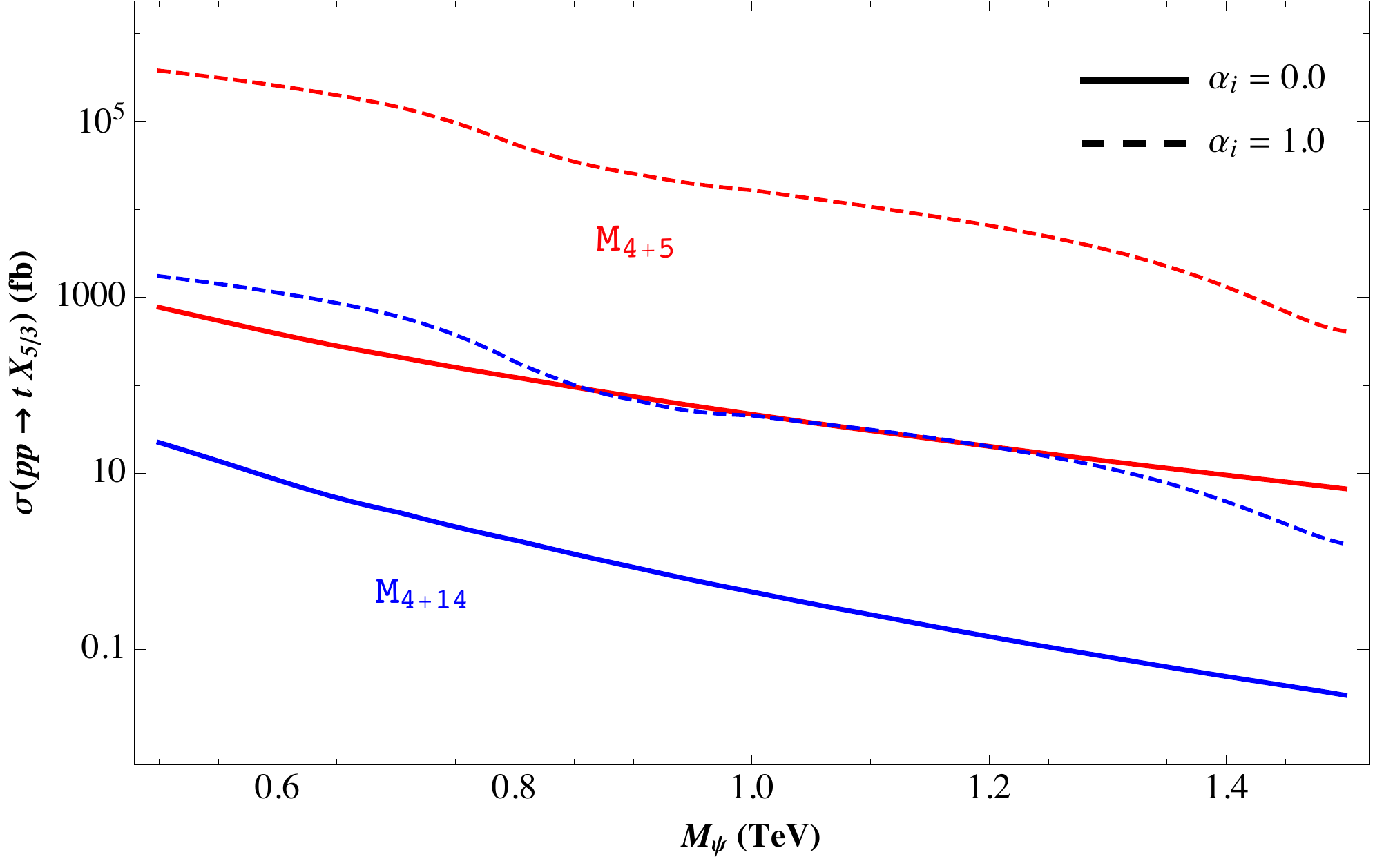}
\includegraphics[scale=0.41]{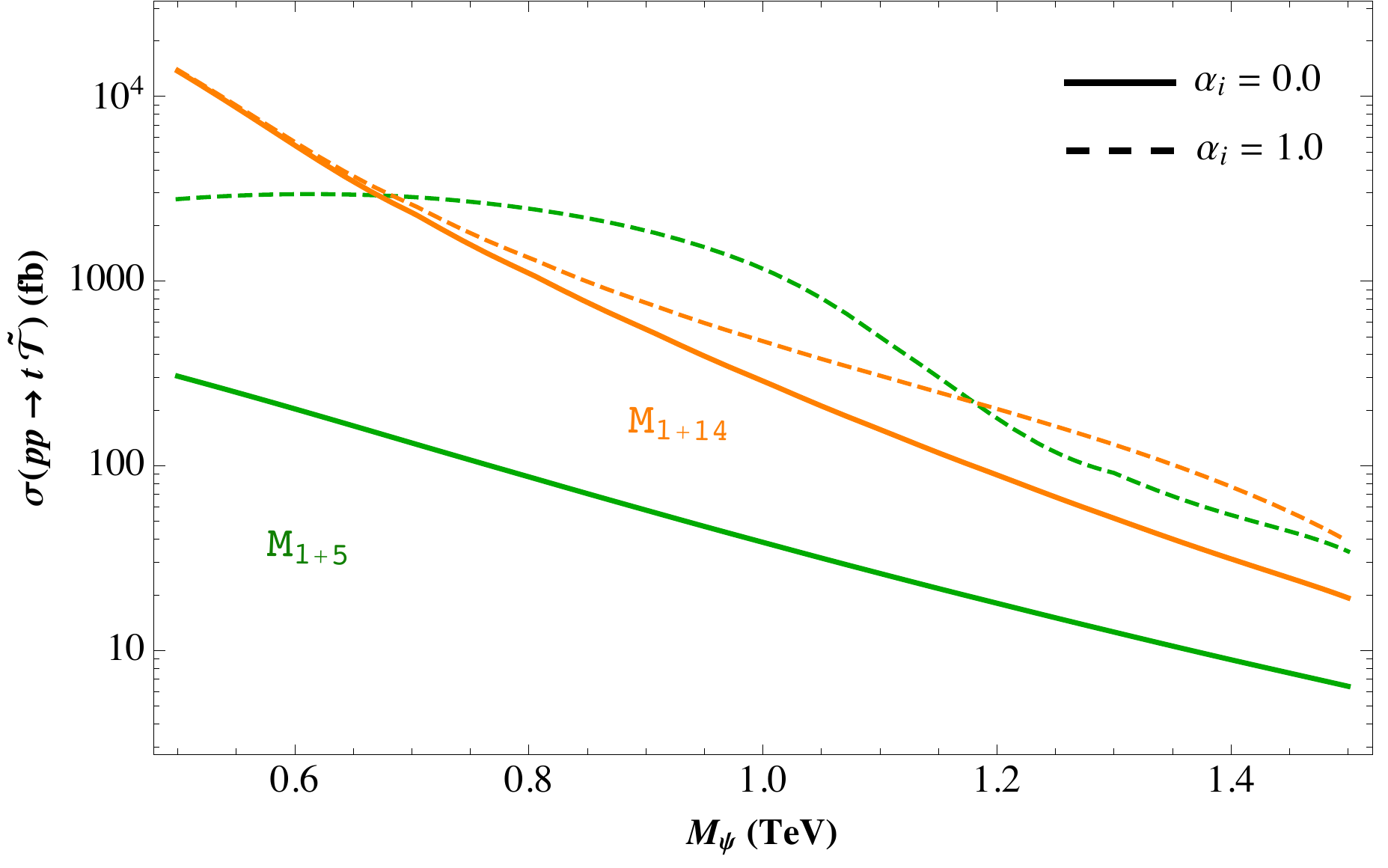}
\hspace*{0.24cm}
\includegraphics[scale=0.385]{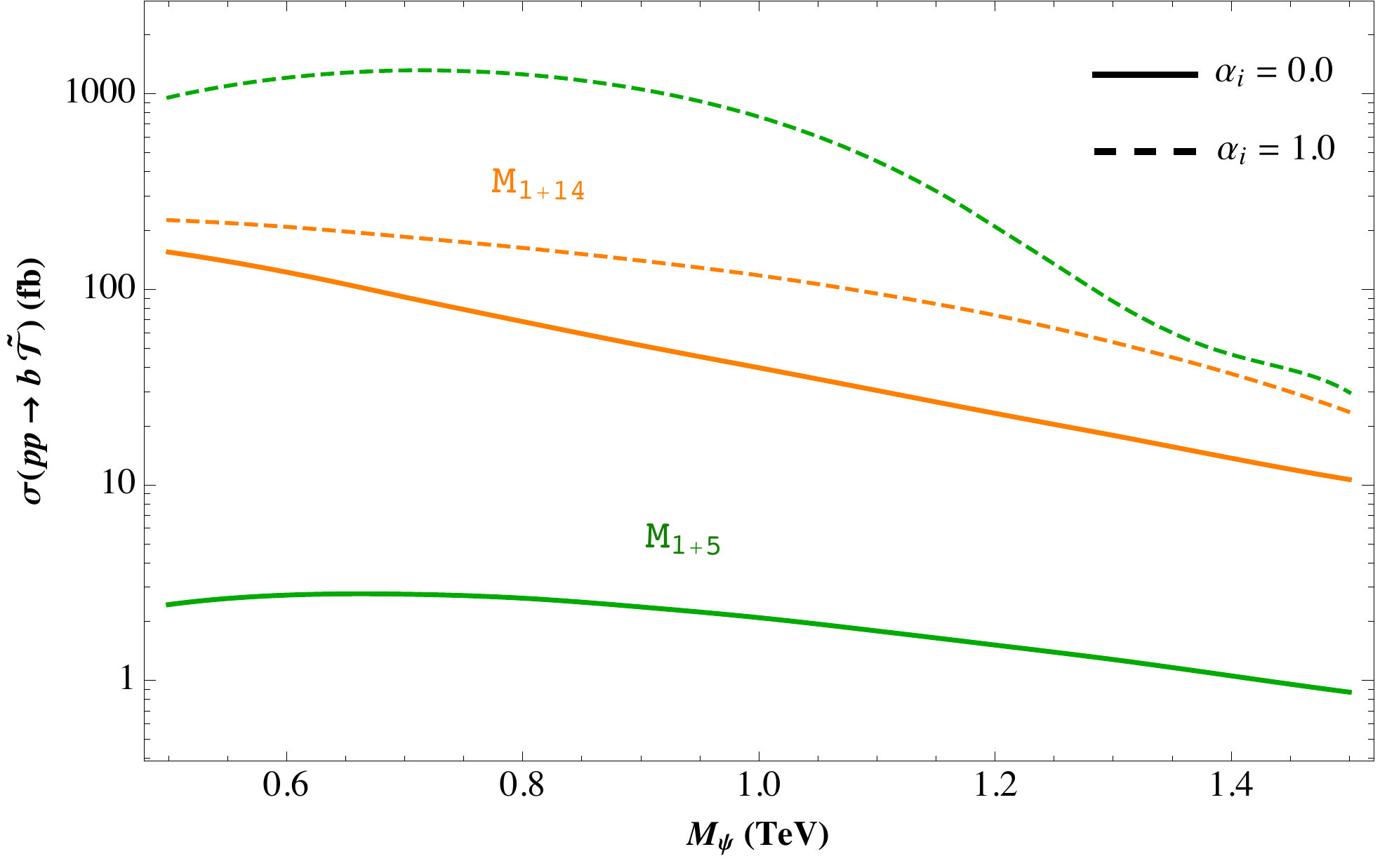}
\caption{\sf Single-partner production cross sections at 14 TeV  for $\xi=0.2$, either for the neutral final states (left) or the charged states (right). Two different situations $\alpha=0,1$ (thick-dashed curves) are plotted to compare the impact on the production from the fermion-vector resonance Lagrangian $\LL_{\bf M\,+\,\rho_\chi}$ of~\eqref{Currents-L-R}.}
\label{Single-partner-production}
\end{center}
\end{figure}

\nt QCD may trigger the production of single-partner final states, together with SM gauge and $\rho^\pm$, $\rho^0$-mediated processes for the case of charged-neutral final states respectively (Fig.~\ref{Single-Production-diagrams}). These channels are gathered in Fig.~\ref{Single-partner-production}, where the charged final state $b\cT$ has been omitted for briefness reasons. Important enhancements occur for the single partner production at the fourplet model (Fig.~\ref{Single-partner-production} 1st row) as the kinematic of less massive final states is implied. The larger number of fermionic current entering in the stage also determines such increasing. Although some cases do not obey this, like the neutral final mode $t\Xtt$ at $\fA$ (3rd row left) and the charged channel at $b\Xtt$ at $\fB$ (1st row right), where the combined effect of fermion-vector resonance diagonalization effects roughly suppresses the induced contributions from the additional interactions of~\eqref{Currents-L-R}.  The channel $b\cB$ is absent at $\fA$  because flavor-changing neutral couplings are forbidden in the charge -1/3 sector as explained in~\cite{DeSimone:2012fs}. Nonetheless, non-zero contributions arise for it as long as the extra fermion-vector resonance effects regarded here together with diagonalization effects are considered  as shown in Fig.~\ref{Single-partner-production} for $\alpha=1$ (1st row left).

Notice again how the single production of the singlet $\widetilde{\cT}$ is dominant at the $\bf{14}$-elementary embeddings rather than at $\bf{5}$-scenario as the involved masses result smaller at the former model\footnote{Recently it has been analysed the top partner single productions through loops mediated by the scalar singlet in~\cite{Kim:2018mks}. With reasonable coupling strengths, the production rate of a top partner, in association with the SM top, can dominate top partner pair production at top partner masses higher than 1.5 TeV. See the reference for more details.}. However the situation may turn as the extra fermion-vector resonance couplings are included, for instance in the charged mode $b\cT$ (4th row right). Quantitatively,  the final states containing either $\Xtt$ or $\Xft$ will be largely produced as they involve partners whose masses are smaller compared with the others partners, as in the case of $t\Xtt$ (3rd row left) and $b\Xtt$-$t\Xft$ (1st-3rd rows right). 

Concerning the singlet partner and the fourplet ones, their single production in association with a SM quark is partly driven by the Higgs boson, suppressing therefore the single production of the up and charm partner by the square of the SM-like up and charm Yukawa coupling, respectively\footnote{Color octet resonances from the strong dynamics can favour the single production of ${\rm SO}(4)$ singlet partners ~\cite{Carmona:2012jk,Redi:2013eaa}.}. Conversely, the large top mass makes the single production of the  top partner one of the dominant mechanism, especially at large top partner mass~\cite{Contino:2008hi,Mrazek:2009yu}. It is worth to quote that single production in association with an EW gauge boson or a Higgs boson is possible~\cite{Atre:2013ap,Delaunay:2013pwa} and are not regarded here. More additional contribution will play a role once the extra fermion-vector resonance couplings $\cJ\cdot\rho$ are accounted for, generically increasing the cross section production for single partners. Likewise the production in pairs, the single productions will be controlled by the effective interacting terms among the fermions and the SM charged and neutral gauge, as well as by the model-dependent couplings $g_{u d \rho^\pm},\,g_{f f \rho^0},\,g_{X f \rho^+},\,g_{X X' \rho^+},\,g_{X f \rho^0},\,g_{X X \rho^0}$ along~\eqref{rho-charged-currents}-\eqref{top-parterners-rho-neutral-currents}. These are computed analytically in our models, and they arise from 
the interactions reported in Appendices~\ref{Charged-neutral-masses}\ref{Physical-fermion-masses} after performing the rotation to the physical basis of mass eigenstates. Since the rotation matrices can be expressed in a closed form the explicit formulae for the couplings are straightforwardly derived. The result is rather involved and for this reason it will not be reported here, however it is easily implemented in a {\sl{Mathematica}} package. 

Finally, some words concerning top partners decays are worth. The main channels are two-body decays to gauge bosons and third-family quarks. For the partners of charge $2/3$ and $-1/3$ also the decay to the Higgs boson is allowed, and competitive with the others in some cases. This originates after the rotation to the physical basis. The relevant couplings are encapsulated in~\eqref{rho-charged-currents}-\eqref{top-parterners-rho-neutral-currents} and by the analogous ones involving SM charged and neutral gauge fields correspondingly. They can be computed analytically, and therefore exact tree-level expressions for the partial widths and eventually for the branching fractions are obtained. In principle, cascade decays $X\rightarrow X' V$ or $X'H$ are also allowed. Exotic channels like $X\rightarrow X' \rho^\pm$ or $X\rightarrow X \rho^0$ are theoretically allowed but less relevant though, as they involve higher masses in the final states\footnote{For a more detailed discussion on relevant decays see~\cite{DeSimone:2012fs} and  for a more recent update check~\cite{Chala:2017xgc,Bizot:2018tds,Banerjee:2017wmg}. Early discussions on the discovery potential of top-partners in a realistic composite Higgs model with LHC data can be found in~\cite{Dissertori:2010ug,Vignaroli:2012nf}.}. Such decays arise in our models, and depending on the chosen parameters, they would either enhance or decrease some standard SM final states, and would strongly depend on the resonance mass spectrum as well as on the decaying partner mass. In a future work, we will explore these issues and the flexibility entailed by the parametric dependence for the feasibility of exotic partner decay channels.

The constraints on the top partners that are inferred from available LHC searches of similar particles, have been recently explored in~\cite{Chala:2017xgc,Bizot:2018tds} by imposing direct bounds on heavy top-like quarks with standard and exotic decays. Constraints on the allowed parameter space of our models are obtained by the imposition of recent LHC partner searches. Specifically, we excluded regions of the parameter space in terms of $\xi$ and the mass scales $M_\rho$ and $M_\Psi$.

\section{Parameter spaces and constraints} 
\label{Some-parameter-spaces}

\nt The most stringent experimental constraints on $\fourplet$ and $\singlet$ from the direct searches had been derived in~\cite{Chatrchyan:2013wfa,Chatrchyan:2013uxa}. In fact, by means of pair production mechanism driven mostly by QCD interactions, rough limits on $m_{\Xft}$ and $m_{\Xtt}$ were respectively established as $800$~GeV and $700$~GeV. Experimental searches for the singly produced partners~\cite{ATLAS:2014pga} and searches for pair production into the bounds on singly produced partners~\cite{DeSimone:2012fs,Azatov:2013hya,Matsedonskyi:2014mna,Matsedonskyi:2015dns} have been considered. Additionally, the nineplet case has been analysed yielding $m_9 \gtrsim 1$~TeV~\cite{Matsedonskyi:2014lla}. These bounds have been updated and refined following the latest ATLAS and CMS results~\cite{Aaboud:2017qpr,Sirunyan:2017pks}. The search for the pair production of vector-like top quarks in final states with exactly one lepton, at least four jets and high missing transverse momentum has allowed to exclude masses below 870~GeV (890~GeV expected) and 1.05~TeV (1.06~TeV expected), for the singlet and doublet models respectively. The search was based on $36.1\,\text{fb}^{-1}$ of $\sqrt{s}=13$~TeV LHC $pp$ collision data recorded by ATLAS in 2015 and 2016 (see~\cite{Aaboud:2017qpr} for more details).  

More recently, CMS has released~\cite{Sirunyan:2017pks} the results of a search for vector-like quarks, with electric charge of 2/3 and -4/3, respectively, that are pair produced in $pp$ interactions at $\sqrt{s} = 13$TeV and decaying exclusively via the $Wb$ channel. Events were selected requiring a lepton and neutrino from one $W$, and a quark-antiquark pair from the other boson gauge. The selection requires a muon or electron, significant missing transverse momentum, and at least four jets. A kinematic fit assuming a pair production of 2/3 or -4/3 electrically charged vector-like quarks  was performed and for every event a corresponding candidate quark mass was reconstructed.  

Upper limits were set in~\cite{Sirunyan:2017pks} for the pair production cross sections as a function of the implied vector-like quark masses. By comparing these limits with the predicted theoretical cross section of the pair production, the production of 2/3  or -4/3 electrically charged vector-like quarks is excluded at 95\% confidence level for masses below 1295 GeV (1275 GeV expected). More generally, the results set upper limits on the product of the production cross section and branching fraction for the $Wb$-channel of any new heavy quark decaying to this mode. Such limits have been imposed in $\sigma \times Br$ for all of our models and are translated into exclusion regions for the parameter spaces involved by $\xi$, $M_\rho$ and $M_\Psi$. We have analytically computed $Br\left(\cT\to Wb\right)$ and $Br(\widetilde{\cT}\to Wb)$ including a heavy resonance in the final states for the total width, and also simulated through MadGraph 5 the pair production cross section of $\cT\cT$ and $\widetilde{\cT}\widetilde{\cT}$ at $\sqrt{s} = 13$ TeV for the fourplet and singlet models respectively. Fig.~\ref{MPsi-Mrho-Xi} gathers the allowed parameter spaces $\left(M_\Psi,\,\xi\right)$ (1st-2nd plots) and $\left(M_\rho,\,\xi\right)$ (3rd-4th) for all the fourplet and singlet models, with a total decay width summing the standard modes $W b$, $Z t$ and  $h t$ up, and augmented by $\rho^0_{1,2}\,t$ and $\rho^\pm_{1,2}\,b$. Consequently, the branching ratio for any channel will be also $M_\rho$-dependent and will entail a parametric dependence on the extra fermion-vector resonance interactions regarded here in~\eqref{Currents-L-R}. Their impact is explored along the range 0.5~TeV$\lesssim M_\Psi \lesssim 1.5$~TeV, as it is the most favoured by concrete models (see~\cite{DeSimone:2012fs} and references therein), while $\xi$ spans up to the value 0.3 allowed by EWPT parameters~\cite{Agashe:2005dk,Contino:2013gna}. Two different situations are displayed: the dashed border regions stand for the allowed parameter spaces assuming extra fermion-vector resonance couplings $\cJ\cdot\rho$ weighted by $\alpha=1$, whilst the others zones denote zero additional interactions, \ie $\alpha=0$. The heavy resonance mass has been set as $M_\rho=3000$ GeV at the first and second plots, whereas the partner mass scale is fixed as $M_\Psi=1250$ GeV at the third and fourth graphs. Some comments are in order:
\begin{figure}
\begin{center}
\includegraphics[scale=0.192]{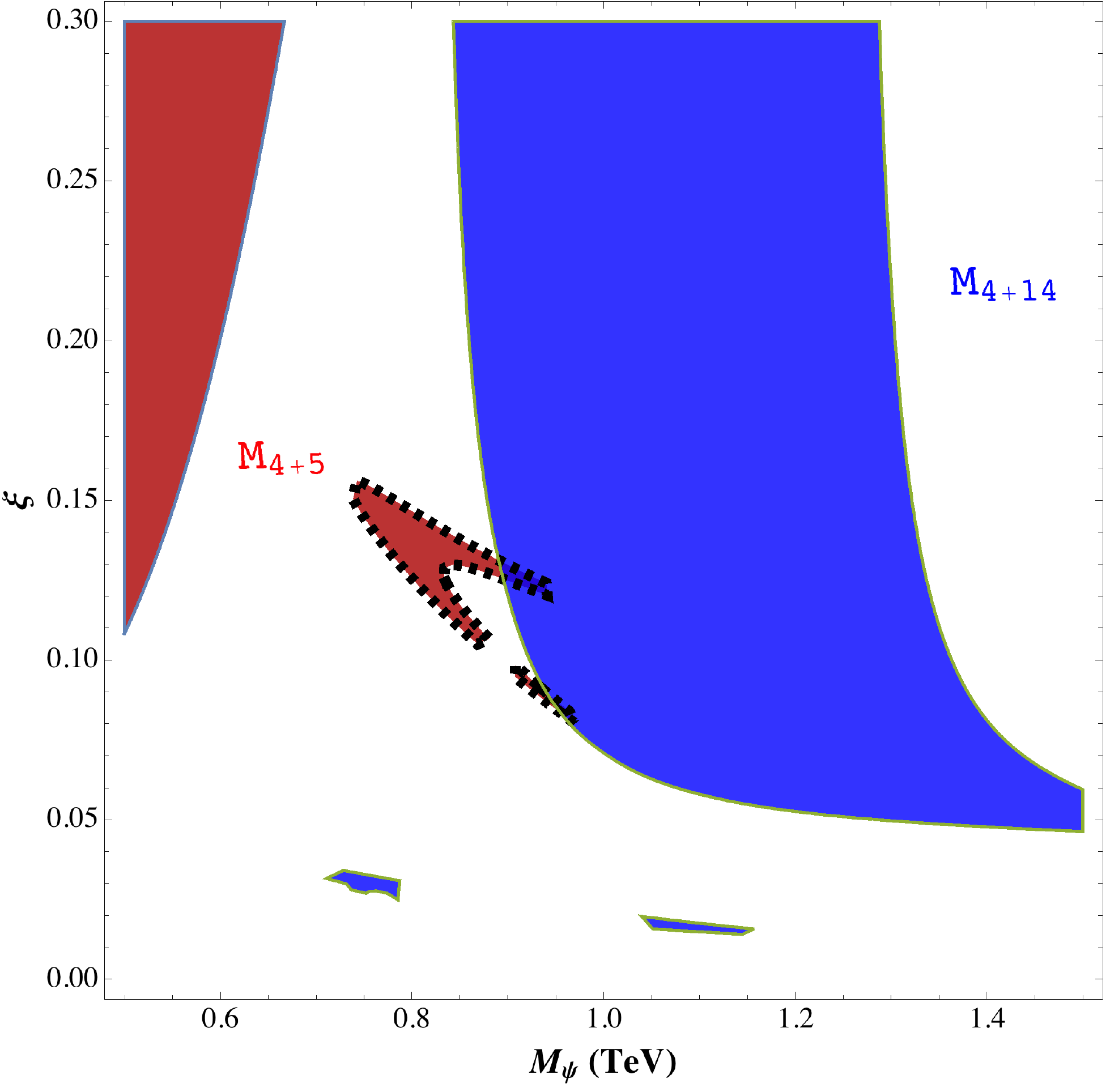}
\hspace*{0.08cm}
\includegraphics[scale=0.192]{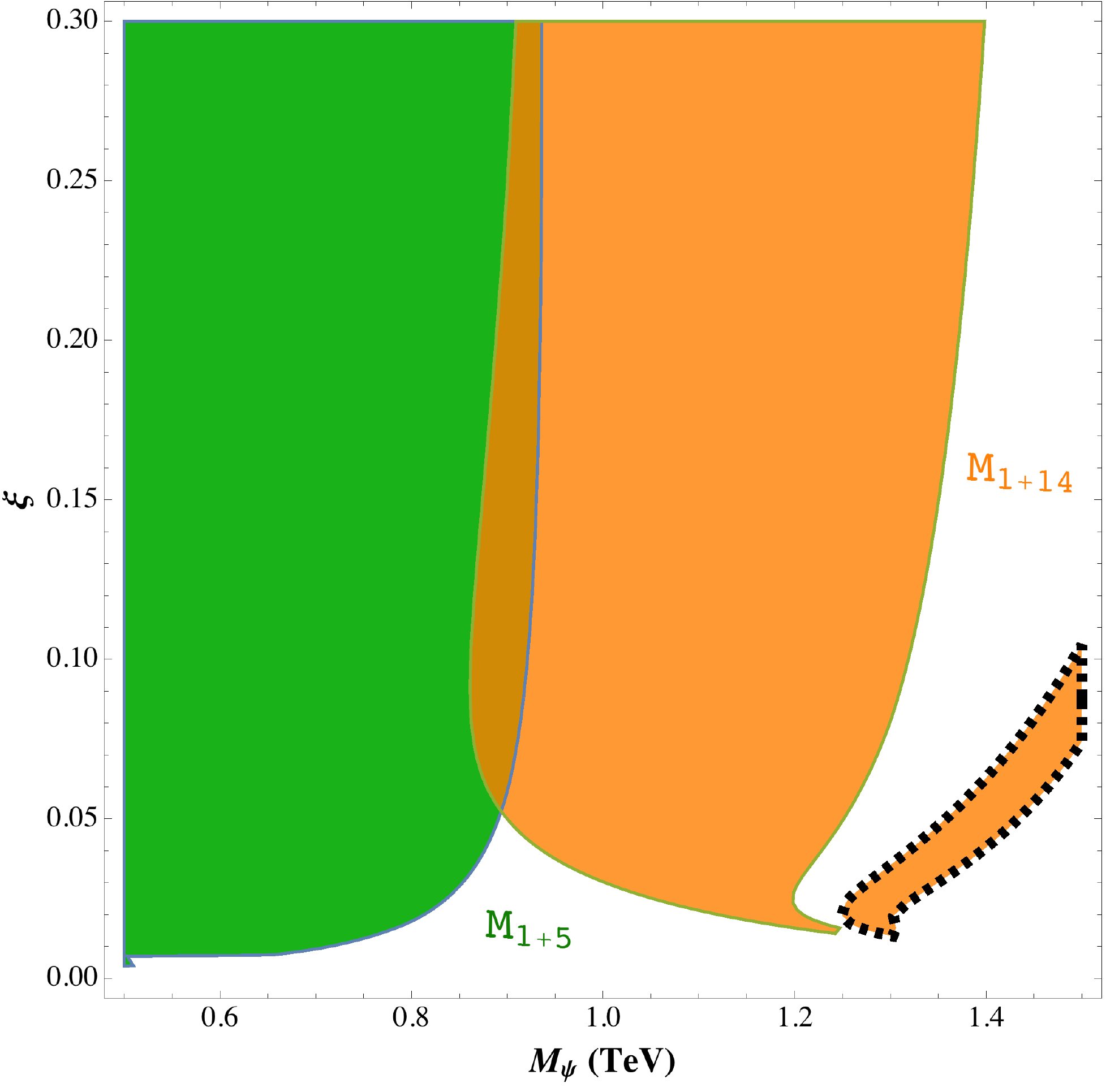}
\hspace*{0.08cm}
\includegraphics[scale=0.192]{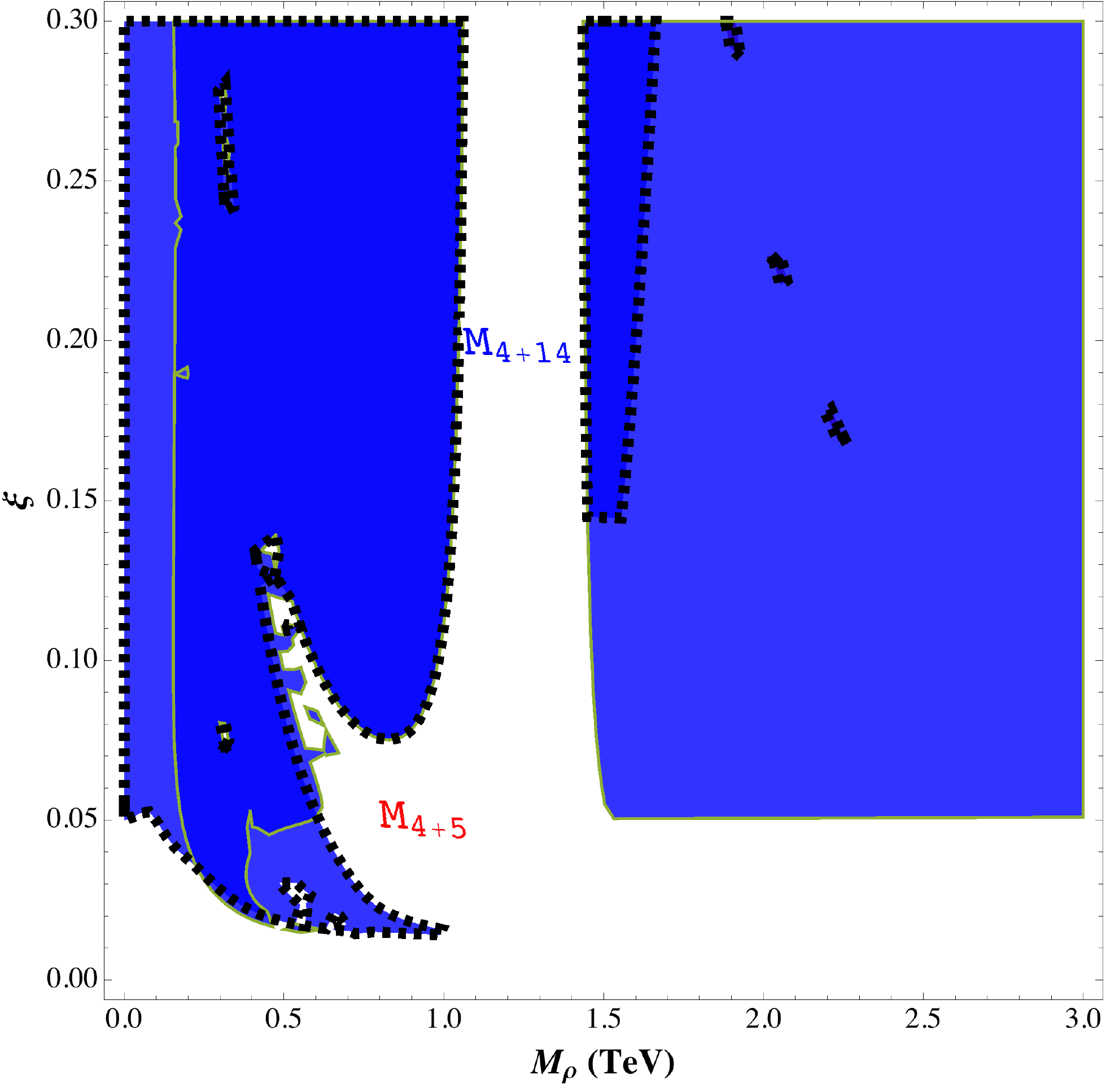}
\hspace*{0.08cm}
\includegraphics[scale=0.192]{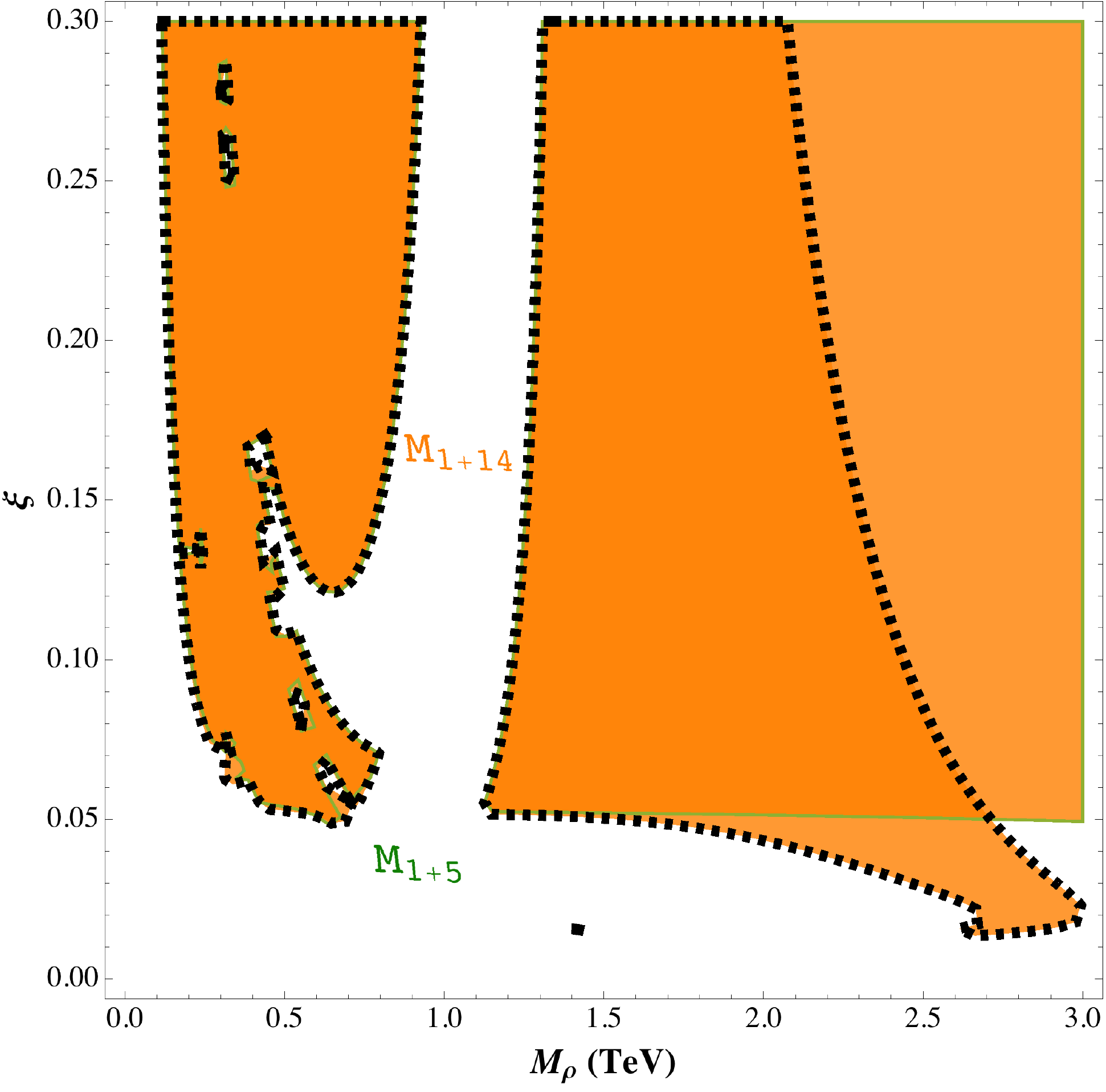}
\caption{\sf Parameter space $\left(M_\Psi,\,\xi\right)$ by setting $M_\rho=3000$ GeV (1st-2nd plots),  and $\left(M_\rho,\,\xi\right)$ by fixing $M_\Psi=1250$ GeV (3rd-4th graphs). Recent bounds~\cite{Sirunyan:2017pks} on top partner searches in the range 0.8-1.6 TeV, through top-like decays into $W b$ final states have been imposed at all models. Two situations have been explored $\alpha=0,1$ (thick-dashed border). The scale $M_\Psi$ has been scanned along 0.5-1.5 TeV, as it is the most favoured range in concrete models (see~\cite{DeSimone:2012fs}), while $\xi$ is explored till the value 0.3 allowed by the EWPT parameters (see~\cite{Agashe:2005dk,Contino:2013gna}).}
\label{MPsi-Mrho-Xi}
\end{center}
\end{figure}

\begin{itemize}

\item When accounting for extra couplings $\cJ\cdot\rho$ in~\eqref{Currents-L-R}, the parameter space in the explored mass region becomes strongly constrained to the tiny areas 
\begin{align*}
\fA:&\quad\xi\sim [0.1,\,0.15]\quad \text{for}\quad M_\Psi\sim [750,\,900]\,\text{GeV}\quad\Rightarrow\quad m_\cT\sim [1130,\,1366]\,\text{GeV}\\
\oB:&\quad\xi\sim [0.05,\,0.1]\quad \text{for}\quad M_\Psi\sim [1300,\,1500]\,\text{GeV}\Rightarrow\quad m_{\widetilde{\cT}}\sim [1055,\,1319]\,\text{GeV}
\end{align*}

The latter mass ranges are partly allowed by the recent limits~\cite{Sirunyan:2017pks} on the exclusion at 95\% confidence level for masses below 1295 GeV. The $\xi$-range at both models are compatible with EWPT bounds, the vector resonance direct production bounds at LHC, as well as the expected LHC single Higgs production, and the double Higgs production at CLIC (see discussion in~\ref{Interplay}). 

\item Conversely, by switching off the extra $\cJ\cdot\rho$-couplings, a broader parameter space is allowed and the previous ranges become relaxed. Certainly, intervals for $\xi$ compatible with experimental expectations are possible at both fourplet models, becoming ruled out at $\fA$ as they entirely fall inside the exclusion limit of~\cite{Sirunyan:2017pks}. By directly excluding masses below 1295~GeV~\cite{Sirunyan:2017pks} on the theoretical mass $m_\cT$ of Eq.~\eqref{Masses-expanded-5}, and combining the latter mass regions with the initial parameter spaces shown in Fig.~\ref{MPsi-Mrho-Xi}, we finally obtain the lower bound at $\fB$ (see Fig.~\ref{4plet-singlet-masses})
\be
\fB:\quad 0.05\lesssim\xi\lesssim 0.30\quad \text{for}\quad M_\Psi\gtrsim 1150\,\text{GeV}
\ee
favouring a extreme part of the obtained parameter space in Fig.~\ref{MPsi-Mrho-Xi}. Likewise, the exclusion limit in~\cite{Sirunyan:2017pks} on the theoretical masses $m_{\widetilde{\cT}}$ of Eq.~\eqref{Masses-expanded-14} leads to
\begin{align*}
\oA:&\quad 0.01\lesssim\xi\lesssim 0.15\quad \text{for}\quad M_\Psi\sim [500,\,930]\,\text{GeV},\\
\oB:&\qquad\qquad \xi\gtrsim 0.3\quad\,\,\,\,\text{for}\quad M_\Psi\sim [1350,\,1400]\,\text{GeV}
\end{align*}
approving a small region for the associated parameter spaces at both models. Tiny $\xi$-values are allowed and still compatible with experimental constraints at both singlet models. $\oB$ gets disfavoured as $\xi$ falls outside the upper bound allowed by EW precision parameters.

\end{itemize}

\nt As a conclusion, the recent upper limits on top-like partners production permit part of the parameter spaces from $\fB$ and from the singlet models as well, and along the explored top partner mass region 0.5-1.5 TeV\footnote{Having scanned along such mass range does not exclude the possible existence of heavier top partner masses at higher energy scales beyond the LHC reach. The aim has been to focus our parameter space along the mass ranges explored by~\cite{Sirunyan:2017pks}, motivated by~\cite{DeSimone:2012fs} and within the current LHC reach.}. By including additional fermion-vector resonance couplings a strongly bounded region at $\fA$ remains only ($\oB$ disfavoured by EWPT). In this sense, those extra couplings are helpful in disregarding-selecting models and refining further their involved parameter space. 

An additional insight into the parametric freedom can be perfomed by fixing now the partner mass scale and letting the resonance mass to vary. This is illustrated in Fig.~\ref{MPsi-Mrho-Xi} (3rd-4th plots) where it has been set $M_\Psi=1250$ GeV (a bit below the threshold for the exclusion limit~\cite{Sirunyan:2017pks}). We can infer:
\begin{figure}
\begin{center}\includegraphics[scale=0.32]{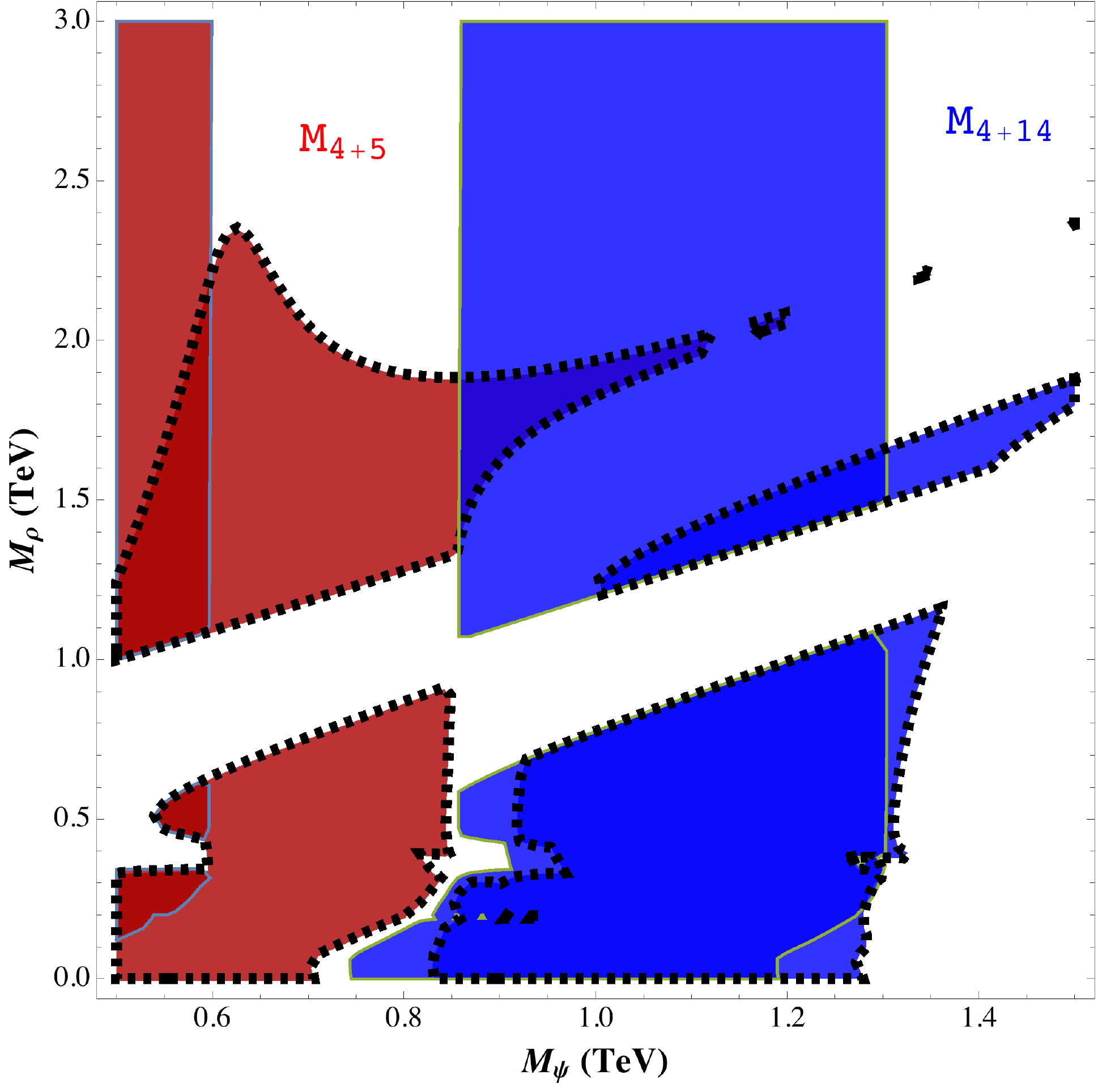}
\hspace*{1.5cm}
\includegraphics[scale=0.315]{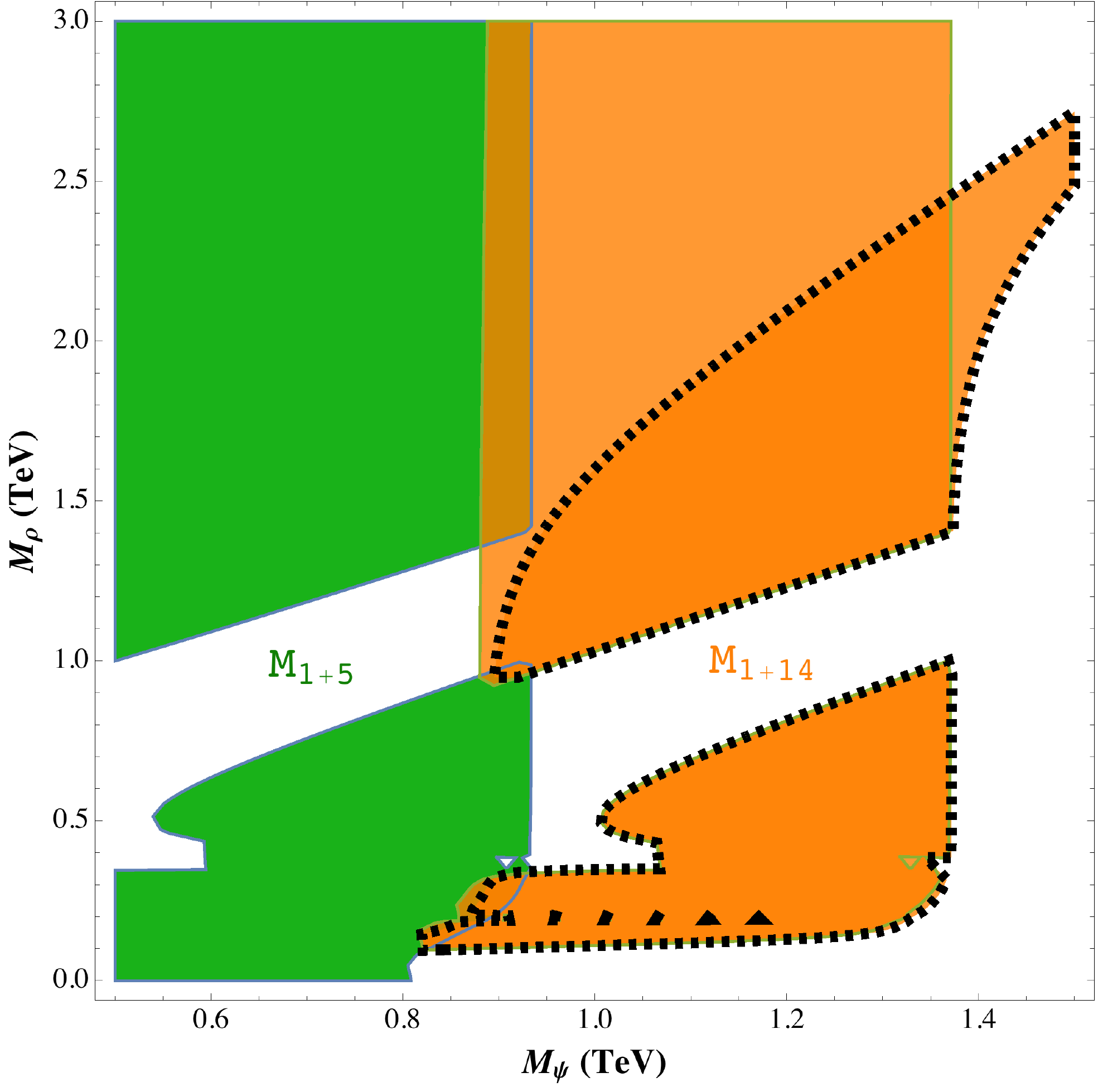}
\caption{\sf Parameter space $\left(M_\Psi,\,M_\rho\right)$ by fixing $\xi=0.2$ at the fourplet (left) and singlet models (right). Two situations have been scanned, $\alpha=0,1$ (thick-dashed border) by imposing recent bounds~\cite{Sirunyan:2017pks} on top partner searches through top-like decays into $W b$ final states at all models. }
\label{MPsi-Mrho}
\end{center}
\end{figure}

\begin{itemize}

\item Before including extra couplings, the parameter spaces are similar at both fourplet and singlet models, notoriously split into a left and right-handed regions, and with slight differences at low and high resonance mass. All the left-handed regions are ruled out by the analysis in~\cite{CarcamoHernandez:2017pei,Shu:2016exh,Shu:2015cxm} and the experimental searches in~\cite{Khachatryan:2016cfx} as they fall well below the lower limit of $\sim$ 2000 GeV for the resonance mass. On the other hand, the right-hand regions fall partly inside the resonance mass excluded region, while permitting a relatively large area for $\fB$ and $\oB$ consistent with the feasible $\xi$-values.

\item After turning on the additional couplings of~\eqref{Currents-L-R}, the parameter space for $\fB$ is slightly enlarged at the left-handed side towards low resonance mass, incompatible with experimental expectations. For the same model, the right-handed region shortens, leaving a small corner compatible with the range $M_\rho\sim$ 2000-2100 GeV. Numerically, it is also allowed the pretty small area around $M_\rho\sim 2850$ GeV and $\xi\sim 0.1$. For the singlet scenarios, $\oB$ allows resonance masses compatible with the expectations, and even for small $\xi\sim 0.02$. In this case, the final permitted area is larger compared with the fourplet case. 
\end{itemize}

\nt Finally, a deeper insight into the parametric dependence is gained by fixing the EW and GB scales while letting the resonance and quark partner mass scales to simultaneously vary. Fig.~\ref{MPsi-Mrho} displays the involved allowed areas for $\xi=0.2$. The influence of the extra fermion-vector resonance interactions proposed in this work is remarkably observed, specially when it tends to drive the permitted regions outside the excluded vector-like quark masses range. Although they become ruled out at $\fB$ by the expected resonance mass, a small window it is still feasible in the case of $\fA$. Likewise, a small region remains for $\oB$. Conversely, when turning the extra interactions off, the latter model does not allow any region compatible with~\cite{Sirunyan:2017pks} (for $\xi=0.2$), though compatible with resonance mass expectations.  At the $\bf{5}$-embeddings scenario no region remains, whilst at $\fB$ a small band results compatible with the expectations. In summary, including extra $\cJ\cdot\rho$-couplings will allow $\fA$ and $\oB$ at small windows, while the removal of those couplings will rule out them for $\fB$ at a relatively small band.

\section{Summary}
\label{Summary}

\nt We have explored in this work the phenomenological signals arising from the interplay among three matter sectors: elementary, top partners and vector resonances in a $SO(5)$ composite Higgs Model. The vector resonance $\rho$, here assumed to be spin-1 triplet of $SU(2)_L\times SU(2)_R$, is coupled to the $SO(5)$-invariant fermionic currents and 2nd rank tensors listed in Table~\ref{Fermion-currents-set} that were proposed in~\cite{Yepes:2017pjr}. The top partners permitted by the unbroken $SO(4)$ are here restricted to the fourplet $\fourplet$ and singlet $\singlet$ embeddings. Such matter content spans four models in~\eqref{Models}, coupled each of them to the $\rho$-resonance via the prescription~\eqref{Currents-L-R} and subsequently scanned along their involved parametric dependence.

Heavy spin-1 production and their decays have been thoroughly studied along some range for the resonance mass scale $M_{\rho}$ and for a given model parameters election in Fig.~\ref{rho-production-cross-sections}. The model $\fA$ is the most predominant one in yielding either charged or neutral heavy resonances, while a higher $\xi=v^2/f^2$ enhances all the productions, but the one for the $\rho^0_2$ at $\fB$ where its production is diminished. Whether the top partner is a fourplet or singlet, the $\bf{5}$-elementary fermions scenario favours higher production values rather than the $\bf{14}$-ones. Among the charged and neutral resonances, $\rho^+_2$ and $\rho^0_2$ are predominantly yielded at $\fA$, reaching rough cross section values of $\sim 400$ pb (20 pb) and $\sim 600$ pb (10 pb)  at $M_\rho \sim 0.6$ TeV (3 TeV) for $\xi=0.2$.  The resonance production is generically increased by the presence of the fermion-vector resonance $\alpha\,\cJ\cdot\rho$-interactions of~\eqref{Currents-L-R}. In some cases such enhancement occurs by some orders of magnitude and are depending of course on the strength of the weighting coefficients $\alpha$. In this work we have tested two situations, $\alpha=0,1$, aimed at exploring the impact of the aforementioned additional couplings upon the implied phenomenology.

Prior to the inclusion of the extra couplings,  the pair gauge $W W$, $W Z$ and the gauge-Higgs $Wh$, $Zh$ final states become dominant for the resonance decay channels, with extremely suppressed-subdominant fermionic modes for the charged-neutral resonances. After adding the extra fermion-vector resonance terms, the pair gauge and the gauge-Higgs final states are still the relevant ones at lower $M_\rho$, becoming subdominant with respect to $\rho^\pm_1\to\{\,t\Xft,\,b\Xtt\}$ and $\rho^\pm_2\to \{t\Xft,\,\Xft\Xtt\}$ at $M_\rho\gtrsim 1$ TeV. The channel $tb$ turns out to be dominant along the explored mass range for the charged resonance decays, as well as the modes $\rho^0_{1,2}\to\{bb,\,tt,\,jj\}$. Higher $M_\rho$-values will trigger other exotic channels, dominant compared with the non-fermionic final states, \eg $\rho^\pm_1\to\{\cT\Xft,\,\cB\Xtt,\,\Xft\Xtt\}$, $\rho^\pm_2\to \{tB,\,\cT\cB\}$ and  $\rho^0_{1,2}\to\{\cB\cB,\,\cT\cT\}$. Even for no fermion-vector resonance currents there will be exotic fermionic modes still active, although less relevant as the gauge and gauge-Higgs channels. Important contributions emerge for such exotic modes after bringing $\cJ \cdot \rho$ onto the stage, some of them being enhanced by one-two orders of magnitude, or even three orders as in the case for $\rho^\pm_2\to b\Xtt$ and  $\rho^0_{1,2}\to\{b\cB,\,t\cT\}$ at higher regime mass $M_\rho$.

Complementarily, the production of  double and single-partner final states has been scanned along the partner mass scale $M_\Psi$ in this work, and they turn out to be controlled by the model-dependent couplings $g_{u d \rho^\pm},\,g_{f f \rho^0},\,g_{X f \rho^+},\,g_{X X' \rho^+},\,g_{X f \rho^0},\,g_{X X \rho^0}$ through~\eqref{rho-charged-currents}-\eqref{top-parterners-rho-neutral-currents} and by the analogous ones involving SM charged and neutral gauge fields correspondingly. QCD drives the double production, as well as the Higgs and $\rho^0$-mediated processes. The $\rho^\pm$-mediated processes also appear for the single production in the case of charged final states. QCD pair production is completely model-independent, although non-zero model-dependent modifications are induced as soon as extra fermion-vector resonance effects are accounted for (Figs.~\ref{Double-partner-production}-\ref{Single-partner-production}). They may enhance double-partner production by one order of magnitude at fourplet models, whereas vanishing contributions and tiny ones are obtained at the singlet. The combined effect of fermion-vector resonance rotation as well as the less number of additional fermionic currents determine such behaviour for these models. Generically, producing pairs either of $\Xtt$ or $\Xft$ will be kinematically favoured at fourplet scenarios with respect double production of both $\cT$ and $\cB$. Similar arguments alike to the fourplet case work for the pair production of the singlet $\widetilde{\cT}$, where the involved masses result smaller at $\bf{14}$-elementary embeddings compared with the one at $\bf{5}$-scenario (Fig.~\ref{4plet-singlet-masses}), favouring the former scenario for its production in pairs. 

Finally, implementing the recent LHC searches for vector-like quarks production in $pp$-collisions at 13 TeV, we were able to exclude regions of the parameter space entailed by our framework (Figs.~\ref{MPsi-Mrho-Xi}-\ref{MPsi-Mrho}). Specifically, part of the parameter space $\left(M_\Psi,\,\xi\right)$ at the singlet models and at $\fB$ are permitted, while being totally ruled out at $\fA$. By including additional fermion-vector resonance couplings a strongly bounded region at $\fA$ and $\oB$ remains only, being compatible with recent experimental LHC exclusion limits. Analogous comments apply for $\left(M_\rho,\,\xi\right)$, while a small window for the region $\left(M_\Psi,\,M_\rho\right)$ is allowed at $\fA$ and $\oB$ if the extra $\cJ\cdot\rho$-couplings are summed up. Otherwise, they become ruled out instead for $\fB$ at a relatively small band compatible with heavy resonance searches.

Concerning all these remarks, those extra fermion-vector resonance couplings proposed in~\cite{Yepes:2017pjr} are helpful in disregarding-selecting models and refining further their involved parameter space.  All this clearly signals a feasible scenario for explaining future observations of heavy resonance and top partner production, as well as their subsequent posterior decays. In this sense, the interactions encoded by~\eqref{Currents-L-R} and Table~\ref{Fermion-currents-set}, might help in determining the model and the strength for the involved effective terms by contrasting their predictions with the experimental signals arising at higher energies. All in all, the EFT approach proposed in this work could be powerful in dealing with BSM frameworks, specifically in coping with new interactions that might underlie the existence of exotic matter in our nature, and potentially discoverable at future high energy colliders.

\section*{Acknowledgements}

\nt J.Y. acknowledges valuable comments from  Dom\`{e}nec Espriu, Christophe Grojean, Oleksii Matsedonskyi and Sebastian Norero, as well as the hospitality at the Universitat de Barcelona and DESY where part of this work was carried out. J.Y. thanks  the support of Fondecyt (Chile) grant No. 3170480. The work of A.Z. is supported by Conicyt (Chile) grants ACT1406 and PIA/Basal FB0821, and by  Fondecyt (Chile) grant 1160423.

\appendix
\small

\section{CCWZ formalism}
\label{CCWZ}

\nt The $SO(4) \simeq SU(2)_L \times SU(2)_R$ unbroken generators and the broken ones parametrizing the coset $\textrm{SO}(5)/SO(4)$ in the fundamental representation are
\be
(T^a_\chi)_{IJ} = -\frac{i}{2}\left[\frac{1}{2}\varepsilon^{abc}
\left(\delta_I^b \delta_J^c - \delta_J^b \delta_I^c\right) \pm
\left(\delta_I^a \delta_J^4 - \delta_J^a \delta_I^4\right)\right],\qquad
T^{i}_{IJ} = -\frac{i}{\sqrt{2}}\left(\delta_I^{i} \delta_J^5 - \delta_J^{i} \delta_I^5\right)\,,
\label{eq:SO4_gen-SO5/SO4_gen}
\ee

\nt with $\chi= L,\,R$ and $a= 1,2,3$, while $i = 1, \ldots, 4$. The normalization of $T^{A}$'s is chosen as ${\rm Tr}[T^A, T^B] = \delta^{AB}$. The $4\times 4$ matrices $\tau^a$ appearing in the bilinear fourplets at $\fA$ and $\fB$ in Table~\ref{Fermion-currents-set} are defined as
\be
\left[T^{a},T^{i}\right]=\left(t^{a}\right)_{{j}{i}}T^{j}\,.
\label{taus}
\ee

\nt Gauging the SM subgroup of $SO(5)$ requires us to introduce local transformations via $U$ matrices that will couple the SM gauge fields to the composite resonances. The CCWZ $d$ and $e$ symbols are in order to do so 
\be
-U^t [A_\mu + i \partial_\mu ] U = d_\mu^{\hat a} T^{\hat a} + e^{a}_{\mu} T^a + e^{X}_{\mu}
\label{d-e}
\ee

\nt where $A_{\mu}$ stands for ${\cal G}_{\text{SM}}$ gauge fields 
\be
\begin{aligned}
A_\mu &= \frac{g}{\sqrt{2}}W^+_\mu T^-_L +\frac{g}{\sqrt{2}}W^-_\mu T^+_L+ g \left(\cw Z_\mu+\sw A_\mu \right)T_L^3+g' \left(\cw A_\mu-\sw Z_\mu \right)(T_R^3+Q_X)
\label{gfd}
\end{aligned}
\ee

\nt with $T^\pm_\chi=\left(T^1_\chi\mp i T^2_\chi\right)/\sqrt{2}$, the implied notation $(\cw,\,\sw)\equiv(\cos\theta_{\rm w},\,\sin\theta_{\rm w})$, and the SM couplings of $SU(2)_L$ and $\textrm{U}(1)_Y$, $g$, $g'$ respectively, where $Q_X$ is the $X$-charge matrix. The definition~\eqref{d-e} can be expanded in fields as
\be
d_\mu^i=\frac{\sqrt{2}}{f}(D_\mu h)^i+{\cal O} (h^3),\qquad
e^{a}_\mu = -A^{a}_\mu-\frac{i}{f^2}(h{\lra{D}_\mu}h)^a+{\cal O} (h^4),\qquad
e^{X}_\mu = -g^{\prime}  Q_X B_\mu \,,
\label{CCWZ-symbols}
\ee

\nt with $B_{\mu}$ the $U(1)_Y$ gauge boson. Covariant derivatives acting on the composite sector fields are built out of $e$ symbols. For the $\Psi$ field transforming in the fundamental representation of $SO(4)$ one has
\be
\nabla_\mu\Psi \,=\,D_\mu\Psi+i\,e_{\mu}^at^a\Psi\,.
\label{covariant-derivative}
\ee
The term $\slashed{e}=e_\mu\gamma^\mu$ is included in $\LL_{\text{comp}}$ to fully guarantee the $SO(5)$ invariance. Strength field tensors are straightforwardly introduced as
\be
e_{\mu \nu} = \partial_{\mu} e_{\nu} - \partial_{\nu} e_{\mu} + i g_{\rho} [e_{\mu},e_{\nu}], \qquad\qquad
e_{\mu \nu}^{X} = \partial_{\mu} e^X_{\nu} - \partial_{\nu} e^X_{\mu}.
\ee

\nt Finally, the covariant derivatives $D_\mu$ associated to each one of the elementary fields as well to the corresponding top partner are given by

\bea
D_\mu\,q_L&=&\left(\partial_\mu-ig_W W_\mu^i {\sigma^i\over 2}-i{1\over 6}g_B B_\mu-i\,g_SG_\mu\right)q_L\,, \\
D_\mu\,u_R&=&\left(\partial_\mu-i{2\over 3}g_B B_\mu-i\,g_SG_\mu\right)u_R \,,\\
D_\mu\fourplet &=& \left(\partial_\mu-i {2\over 3} g_B B_\mu -i\,g_SG_\mu\right)\fourplet\,.
\label{cder}
\eea

\nt The SM gauge couplings $g$ and $g'$ will be generated via mixing effect with the couplings $\grhoL$ and $\grhoR$ entailed by the heavy vector resonances sector.

\section{Mass spectrum}
\label{Physical-basis}
\subsection{Charged-neutral gauge masses}
\label{Charged-neutral-masses}

\nt The gauge basis is defined by $\left(W'^\pm,\,\,\rho^\pm_L,\,\,\rho^\pm_R\right)^T$ and $\left(W^3,\,\,\rho^3_L,\,\, B,\,\,\rho^3_R\right)^T$,  with the charged fields $W_{\mu,\chi}^\pm$ defined as usual $W'^\pm\equiv \left(W^1\,\mp\,i\,W^2\right)/\sqrt{2}$ and 
$\rho_\chi^\pm\equiv \left(\rho_\chi^1\,\mp\,i\,\rho_\chi^2\right)/\sqrt{2}$. The mass eigenstate basis is defined by $\left(
W^\pm,\,\,
\rho^\pm_1,\,\,
\rho^\pm_2
\right)^T$ and $\left(
A,\,\,
Z,\,\,
\rho^0_1,\,\,
\rho^0_2
\right)^T$ and it can be linked to the gauge basis through field transformations. Considering canonical $\rho$-fields and linear EWSB effects, the mass matrices for the charged sector turn out to be in the gauge basis as
\be
\cM_{\cW}=
\left(
\begin{array}{ccc}
 \frac{g^2\,\mrhoL^2}{\grhoL^2}+ \frac{g^2\,\xi}{2} \left(\frac{\mathit{f}^2}{2}-\frac{\mrhoL^2}{\grhoL^2}\right) & -\frac{g\,\mrhoL^2}{\grhoL}\left(1-\frac{\xi}{4}\right) & -\frac{g\,\xi\,\mrhoR^2}{4\,\grhoR} \\[3mm]
 -\frac{g\,\mrhoL^2}{\grhoL}\left(1-\frac{\xi}{4}\right) & \mrhoL^2 & 0 \\[3mm]
 -\frac{g\,\xi\,\mrhoR^2}{4\,\grhoR} & 0 & \mrhoR^2 \\
\end{array}
\right)
\label{Charged-gauge-matrix}	
\ee

\nt while for the neutral sector 
\be
\hspace*{-0.6cm}
\small{
\cM_{\cN}=\left(
\begin{array}{cccc}
 \frac{g^2 \mrhoL^2}{\grhoL^2}+ \frac{g^2\,\xi}{2} \left(\frac{\mathit{f}^2}{2}-\frac{\mrhoL^2}{\grhoL^2}\right) &  -\frac{g\,\mrhoL^2}{\grhoL}\left(1-\frac{\xi}{4}\right) & -\frac{g\,g'\,\xi}{4}   \left(\mathit{f}^2-\frac{\mrhoL^2 }{\grhoL^2}-\frac{\mrhoR^2}{\grhoR^2}\right) & -\frac{g \xi  \mrhoR^2}{4 \grhoR} \\[3mm]
-\frac{g\,\mrhoL^2}{\grhoL}\left(1-\frac{\xi}{4}\right)& \mrhoL^2 & -\frac{\xi  \mrhoL^2 g'}{4 \grhoL} & 0 \\[3mm]
-\frac{g\,g'\,\xi}{4}   \left(\mathit{f}^2-\frac{\mrhoL^2 }{\grhoL^2}-\frac{\mrhoR^2}{\grhoR^2}\right) & -\frac{\xi  \mrhoL^2 g'}{4 \grhoL} & \frac{\mrhoR^2 g'^2}{\grhoR^2}+\frac{g'^2\,\xi}{2} \left(\frac{\mathit{f}^2}{2}-\frac{\mrhoR^2}{\grhoR^2}\right) & -\frac{g'\,\mrhoR^2}{\grhoR}\left(1-\frac{\xi}{4}\right) \\[3mm]
 -\frac{g \xi  \mrhoR^2}{4 \grhoR} & 0 &  -\frac{g'\,\mrhoR^2}{\grhoR}\left(1-\frac{\xi}{4}\right) & \mrhoR^2 \\
\end{array}
\right).
\label{Neutral-gauge-matrix}
}
\ee

\nt The rotation matrices are rather involved to be presented here. Expanding up to the order $\cO(\xi)$, the charged gauge masses are
\be
M^2_W=\frac{1}{4}\,g^2\,f^2\,\xi,\,\qquad M^2_{\rho^\pm_1}=m_{\rho_R}^2,\,\qquad M^2_{\rho^\pm_2}=\mrhoL^2\,\frac{\grhoL^2}{\grhoL^2-g^2} \left[1-\frac{\xi}{2}\,\frac{g^2}{\grhoL^2}\left(1-\frac{g^2\mathit{f}^2}{2\,\mrhoL^2}\right)\right]
\label{Charged-masses-expanded}
\ee

\nt whereas the neutral ones are
\be
M^2_Z=\frac{1}{4}  \left(g^2+g'^2\right)\mathit{f}^2 \xi,
\label{Z-mass}
\ee

\be
\begin{aligned}
 M^2_{\rho^0_1}&=\mrhoL^2\,\frac{\grhoL^2}{\grhoL^2-g^2} \left[1-\frac{\xi}{2}\,\frac{g^2}{\grhoL^2}\left(1-\frac{g^2\mathit{f}^2}{2\,\mrhoL^2}\right)\right],\\
 \\
 M^2_{\rho^0_2}&=\mrhoR^2\,\frac{\grhoR^2}{\grhoR^2-g'^2} \left[1-\frac{\xi}{2}\,\frac{g'^2}{\grhoR^2}\left(1-\frac{g'^2\mathit{f}^2}{2\,\mrhoR^2}\right)\right]\,,
\label{rho-Neutral-masses-expanded}
\end{aligned}
\ee

\nt where the link among the $SU(2)_{L,R}$ and the SM weak and hypercharge gauge couplings
\be
\frac{1}{\grhoL^2} + \frac{1}{g^2_W}= \frac{1}{g^2},\qquad\qquad\frac{1}{\grhoR^2} + \frac{1}{g^2_B}= \frac{1}{g'^2}
\label{LR-gauge-couplings}
\ee 

\nt have been implemented and will be thoroughly used hereinafter.

\subsection{Physical fermion masses}
\label{Physical-fermion-masses}

\subsubsection{${\bf 5}$-plets embeddings}
\label{Physical-fermion-masses-5}

\nt Considering both the fourplet and singlet models simultaneously, the mass matrices for the top-like and bottom-like sectors become
\be
\left(
\begin{array}{cccc}
 0 & \frac{1}{2} \mathit{f}\,\xi\, y_L & \mathit{f} (1-\frac{\xi}{2}) y_L & -\frac{\mathit{f} \sqrt{\xi }\,\tilde{y}_L}{\sqrt{2}} \\ [3mm]
 -\frac{\mathit{f} \sqrt{\xi }\,y_R}{\sqrt{2}} & -M_{\bf{4}} & 0 & 0 \\ [3mm]
 \frac{\mathit{f} \sqrt{\xi }\,y_R}{\sqrt{2}} & 0 & -M_{\bf{4}} & 0 \\ [3mm]
 \mathit{f} \sqrt{1-\xi }\,\tilde{y}_R & 0 & 0 & -M_{\bf{1}} \\
\end{array}
\right),\qquad\qquad \left(
\begin{array}{cc}
 0 & \mathit{f} y_L \\
 0 & -M_{\bf{4}} \\
\end{array}
\right)
\label{Top-bottom-matrices-5}
\ee

\nt being defined in the fermion field basis $\left(t,\,\,\Xtt,\,\,\cT,\,\,\widetilde{\cT}\right)^T$ and $\left(b,\,\,\cB\right)^T$ respectively. After diagonalization the physical masses are
\beq
\begin{aligned}
&m_t=\sqrt{\frac{\xi \left(\tilde{\eta }_L\,\tilde{\eta }_R M_{\bf{1}} -\eta _L\eta _R M_{\bf{4}}\right)^2}{2\left(\eta _L^2+1\right) \left(\tilde{\eta }_R^2+1\right)}}\,,\quad\quad
&&\quad m_{\widetilde{\cT}}=M_{\bf{1}} \sqrt{\tilde{\eta }_R^2+1}\,,\quad\quad 
&&m_\cB=M_{\bf{4}} \sqrt{\frac{\eta _L^2+1}{\tilde{\eta }_R^2+1}}\,,  \\[4mm]
&m_\cT=M_{\bf{4}}\sqrt{\eta _L^2+1} \,,\quad\qquad
&&m_\Xtt=m_\Xft=M_{\bf{4}}\,,\quad\quad 
&&  \\
\end{aligned}
\label{Masses-expanded-5}
\eeq

\nt where the parameters $\eta_{L(R)}$ are defined through
\be
\eta _{L(R)} \equiv  \frac{y_{L(R)} \mathit{f}}{M_{\bf{4}}},\qquad\qquad
\tilde{\eta}_{L(R)}\equiv  \frac{\tilde{y}_{L(R)} \mathit{f}}{M_{\bf{1}}}.
\label{eta-parameters}
\ee

\subsubsection{${\bf 14}$-plets embeddings}
\label{Physical-fermion-masses-14}

\nt Considering both the fourplet and singlet models simultaneously, the mass matrix for the top-like sector is
\be
\small{
\left(
\begin{array}{cccc}
 -\mathit{f} \sqrt{2\,(1-\xi)\,\xi }\,y_R & \frac{1}{2} \mathit{f} \left(2 \xi +\sqrt{1-\xi }-1\right) y_L & \frac{1}{2} \mathit{f} \left(-2 \xi +\sqrt{1-\xi }+1\right) y_L & -\mathit{f} \sqrt{2\,(1-\xi)\,\xi }\,\tilde{y}_L \\
 0 & -M_{\bf{4}} & 0 & 0 \\
 0 & 0 & -M_{\bf{4}} & 0 \\
 0 & 0 & 0 & -M_{\bf{1}} \\
\end{array}
\right)
\label{Top-matrix-14}
}
\ee

\nt while the one for the bottom-like sector becomes
\be
\left(
\begin{array}{cc}
 0 & \mathit{f} \sqrt{1-\xi }\,y_L \\
 0 & -M_{\bf{4}} \\
\end{array}
\right)
\label{Bottom-matrix-14}
\ee

\nt The corresponding physical masses are
\beq
\begin{aligned}
&m_t=\sqrt{2} \sqrt{\frac{\xi \left(M_{\bf{1}} \tilde{\eta }_L \tilde{\eta }_R+M_{\bf{4}} \eta _R\right)^2}{\left(\eta _L^2+1\right) \left(\tilde{\eta }_R^2+1\right)}}\,,\quad\quad
&&\quad m_{\widetilde{\cT}}=\frac{M_{\bf{1}}}{\sqrt{\tilde{\eta }_R^2+1}}\,,\quad\quad 
&&m_\cB=M_{\bf{4}} \sqrt{\frac{\eta _L^2+1}{\tilde{\eta }_R^2+1}}\,,  \\[4mm]
&m_\cT=M_{\bf{4}}\sqrt{\eta _L^2+1} \,,\quad\qquad
&&m_\Xtt=m_\Xft=M_{\bf{4}}\,,\quad\quad 
&&  \\
\end{aligned}
\label{Masses-expanded-14}
\eeq

\begin{figure}
\begin{center}
\includegraphics[scale=0.32]{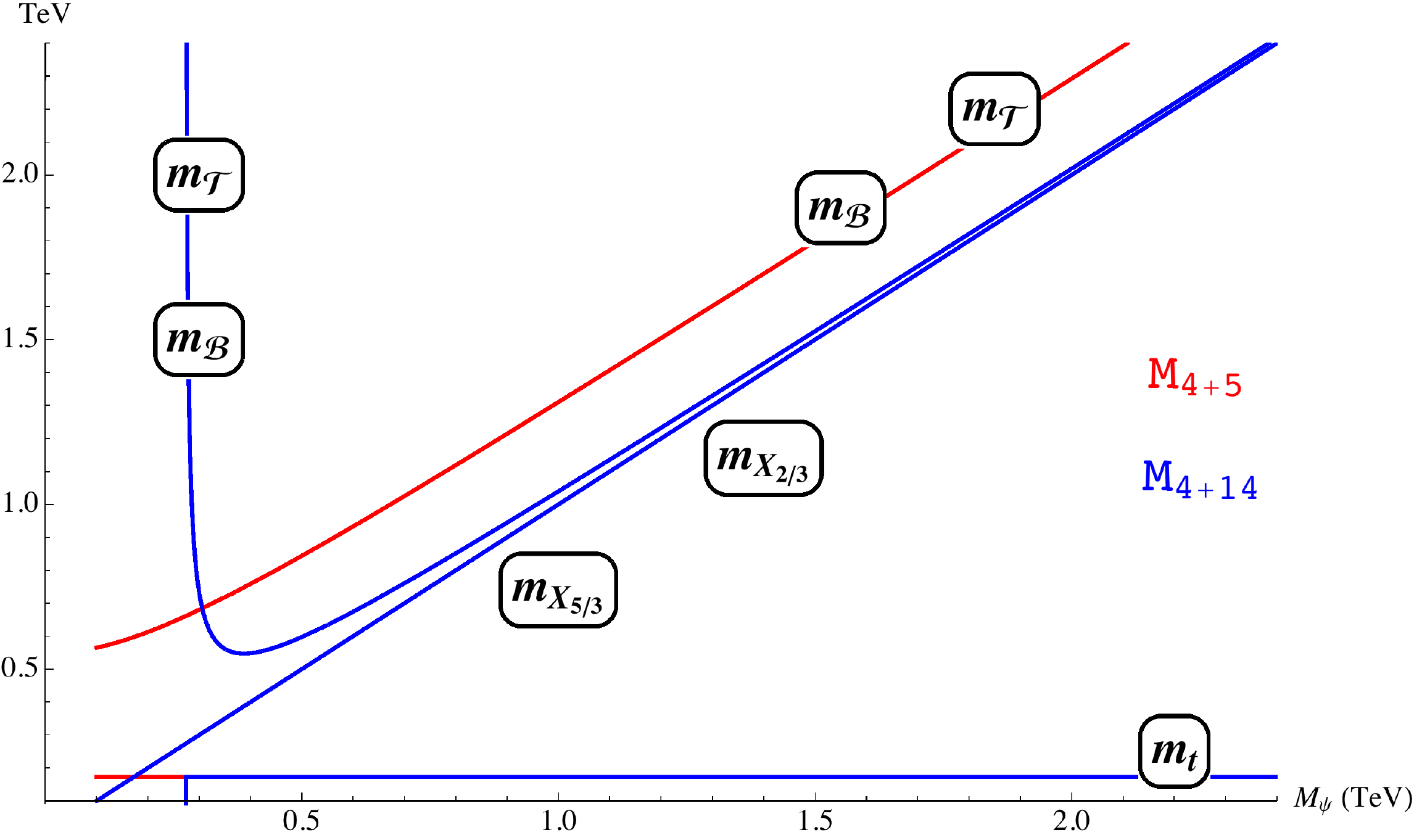}
\hspace*{1.3cm}
\includegraphics[scale=0.31]{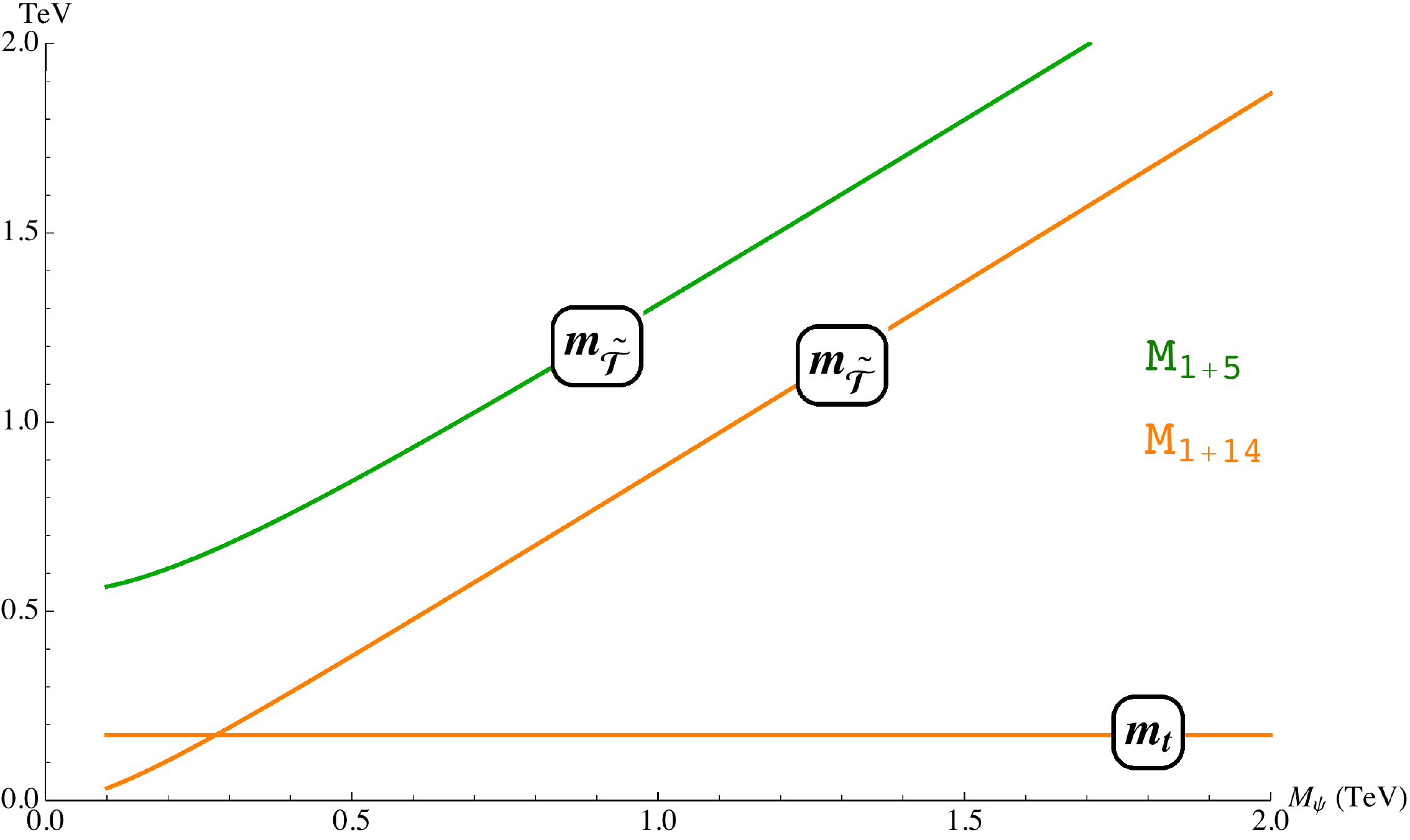}
\caption{\sf Spectrum of masses and their dependence on the NP scale $M_{\bf{4}}=M_{\bf{1}}=M_\Psi$ for the fourplet (left) and singlet cases (right), for $\xi=0.2$ and setting $\eta_L=\eta_R$, $\tilde{\eta }_L=\tilde{\eta }_R$.}
\label{4plet-singlet-masses}
\end{center}
\end{figure}

\subsubsection{Effective couplings}
\label{Effective-couplings}

\nt After rotating the gauge-resonance sector to the mass eigenstates basis and implementing the relations in~\eqref{LR-gauge-couplings}, the effective couplings parametrising the cubic interactions of~\eqref{rho-W-Z} are
\beq
\begin{aligned}
&\mathit{g}_{\text{\textit{$\rho^+_1 W Z$}}}^{(1)}
=\mathit{g}_{\text{\textit{$\rho^+_1 W Z$}}}^{(2)}
=\mathit{g}_{\text{\textit{$\rho^+_1 W Z$}}}^{(3)}
=-\frac{g \xi  \sqrt{g^2+g'^2}}{4 \mathit{g}_{\rho _R}}\,,  \\[4mm]
&\mathit{g}_{\text{\textit{$\rho^+_2 W Z$}}}^{(1)}
=\mathit{g}_{\text{\textit{$\rho^+_2 W Z$}}}^{(2)}
=\mathit{g}_{\text{\textit{$\rho^+_2 W Z$}}}^{(3)}
=-\frac{g \xi  \left(\mathit{g}_{\rho _L}^2-g^2\right) \sqrt{\left(g^2+g'^2\right) \left(\mathit{g}_{\rho _L}^2-g^2\right)}}{4 \mathit{g}_{\rho _L}^4}\,,
\end{aligned}
\label{Effective-couplings-rho-W-Z}
\eeq

\nt while those of~\eqref{rho-WW}
\beq
\begin{aligned}
&\mathit{g}_{\text{\textit{$\rho^0_1 W W$}}}^{(1)}
=\mathit{g}_{\text{\textit{$\rho^0_1 W W$}}}^{(2)}=-\frac{g^2 \xi  \left(\mathit{g}_{\rho _L}^2-g^2\right){}^{3/2}}{4 \mathit{g}_{\rho _L}^4}\,,  \\[4mm]
&\mathit{g}_{\text{\textit{$\rho^0_2 W W$}}}^{(1)}
=\mathit{g}_{\text{\textit{$\rho^0_2 W W$}}}^{(2)}=-\frac{g^2 \xi  \left(\mathit{g}_{\rho _R}^2-g'^2\right){}^{3/2}}{4 \mathit{g}_{\rho _R}^4}
\end{aligned}
\label{Effective-couplings-rho-W-W}
\eeq

\nt and the ones in~\eqref{rho-W-Z-h}
\beq
\begin{aligned}
&g_{\rho^+_1 W h}=-\frac{1}{2} \mathit{f} g \mathit{g}_{\rho _R}\sqrt{\xi } \,,\quad\quad
&&\quad g_{\rho^0 Z_1 h}=\frac{1}{2} \mathit{f} \sqrt{\xi  \left(g^2+g'^2\right) \left(\mathit{g}_{\rho _L}^2-g^2\right)}\,,  \\[4mm]
&g_{\rho^+_2 W h}=\frac{1}{2} \mathit{f} g \sqrt{\xi  \left(\mathit{g}_{\rho _L}^2-g^2\right)}\,,\quad\quad
&&\quad g_{\rho^0 Z_2 h}=-\frac{1}{2} \mathit{f} \sqrt{\xi  \left(g^2+g'^2\right) \left(\mathit{g}_{\rho _R}^2-g'^2\right)}\,.
\end{aligned}
\label{Effective-couplings-rho-W-Z-h}
\eeq


\providecommand{\href}[2]{#2}\begingroup\raggedright\endgroup

\end{document}